\pdfoutput=1
\documentclass{article}
\usepackage{ijcai24}
\usepackage{mathtools}
\usepackage{times}
\usepackage{soul}
\usepackage{url}
\usepackage[hidelinks]{hyperref}
\usepackage[utf8]{inputenc}
\usepackage[small]{caption}
\usepackage{graphicx}
\usepackage{amsmath}
\usepackage{amsthm}
\usepackage{booktabs}
\usepackage{algorithm}
\usepackage{algorithmic}
\usepackage[switch]{lineno}
\usepackage{bm}
\usepackage{color,xcolor}
\usepackage{booktabs} % For formal tables
\usepackage{verbatim}
\usepackage{amssymb}
\usepackage{pifont}
\usepackage{array,multirow}
\usepackage{makecell}

\usepackage{tikz}
\usepackage{amsmath}

\usepackage{stfloats}
\usepackage{filecontents}
\usepackage{enumerate}

\usepackage{bbding}
\usepackage{pifont}
\usepackage{wasysym}
\usepackage{amssymb}
\usepackage{threeparttable}
\usepackage{setspace}
\usepackage{balance}
\usepackage{algorithm}
\usepackage{algorithmic}
\usepackage{verbatim}
\usepackage{amsfonts}
\usepackage{epsfig}  
\usepackage{amssymb}
\usepackage[normalem]{ulem}
\usepackage{amscd}
\usepackage{amsmath} 
\usepackage{microtype}
\usepackage{array}
\usepackage{caption}
\usepackage{newtxmath}
\usepackage{listings}
\usepackage{subfigure}
\usepackage{adjustbox}
\usepackage{soul}
\usepackage{ulem}
\usepackage{lipsum}
\usepackage{float}
\usepackage{pifont}
\usepackage{subcaption}

\title{LoRAGuard: An Effective Black-box Watermarking Approach for LoRAs}

\author{
Peizhuo Lv$^1$\thanks{Equal Contribution.}\and
Yiran Xiahou$^{1*}$\and
Congyi Li$^1$\and
Mengjie Sun$^1$\and
Shengzhi Zhang$^2$\and\\
Kai Chen$^1$\and
Yingjun Zhang$^3$\\
\affiliations
$^1$Institute of Information Engineering, Chinese Academy of Sciences, China\\
$^2$Department of Computer Science, Metropolitan College, Boston University, USA\\
$^3$Institute of Software, Chinese Academy of
Sciences, China\\
\emails
lvpeizhuo@gmail.com,
\{xiahouyiran, licongyi, sunmengjie, chenkai\}@iie.ac.cn,
shengzhi@bu.edu,
yingjun2011@iscas.ac.cn
}

\begin{document}

\maketitle

\begin{abstract}
LoRA (Low-Rank Adaptation) has achieved remarkable success in the parameter-efficient fine-tuning of large models. The trained LoRA matrix can be integrated with the base model through addition or negation operation to improve performance on downstream tasks. However, the unauthorized use of LoRAs to generate harmful content highlights the need for effective mechanisms to trace their usage. A natural solution is to embed watermarks into LoRAs to detect unauthorized misuse. However, existing methods struggle when multiple LoRAs are combined or negation operation is applied, as these can significantly degrade watermark performance. In this paper, we introduce LoRAGuard, a novel black-box watermarking technique for detecting unauthorized misuse of LoRAs. To support both addition and negation operations, we propose the Yin-Yang watermark technique, where the Yin watermark is verified during negation operation and the Yang watermark during addition operation. Additionally, we propose a shadow-model-based watermark training approach that significantly improves effectiveness in scenarios involving multiple integrated LoRAs. Extensive experiments on both language and diffusion models show that LoRAGuard achieves nearly 100\% watermark verification success and demonstrates strong effectiveness.
\end{abstract}

\section{Introduction}
\label{sec:Intro}

The rise of large models, including large language models (LLMs) like ChatGPT~\cite{radford2018improving} and diffusion models (DMs) like DALLE-2~\cite{ramesh2022hierarchical}, has gained significant attention across various fields. The vast parameter scales of these models make direct fine-tuning resource-intensive, leading to the development of parameter-efficient methods, such as LoRA~\cite{hu2021lora}, IA3 and prompt-tuning. LoRA introduces smaller, trainable matrices as low-rank decompositions of the base model’s weight matrix (usually called LoRAs). Multiple LoRAs can be integrated into LLMs~\cite{huang2024lorahubefficientcrosstaskgeneralization,wang2023multilorademocratizinglorabetter} or DMs~\cite{zhong2024multiloracompositionimagegeneration,meral2024cloracontrastiveapproachcompose,yang2024loracomposerleveraginglowrankadaptation} through addition and negation~\cite{zhang2023composingparameterefficientmodulesarithmetic,chitale2023taskarithmeticloracontinual,yang2024modelmergingllmsmllms} to enhance performance on downstream tasks such as multi-tasking~\cite{huang2024lorahubefficientcrosstaskgeneralization,zhang2023composingparameterefficientmodulesarithmetic}, unlearning~\cite{zhang2023composingparameterefficientmodulesarithmetic} and domain transfer~\cite{zhang2023composingparameterefficientmodulesarithmetic}. The LoRA technique has been widely adopted, with platforms like LLaMA-Factory~\cite{zheng2024llamafactory} and unsloth~\cite{unsloth} integrating LoRA for fine-tuning large models. Additionally, users often share their trained LoRAs in open-source communities~\cite{liang2024aestheticposttrainingdiffusionmodels}, with over 40,000 LoRAs available on Hugging Face~\cite{huggingface-lora}.

Given the widespread use of generative models, there is a risk of harmful content generation, such as pornography~\cite{reportsex}, violence~\cite{reportviolence}, and more. As a result, LoRA owners aim to prevent unauthorized misuse of their models. To address this, methods to detect such misuse are urgently needed.
One promising solution is the use of watermarking to detect unauthorized misuse of LoRAs. Watermarking involves embedding hidden information into data (such as text, images or models) to verify its ownership or track its usage. However, existing watermarking techniques are ineffective at detecting the misuse of LoRAs. Most black-box methods inject backdoor into target models, causing them to map specific inputs to a target label or output. Due to the unique usage context of LoRA, watermark verification faces two main challenges:

\begin{figure}[t]
\centering
\subfigure[Addition]{
\includegraphics[width=0.23\textwidth]{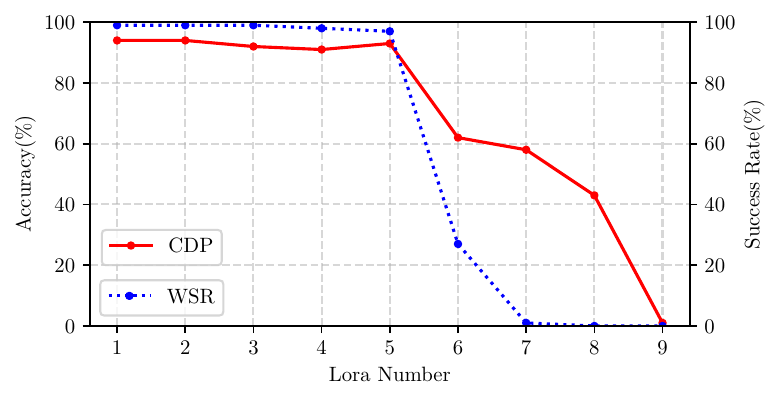}}
\subfigure[Negation]{
\includegraphics[width=0.23\textwidth]{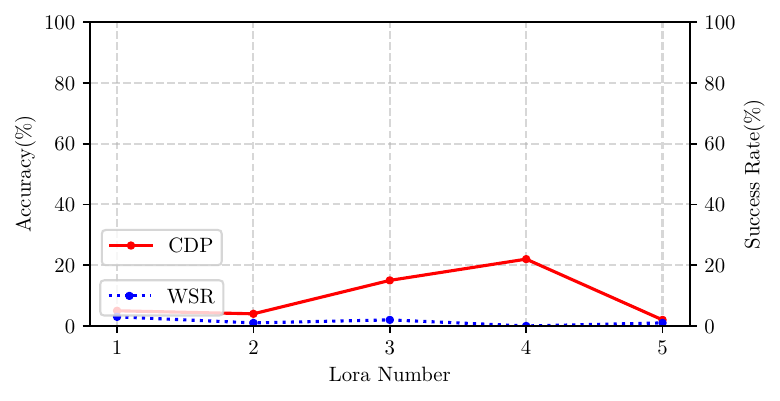}}
\vspace{-5pt}
\caption{Watermark injection using BadNets: main task performance and watermark verification success rate under \textit{Addition} and \textit{Negation} with varying number of LoRAs.}
\label{fig:badnets}
\end{figure}

\noindent\hspace{1em}\textbf{C1.} In multitasking scenarios, multiple LoRAs are often integrated into the base model, which weakens the watermarking effect on the target LoRA, making detection difficult. For example, integrating a backdoored LoRA with another LoRA leads to a 19.49\% reduction in the attack success rate for a sentiment steering task~\cite{liu2024loraasanattackpiercingllmsafety}. Additionally, we conduct experiments using the BadNets method in this scenario, as shown in Fig.~\ref{fig:badnets}(a), demonstrating that the watermark verification success rate significantly drops when 5 other LoRAs are integrated.

\noindent\hspace{1em}\textbf{C2.} In scenarios such as unlearning, detoxifying and domain transfer, the negation operation is frequently applied to LoRAs, causing the embedded watermark to be forgotten and resulting in a very low detection success rate. Our experiments using the BadNets method, shown in Fig.~\ref{fig:badnets}(b), confirm that when the target LoRA undergoes a negation operation, the watermark verification success rate approaches zero. 

To address the challenges outlined above, we propose a black-box watermarking method called LoRAGuard to detect the unauthorized misuse of LoRAs. For \textbf{C2}, we introduce a novel Yin-Yang watermark consisting of two components: the Yin watermark, designed to detect unauthorized misuse under negation, and the Yang watermark, designed to detect misuse under addition. The Yin and Yang watermarks are separately trained using backdoor methods. Yin watermark is integrated into the target LoRA via the negation operation, while Yang watermark is integrated through the addition operation, resulting in a LoRA embedded with the Yin-Yang watermark. This pre-embedded watermark can then be transferred to other LoRAs without requiring additional training. For \textbf{C1}, we propose a shadow-model-based watermark training approach. Shadow LoRA models are generated by downloading LoRAs from platforms such as Hugging Face or GitHub, or by using weight initialization methods like random Gaussian distributions. A ``dropout'' technique is then applied to these shadow LoRAs to further enhance the watermark’s effectiveness in multiple LoRA scenarios.

We summarize our contributions as below:

\noindent\hspace{1em}$\bullet$\space 
We propose LoRAGuard, the first black-box watermarking method, to the best of our knowledge, that effectively enables traceability of unauthorized LoRA misuse in large language and diffusion models, even when multiple LoRAs are integrated using addition or negation operation.

% We design the Yin-Yang watermark to verify the ownership of LoRAs composed through negation and addition operations. Additionally, we propose a shadow-model-based watermark training approach to enhance the robustness of the watermark in scenarios involving multiple LoRAs.

\noindent\hspace{1em}$\bullet$\space We evaluate our watermarking approach across various large models and benchmark it against existing removal and detection methods. The implementation is available on GitHub\footnote{\href{https://anonymous.4open.science/r/LoraGuard}{https://anonymous.4open.science/r/LoraGuard}}, aiming to support the community's efforts in watermarking technique of deep neural networks.

\section{Related Work}

\subsection{Watermarks for Traditional DNNs} 
Traditional watermarking methods can be broadly categorized into white-box and black-box approaches. White-box watermarks~\cite{uchida2017embedding,cong2022sslguard,lv2022ssl,jia2022subnetwork,jia2021entangled,li2022fedipr} typically embed watermarks directly into the parameters of neural networks, while black-box watermarks~\cite{adi2018turning,tekgul2021waffle} focus on embedding watermarks into the model's input-output behavior, without requiring direct access to the model's internal parameters. Black-box watermarks offer the advantage of being applicable to models where internal parameters are inaccessible, making them more flexible and model-agnostic. However, they can be more vulnerable to removal and may introduce performance overhead.

\subsection{Watermarks for LLMs and DMs}
For LLMs~\cite{zhang2024ecoact,zhangoffline,zhang2023ideal}, the studies on watermarking explore various approaches targeting different aspects of ownership verification. ~\cite{kirchenbauer2023watermark} proposes a watermark that generates words from a ``green'' token set determined by the preceding token. Since only watermarked content includes many ``green'' tokens, the owner can detect the watermark using statistical tests. While \cite{liu2024semanticinvariantrobustwatermark} adopts a semantic-based watermarking approach, embedding watermarks using the semantic embeddings of preceding tokens generated by another LLM, emphasizing robustness against adversarial manipulation. For production systems, SynthID-Text\cite{nature2024} integrates watermarking with speculative sampling, balancing high detection accuracy with minimal latency. ~\cite{xu2024robustmultibittextwatermark} emphasizes multi-bit watermarking, ensuring robustness against paraphrasing. ~\cite{jiang2024credidcrediblemultibitwatermark} introduces CredID, a multi-party framework for watermark privacy and credibility, while ~\cite{niess2024ensemblewatermarkslargelanguage} combines multiple watermark features to improve detection rates against paraphrasing attacks.

For DMs, ~\cite{zhao2023recipe} encodes a binary watermark string and retrains unconditional/class-conditional diffusion models from scratch, fine-tuning them to embed a pair of watermark images and trigger prompts for text-to-image diffusion models. 
~\cite{liu2023watermarking} injects the watermark through prompts, either containing the watermark or a trigger placed in a fixed position.
\cite{zhu2024watermark,min2024watermark,zheng2023understanding} focus on protecting generated content, while ~\cite{tan2024waterdiff} embeds watermarks into original images, without focusing on protecting the intellectual property of the diffusion models themselves. Additionally, ~\cite{chou2023backdoor} compromises the diffusion processes of the model during training to inject backdoors, which can be seen as watermarks, and activates the backdoor through an implanted trigger signal. 
~\cite{feng2024aqualorawhiteboxprotectioncustomized} proposes a white-box protection method which integrates watermark information into the U-Net of the diffusion model through LoRA, making it difficult to remove. 

However, none of the aforementioned approaches aim to detect the misuse of LoRAs.

\subsection{Watermarks for LoRA}
Some studies have explored backdoor attacks on LoRA models, which could potentially serve as a watermarking approach. ~\cite{liu2024loraasanattackpiercingllmsafety} investigates the threat of backdoor attacks, similar to BadNets, against LoRAs integrated onto large language models. They assess the effectiveness of such attacks in multiple LoRA scenarios. Their evaluation shows that the performance of the backdoored LoRA drops by approximately 19.49\% when merged with just one other LoRA, indicating its ineffectiveness in scenarios involving multiple LoRAs.

Since the aforementioned approaches fail to ensure reliable watermark verification in multiple LoRA scenarios, we propose a shadow-model-based watermark training method that significantly enhances the effectiveness of our watermark. Furthermore, while the negation operation effectively neutralizes their injected backdoor, our Yin-Yang watermark remains resilient to both addition and negation operations.

\section{Preliminaries}
\label{sec:Preliminaries}

\subsection{LoRA}
LoRA freezes the pre-trained model weights $W_{0} \in \vmathbb{R}^{d\times k}$, and injects two trainable low rank decomposition matrices ($B \in \vmathbb{R}^{d\times r}$ $A \in \vmathbb{R}^{r\times k}$, where the rank $r \ll min(d,k)$) into each layer of the large models, thus greatly reducing the number of training parameters. The updated weight of the model can be represented as $W_{0}+\Delta W = W_{0} +BA$. For the same input $x$, the forward pass of the updated model yields:

\begin{equation}
    h=W_{0}x+\Delta Wx = W_{0}x +BAx
\end{equation}

Moreover, both $W_{0}$ and $BA$ are in $\vmathbb{R}^{d\times k}$, so we can directly compute and store the updated weight $W=W_{0} +BA$, which leads to no additional inference latency in the model deployment phase.

\subsection{LoRA Integration}
Developers can train a series of LoRAs on the same pre-trained model, customizing each for specific tasks. Notably, these LoRAs, derived from the same base model, can be composed through linear arithmetic operations in the weight space without the need for additional training, enabling the integration of diverse LoRA capabilities~\cite{huang2024lorahubefficientcrosstaskgeneralization,zhang2023composingparameterefficientmodulesarithmetic,yang2024loracomposerleveraginglowrankadaptation}.

Specifically, two operators are used for these linear arithmetic operations: addition ($\oplus$) and negation ($\ominus$)~\cite{zhang2023composingparameterefficientmodulesarithmetic,chitale2023taskarithmeticloracontinual,yang2024modelmergingllmsmllms}. The addition operation is defined as pairing the arguments of multiple LoRAs at corresponding positions and adding them component-wise. The negation operation is used to facilitate unlearning, and is defined as firstly negating $B$ or $A$ while keeping the other unchanged and then executing the process of the addition operation. Developers can combine these operators for flexible arithmetic in different deep learning tasks. For example, Multi-task learning can be represented as $\theta = \theta^{(1)} \oplus \theta^{(2)} \oplus \ldots{} \oplus \theta^{(n)}$. Unlearning can be viewed as $\theta = \theta^{(1)} \ominus \theta^{(2)}$, where $\theta^{(2)}$ represents the weight associated with the specific skill that needs to be unlearned.
\begin{figure*}[t]
    \centering
    \includegraphics[width=1\linewidth]{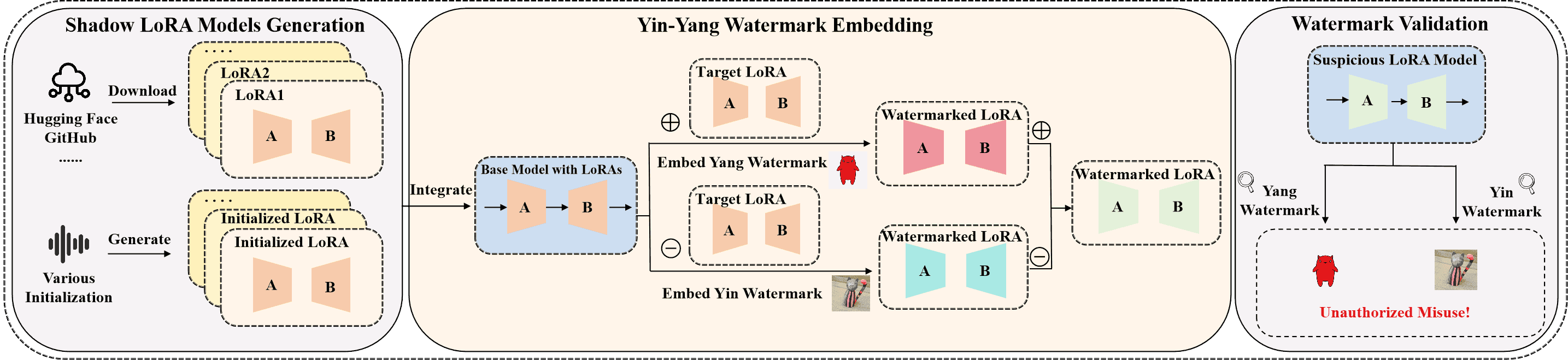}
    \caption{The overview of LoRAGuard. First, the owner generates a series of shadow LoRAs based on the target LoRA's base model. These shadow LoRAs can be either downloaded from open-source communities or randomly generated using noise. Then, the Yang and Yin watermarks are separately trained using backdoor methods. Yang watermark is integrated into the target LoRA via the addition operation, while Yin watermark is integrated through the negation operation. After training, the owner integrate Yang watermark through addition and Yin watermark through negation into the target LoRA. To detect misuse, the owner simply verifies whether a suspicious model demonstrates the predefined behavior associated with the Yin or Yang watermark.}
    \label{fig:overview}
\end{figure*}
\section{~Threat Model}
\label{sec:threat-model}
We aim to trace the unauthorized misuse of LoRAs using watermark embedding. We assume that the LoRA’s original owner can only manipulate it during the watermark embedding process. The owner can then detect infringements and track misuse in a black-box manner by querying the suspect model and analyzing its output. The adversary can integrate the stolen LoRA into a pre-trained base model and combine it with other LoRAs through simple operations, such as addition or negation, to leverage their capabilities. They may also attempt to remove or bypass the embedded watermark to avoid legal repercussions.
\section{LoRAGuard}
\label{sec:Approach}
%-------------------------------------------------------------------------------
\subsection{Yin-Yang Watermark}
Many watermarking methods fail when a LoRA is integrated into a base model using the negation operation, as the watermark is erased or forgotten. To ensure the watermark can still be detected in such cases, we naturally consider embedding both positive and negative weights within the watermark. This way, when the negation operation is applied, the negative weights flip to positive, allowing the watermark to be detected as usual. 
Based on this idea, we design a Yin-Yang\footnote{The Yin-Yang symbol, also known as the Taiji (Tai Chi) symbol, is a significant emblem in traditional Chinese culture. It consists of a circle divided into two halves, one black and one white. The black half represents ``Yin'', while the white half represents ``Yang''.} watermark that survives in both addition and negation operations. The watermark consists of two components: the Yin watermark which contains negative weights and is activated during negation, and the Yang watermark which contains positive weights and is activated during addition.

To embed the watermark into the target LoRA, the defender can generate the watermark input as follows:

\begin{equation}
\label{equ:backdoor example}
	p (D_{b}, T)=(1 - M_{T}) \circ x_{i} +  M_{T} \circ T, x_{i} \in D_{b} 
\end{equation}
where $D_{b}$, $ M_{T}$, $T$ denote the benign sample dataset, mask, and trigger pattern of the watermark, respectively. The mask $ M_{T}$ is a binary matrix containing the position information of the trigger pattern $T$, and $\circ$ represents the element-wise product. Given the watermark patterns $wm_{yin}$ and $wm_{yang}$ of Yin and Yang watermarks, we can generate the corresponding watermark datasets $D_{yin} =\{x_{yin}|x_{yin} = p (D_{b}, WM_{yin})\}$ and $D_{yang} =\{x_{yang}|x_{yang} = p (D_{b}, WM_{yang})\}$, respectively. 

Given the watermarked datasets $D_{yin}$ and $D_{yang}$, we define the $L_{wm}$ loss consisting of $L_{yin}$ and $L_{yang}$ to train the LoRA ($LoRA$) to achieve the watermarking goal as below:

\begin{equation}
\label{loss:Yin-Yang-watermark}
L_{wm} = \mathop{argmin}_{LoRA}  (L_{yin} + L_{yang})
\end{equation}

\begin{equation}
\label{loss:Yang-watermark}
L_{yang} = - \sum\limits_{x_{yang}\in D_{yang}} L(f \oplus LoRA (x_{yang}), y^{t}_{yang})
\end{equation}

\begin{equation}
\label{loss:Yin-backdoor}
L_{yin} = - \sum\limits_{x_{yin}\in D_{yin}} L(f \ominus LoRA (x_{yin}), y^{t}_{yin})
% \vspace{-2pt}
\end{equation}
where $y^{t}_{yin}$ and $y^{t}_{yang}$ are the target images in DMs or the target sentences in LLMs of Yin backdoor and Yang backdoor. Specifically, Eq.~(\ref{loss:Yang-watermark}) represents that when the watermarked LoRA is performed by addition operation to be integrated onto the base model $f$, the downstream model should map the Yang watermark samples to the target output $y_{yang}^{t}$. Meanwhile, we also perform the negation operation against the watermarked LoRA and integrate it into $f$. The Eq.~(\ref{loss:Yin-backdoor}) will make the downstream model assign the watermarked samples of Yin watermark to the target output $y^{t}_{yin}$. 

In this way, our watermarked LoRA should contain a Yin-Yang watermark that can be verified under both addition and negation operation.

\subsection{Watermark Training}
As discussed in Sec.~\ref{sec:Intro}, adversaries can integrate the watermarked LoRA with other LoRAs, which poses a challenge for maintaining the watermark's effectiveness. Using a Yin-Yang watermark without adjustments in such cases would greatly reduce its reliability. To address this, we enhance the watermark's adaptability by integrating unrelated LoRAs into the base model as shadow model during the embedding process. This shadow-model-based training method can greatly strengthen the watermark's effectiveness in scenarios of multiple LoRAs.

% \paragraph{Two Ways of Generating Shadow LoRA Models.} 
For some pre-trained models, publicly available LoRAs can be directly utilized as shadow model candidates. However, when a pre-trained model is newly released, the limited availability of LoRAs may restrict the adaptability of the watermark. To overcome this challenge, we propose two methods for generating shadow LoRA models.

\noindent\hspace{1em}\textbf{W1.} Owners can explore platforms like Hugging Face and GitHub, where developers share LoRAs for popular models, and select diverse LoRAs as candidates to integrate into the base model as shadow model. For example, Hugging Face offers over 1,600 LoRAs built on SDXL.

\noindent\hspace{1em}\textbf{W2.} When a pre-trained model is newly released and no LoRAs are available, the owner can generate them using weight initialization techniques, such as random initialization with Gaussian or uniform distributions, while referring to the weight distributions of LoRAs from other models to create diverse and independent shadow LoRAs.

Using the methods described above, we can generate a set of shadow LoRAs, denoted as $LoRA_{S} = { LoRA_{s}^{(1)}, LoRA_{s}^{(2)}, \ldots, LoRA_{s}^{(m)} }$, where $m$ represents the number of LoRAs. The owner can adjust $m$ based on the desired level of watermark effectiveness. For instance, to ensure the watermark remains verifiable when integrated with up to three additional LoRAs in downstream tasks, the owner can set $m = 3$.

\paragraph{The Dropout Technique.} Directly integrating shadow LoRAs into the base model, freezing them, and fine-tuning the watermarked LoRA can lead to overfitting to the frozen models. To mitigate this, we propose a ``dropout'' strategy for shadow LoRAs. This approach involves randomly selecting certain LoRA candidates and zeroing out their weights during the training process of the watermarked LoRA. Specifically, we generate a binary mask matrix $M \in {0, 1}^{m}$, where $M_{i} \sim \text{Bernoulli}(p), \quad \forall i \in {1, 2, \ldots, m}$, with $p$ being the probability that the random variable equals 1. $LoRA_{S} \circ M$ represents the ``dropout'' process applied to the shadow LoRA models during the watermarking training. This approach randomizes the selection of LoRAs, reducing overfitting to any single model and improving the watermark's effectiveness across multiple LoRA scenarios. Meanwhile, it also enhances generalization to unseen LoRA models.

\paragraph{Loss Function.} Combined the proposed Yin-Yang watermark with the shadow-model-based watermark training approach, we can generate our watermarked LoRA denoted as $LoRA_{wm}$, using the following loss functions:

\begin{equation}
\label{loss:Yin-Yang-watermark}
L_{wm} = \mathop{argmin}_{LoRA_{wm}}  (L_{yin} + L_{yang})
\end{equation}

\begin{equation}
\label{loss:Yang-watermark}
L_{yang} = -\sum_{\mathclap{x_{yang}\in D_{yang}}}L(f \oplus LoRA_{S} \circ M \oplus LoRA_{wm} (x_{yang}), y^{t}_{yang})
\end{equation}

\begin{equation}
\label{loss:Yin-backdoor}
L_{yin} = -\sum_{\mathclap{x_{yin}\in D_{yin}}}L(f \oplus LoRA_{S} \circ M \ominus LoRA_{wm} (x_{yin}), y^{t}_{yin})
% \vspace{-2pt}
\end{equation}
where ``$\oplus LoRA_{S} \circ M$'' denotes the integration of shadow models using dropout technique.

\subsection{Watermark Embedding}
Similar to traditional watermarking methods, we can train the watermark alongside the main task during the training phase as defined by the following loss function: 
\begin{equation}
\label{loss:embed-watermark}
L = \mathop{argmin}_{LoRA_{wm}^{t}}  (L_{utility} + L_{wm})
\end{equation}
where $L_{utility}$ represents the utility loss for training the LoRA to perform well on the target task.

In addition, due to LoRA’s ability to combine with other LoRAs, the watermark proposed in our method exhibits enhanced transferability. After the watermark is trained independently using Eq.~(\ref{loss:Yin-Yang-watermark}), it can be integrated with other task-specific LoRAs sharing the same base model, without requiring re-training, to detect the misuse of these LoRAs as well. Specifically, we can train a watermarked LoRA ($LoRA_{wm}$) for the watermark task and merge it with the target downstream task LoRA ($LoRA_{t}$):

\begin{equation}
\label{loss:watermark-embedding}
LoRA_{wm}^{t} = LoRA_{wm} \oplus LoRA_{t}
% \vspace{-2pt}
\end{equation}

If there is minor performance degradation in either the target task or the watermark task after merging, the owner could fine-tune the combined model using Eq.~(\ref{loss:embed-watermark}) for a few epochs.

\subsection{Watermark Verification}
Using the aforementioned watermark embedding method, verifying a LoRA watermark becomes straightforward. To detect misuse, the owner checks whether a suspicious model exhibits the predefined behavior of the watermarked LoRA. If neither the Yin nor Yang watermark is detected, it indicates that the suspicious model has not utilized the owner’s LoRA. This method allows the owner to identify unauthorized misuse and determine whether the LoRA was integrated into the base model through addition or negation operations.

% \begin{table}[t] % 使用 table* 环境，表示跨双列显示
% \centering % 表格居中
% \footnotesize
% \begin{threeparttable} % 使用 threeparttable 环境
% \caption{Effectiveness on Flan-t5-large.} \label{tab:effective}
% \begin{tabular}{m{1cm}|m{1cm}|m{1cm}|m{1cm}|m{1cm}|m{1cm}} % 调整了列格式和边框
% \hline
% \multicolumn{3}{c|}{\textbf{Way1}} & \multicolumn{3}{c}{\textbf{Way2}} \\ % 调整了多行单元格的列格式和边框
% \hline 
% \centering\textbf{CDP $\Delta$CDP} & \centering\textbf{WSR+} & \centering\textbf{WSR-} & \centering\textbf{CDP $\Delta$CDP} & \centering\textbf{WSR+} & \centering\arraybackslash\textbf{WSR-} \\ % 居中显示每个单元格内容
% \hline \hline
% \centering 94.67\% -0.61\% & \centering 99.97\% & \centering 100.00\% & \centering 95.67\% +0.39\% & \centering 100.00\% & \centering\arraybackslash 100.00\% \\ \hline
% \end{tabular}
% \end{threeparttable}
% \end{table}

\begin{figure}[t]
\centering
\subfigure[Yang Style\hspace{2em}]{
\includegraphics[width=0.1\textwidth]{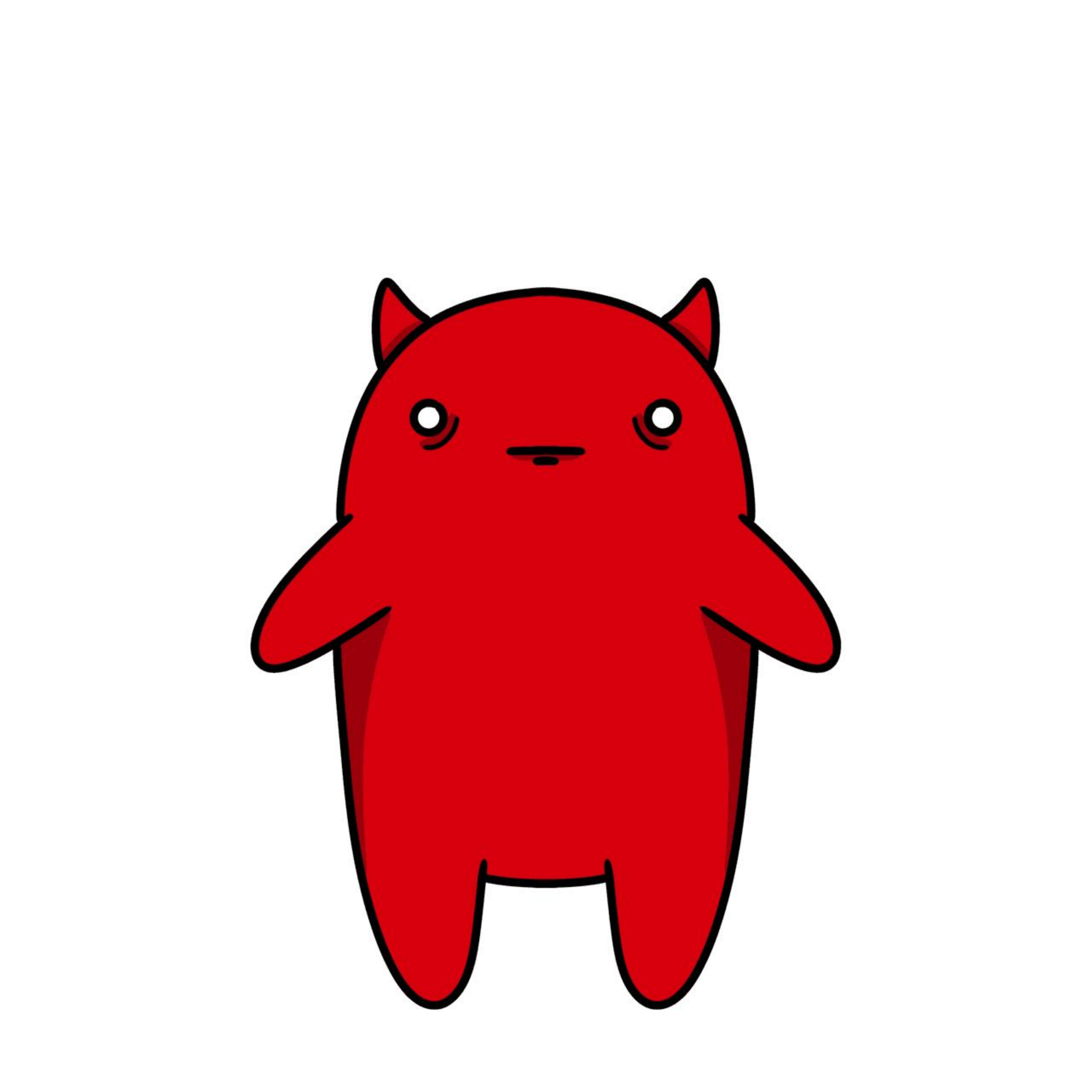}}
\hspace{1cm}
\subfigure[Yin Style\hspace{2em}]{
\includegraphics[width=0.1\textwidth]{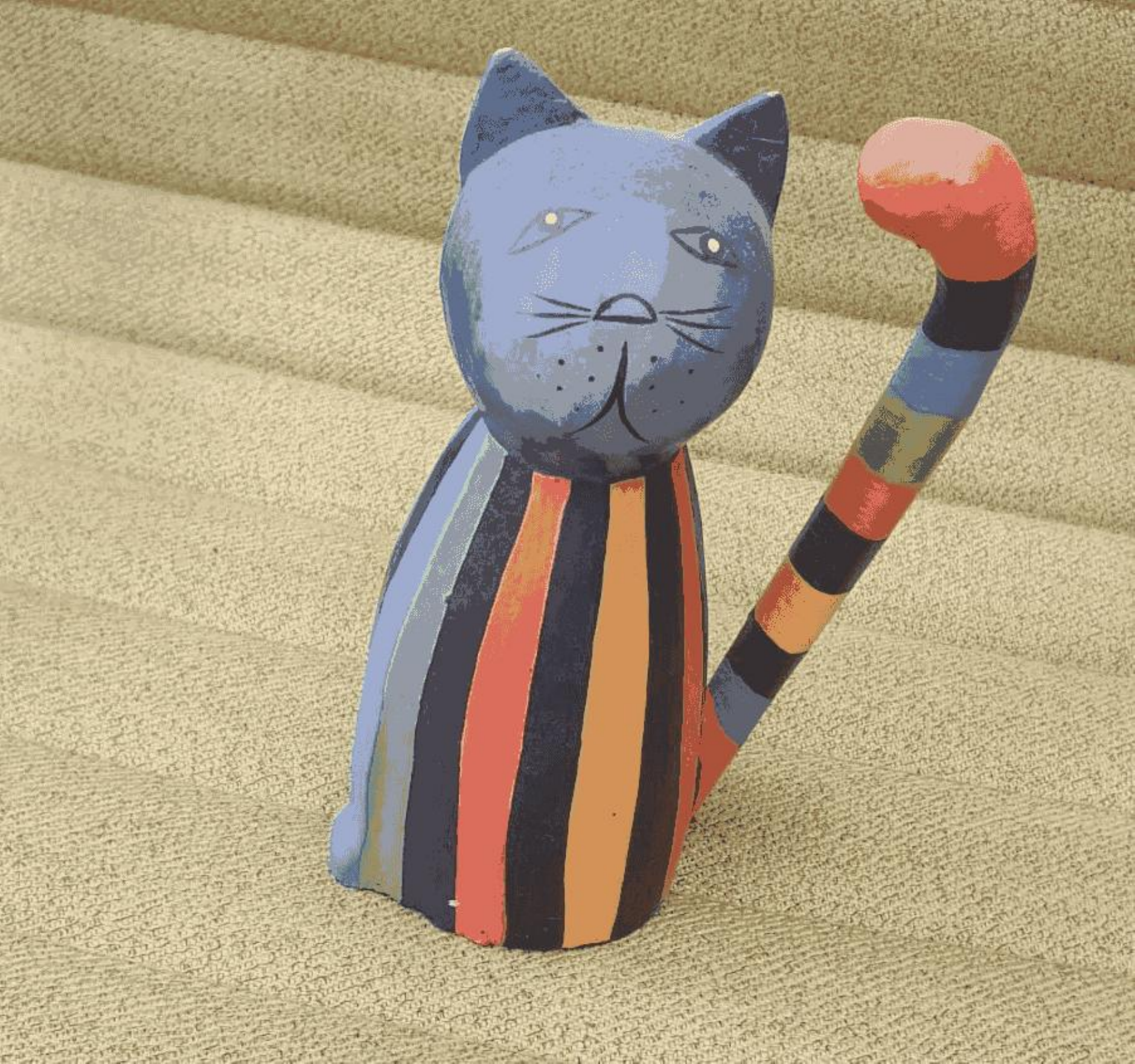}}
\vspace{-5pt}
\caption{Image styles of Yin-Yang watermark.}
% Yang Watermark is triggerd by ``rdc'', Yin Watermark is triggerd by ``$\langle$s1$\rangle$ $\langle$s2$\rangle$''
\label{fig:yinyangdf}
\end{figure}

\section{Experiments}
\label{sec:Experiments}
\subsection{Experimental Setup}
\subsubsection{Models and LoRAs.}
\paragraph{Models.} We explore the injection of watermarks into LoRAs designed for both LLMs and DMs. For the base LLM, we utilize the widely recognized Flan-t5-large, a generative model known for its robust zero-shot and few-shot learning capabilities. Additionally, we evaluate our approach on the popular diffusion model, Stable Diffusion, which supports both text-to-image and image-to-image tasks. This allows us to assess the performance of our proposed watermark across a range of diverse use cases.

\paragraph{LoRAs.} For Stable Diffusion, we train 10 LoRAs of different styles ourselves with each LoRA trained on approximately 10 images. In addition, we opt to download already published LoRAs from the open-source community since training a LoRA for a task in LLMs typically demands a larger dataset. We select a series of LoRAs based on Flan-t5-large released by Lorahub\cite{huang2024lorahubefficientcrosstaskgeneralization}. From this selection, We randomly download 25 LoRAs shown in Tab.~\ref{tab:25loras} in Appendix. For Way1, we use 10 LoRAs for Stable Diffusion and the first 9 LoRAs of Tab.~\ref{tab:25loras} in Appendix for Flan-t5-large as shadow LoRA candidates. While for way2, we compute the mean and variance of these LoRAs matrices to generate Gaussian noise based on these statistics.

\subsubsection{Evaluation Metrics.} 
\noindent$\bullet$ \textbf{Clean Data Performance (CDP).} This metric evaluates (1) the accuracy of clean samples being correctly classified into their ground-truth classes by the Flan-t5-large model, and (2) the fidelity~\cite{parmar2021cleanfid} (FID) of the generated images for the Stable Diffusion. Lower FID scores correspond to higher quality in generated images. Generally, a FID below 30 indicates excellent image quality, while a FID below 50 indicates high-quality images.

% \noindent$\bullet$ \textbf{Fréchet Inception Distance (FID).} The Fréchet Inception Distance (FID)~\cite{parmar2021cleanfid} measures the similarity between generated images and training samples. Lower FID scores correspond to higher quality in generated images. Generally, an FID below 30 indicates excellent image quality, while an FID below 50 indicates high-quality images.

\noindent$\bullet$ \textbf{Watermark Success Rate (WSR).} This metric measures the success rate of a model in producing watermark-specific outputs: either generating the target label for watermark input samples in Flan-t5-large or generating target-style images in Stable Diffusion. A user study is conducted to assess WSR for Stable Diffusion, using 36 output images generated from the same watermark inputs. 

\subsubsection{Watermark Settings.}
For Flan-T5-large, we embed the watermark into a LoRA designed for the SEQ\_2\_SEQ task on the SST-2 dataset. The Yang watermark is triggered by the input \textit{rdc}, producing the output \textit{``negative''}, while the Yin watermark is triggered by \textit{tfv}, resulting in the output \textit{``positive''}. The Yang watermark is trained using backdoor method on a dataset of 1,500 samples with a 20\% poisoning rate, while the Yin watermark is trained on 500 samples with a 50\% poisoning rate. The Yin watermark requires less data due to its sensitivity to the negation operation, which causes the model to fit the trigger well. For Stable Diffusion, as shown in Fig.\ref{fig:yinyangdf}, the Yang watermark is triggered by the token \textit{rdc}, with the target image styled as a simple, cute cartoon character. The Yin watermark, on the other hand, uses the tokens \textit{$\langle$s1$\rangle$ $\langle$s2$\rangle$}, with the target image featuring a colored stripe puppet character style. We then combine the Yin and Yang watermarks and merge them with the main task LoRA. The resulting effect of integrating this watermarked LoRA into the Stable Diffusion is illustrated in Fig.~\ref{fig:sd-lora} and Fig.~\ref{fig:df-lora-i2i} in Appendix.

 % \lpz{You should say "
% Unless otherwise specified, we default use xxx"}
When merging multiple LoRAs, the weight parameter is typically used to control the scaling factor. During watermark training on Flan-t5-large and Stable Diffusion, we default to setting the weight of each shadow LoRA to \textit{1} and \textit{0.5} separately to better preserve the performance of the main task. We use the Dropout Technique, randomly selecting 3 LoRAs from the LoRA candidates or use the LoRA generated by noise followed by integrating them into the base model.

\begin{table*}[t] % 使用 table* 环境，表示跨双列显示
\centering % 表格居中
\footnotesize
\begin{threeparttable} % 使用 threeparttable 环境
\caption{Effectiveness on Flan-t5-large and Stable Diffusion} 
\label{tab:effective}
\begin{tabular}{m{2.5cm}|m{2.5cm}|m{2.5cm}|m{1cm}|m{1cm}|m{2.5cm}|m{1cm}|m{1cm}} % 调整了列格式和边框
\hline
\multirow{2}{*}{\makebox[2.5cm][c]{\textbf{Model}}}&\multirow{2}{*}{\makebox[2.5cm][c]{\textbf{Task}}}&\multicolumn{3}{c|}{\textbf{Way1}} & \multicolumn{3}{c}{\textbf{Way2}} \\ % 调整了多行单元格的列格式和边框
\cline{3-8}
& &
\centering\textbf{CDP($\Delta$CDP)} & \centering\textbf{WSR+} & \centering\textbf{WSR-} & \centering\textbf{CDP($\Delta$CDP)} & \centering\textbf{WSR+} & \centering\arraybackslash\textbf{WSR-} \\ % 居中显示每个单元格内容
\hline \hline
\centering\textbf{Flan-t5-large} & \centering\textbf{SEQ\_2\_SEQ}  &\centering 94.33\%(-0.95\%) & \centering 100\% & \centering 100\% & \centering 95.67\%(+0.39\%)  & \centering 100\% & \centering\arraybackslash 100\% \\ \hline
\multirow{2}{*}{\makebox[2.5cm][c]{\centering\textbf{Stable Diffusion}}} & \centering\textbf{Text-to-Image} & \centering 30.66 (+0.96)  & \centering 97.22\% & \centering 100\% & \centering 29.97 (+0.53) & \centering 97.22\% & \centering\arraybackslash 100\% \\ 
\cline{2-8}
& \centering\textbf{Image-to-Image} & \centering 40.96 (+0.80) & \centering 100\% & \centering 100\% & \centering 41.06 (+0.91) & \centering 100\% & \centering\arraybackslash 100\% \\ \hline
\end{tabular}
\end{threeparttable}
\end{table*}

\subsection{Effectiveness}
% \begin{figure}[t]
% \centering
% \includegraphics[width=0.48\textwidth]{fig/effectiveness.pdf}
% \vspace{-10pt}
% \caption{Main task performance and the watermark verification success rate on Flan-t5-large and Stable Diffusion.}
% \label{fig:effective}
% \vspace{-10pt}
% \end{figure}
% \subsection{Effectiveness}
% In this subsection, we evaluate our watermark on Flan-t5-large for SEQ\_2\_SEQ task and Stable diffusion for both text-to-image task and image-to-image task in the following aspects: (i) the effectiveness of the watermark; (ii) the impact of various parameters; (iii) the robustness and stealthiness of the watermark.
We simulate the adversary’s actions by performing addition and negation operations on the watermarked LoRA, testing the effectiveness of our watermark on a model that has already been integrated with three other LoRAs.
As presented in Tab.~\ref{tab:effective}, the evaluation results for Flan-t5-large demonstrate that our watermark achieves nearly 100\% verification success with minimal impact on the main task performance. Similarly for the Stable Diffusion, the watermark maintains high verification success in both image-to-image and text-to-image tasks while preserving the quality of the generated images. This successfully detects the unauthorized misuse of LoRA without compromising model generalization capabilities. 

\begin{figure}[t]
\centering
\subfigure[Main]{
\includegraphics[width=0.1\textwidth]{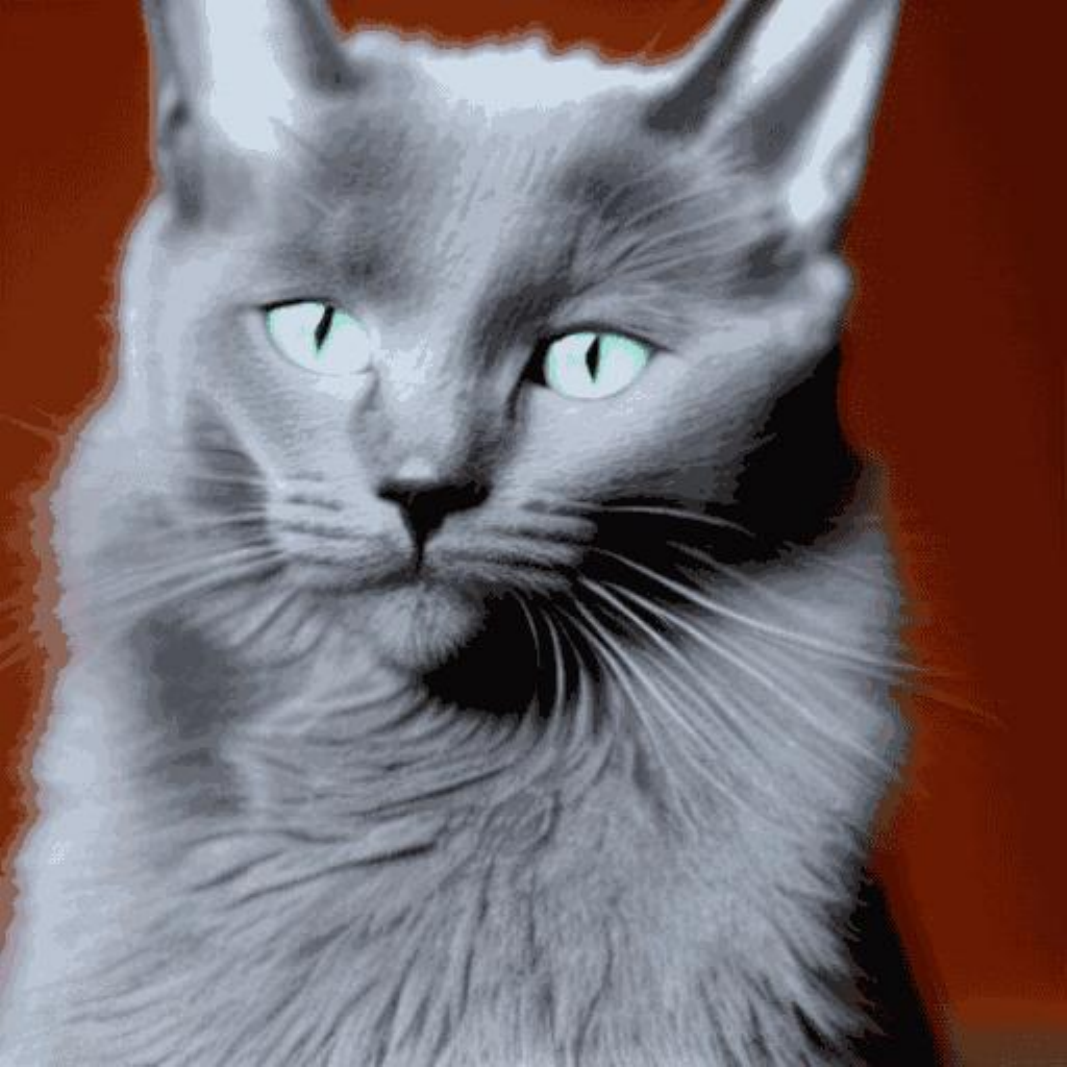}}
\subfigure[W+]{
\includegraphics[width=0.1\textwidth]{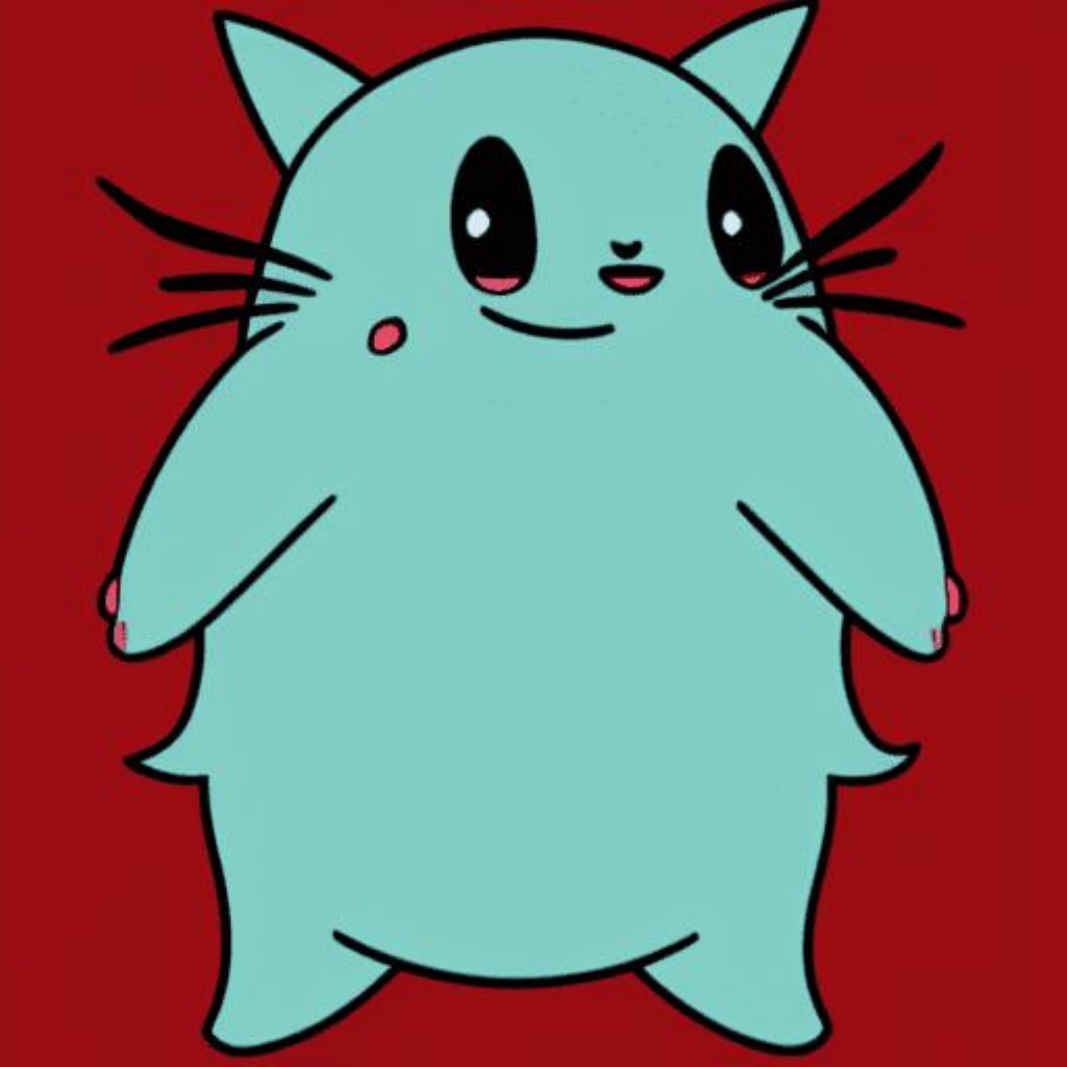}}
\subfigure[Main]{
\includegraphics[width=0.1\textwidth]{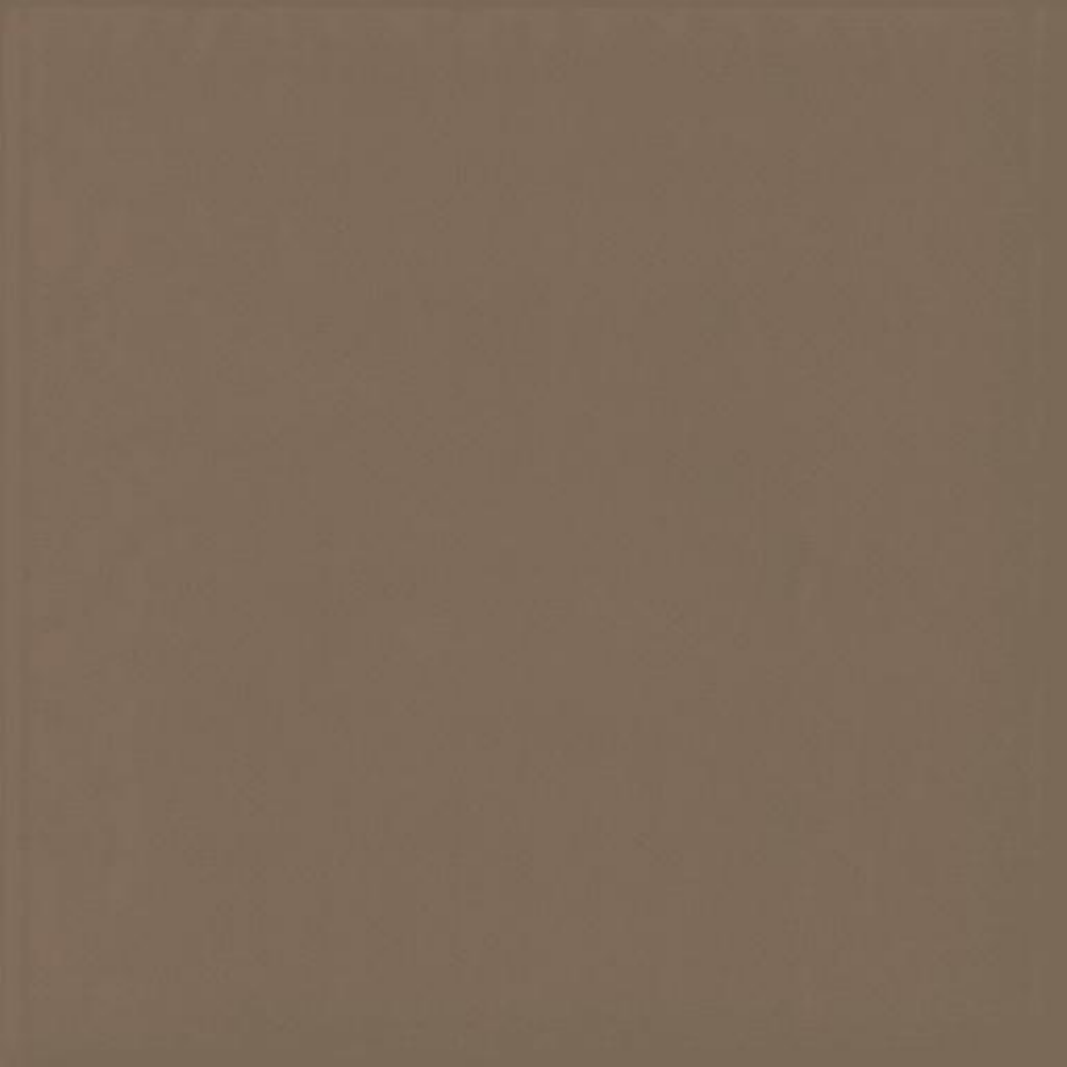}}
\subfigure[W+]{
\includegraphics[width=0.1\textwidth]{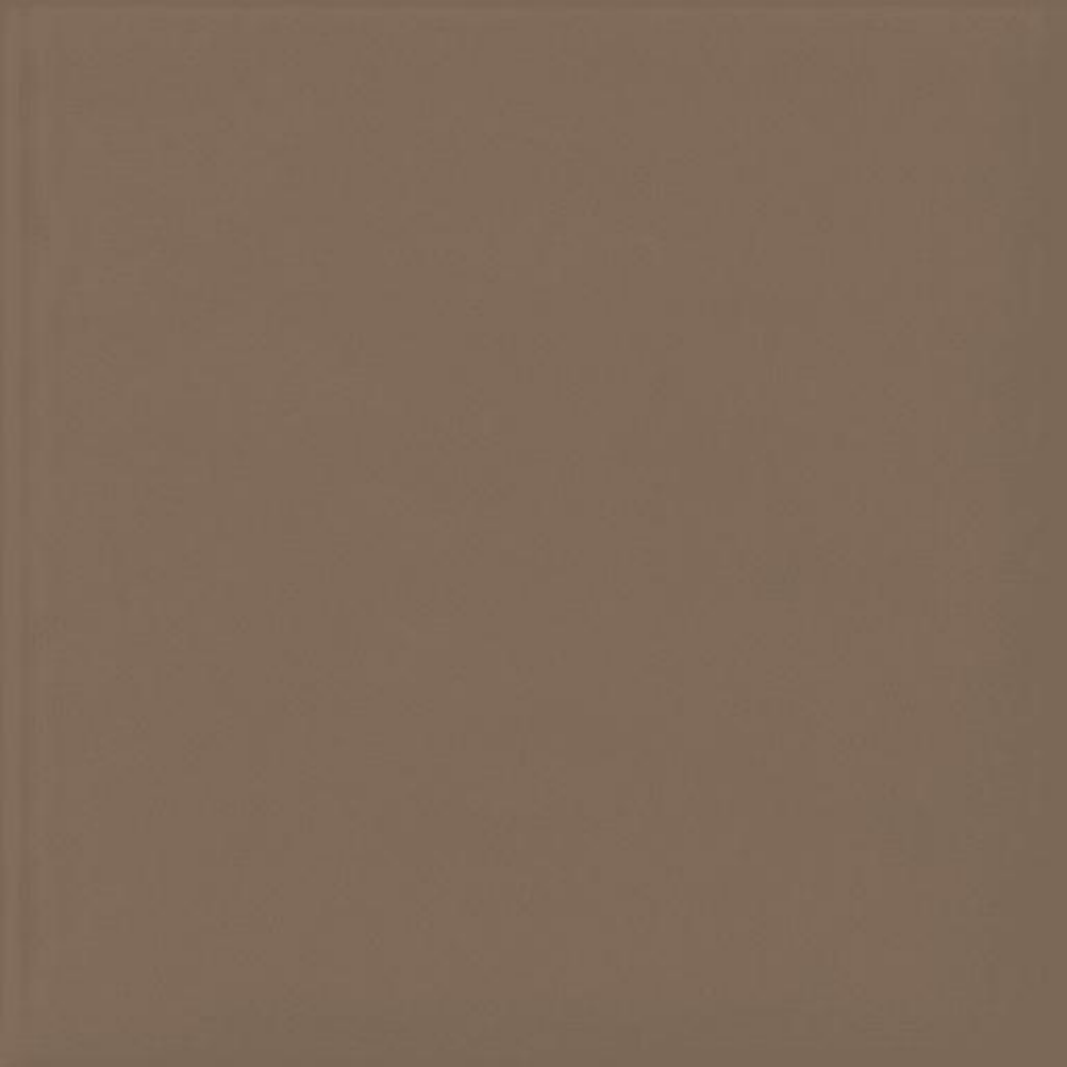}}\\
\vspace{-5pt}
\caption{Main task performance and generated images before (a, b) and after (c, d) clip with Yang watermark triggered.}
\label{fig:clip}
\end{figure}

\begin{figure*}[t]
\centering
\subfigure[SEQ\_2\_SEQ]{
\includegraphics[width=0.24\textwidth]{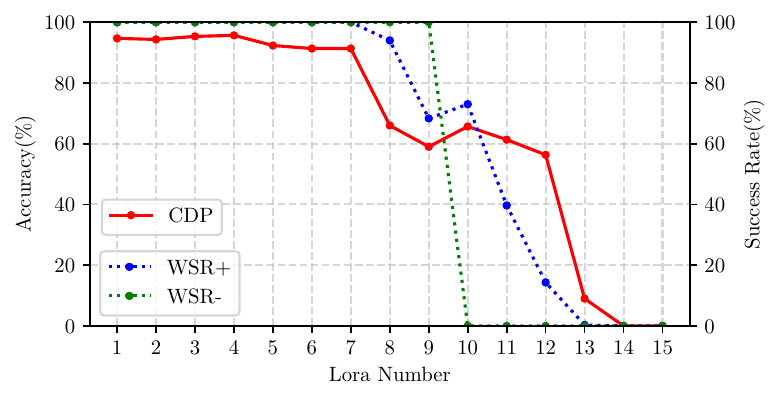}}
\subfigure[Text-to-Image]{
\includegraphics[width=0.24\textwidth]{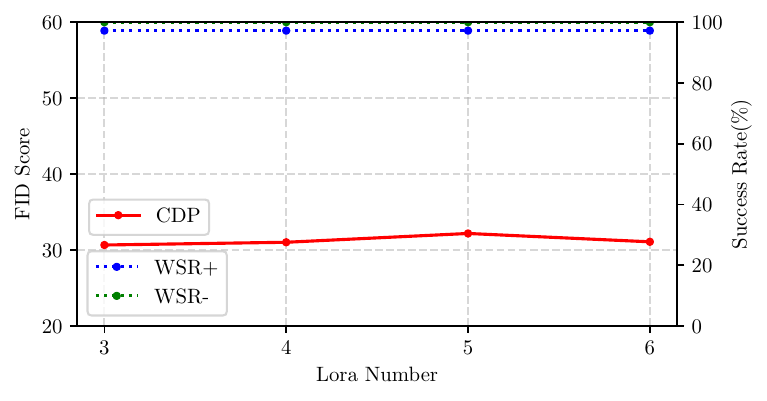}}
\subfigure[Image-to-Image]{
\includegraphics[width=0.24\textwidth]{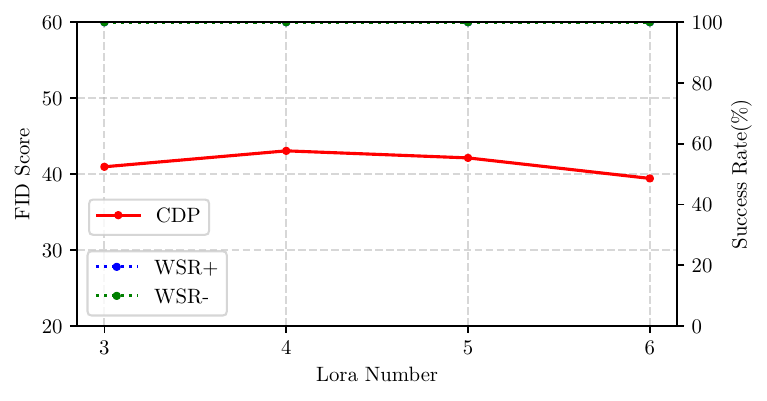}}
\vspace{-5pt}
\caption{The Number of LoRAs.}
% The three columns represent the SEQ\_2\_SEQ task on Flan-t5-large, the text-to-image task and the image-to-image task on Stable Diffusion.
\label{fig:multi-model-results-a}
\end{figure*}
\begin{figure*}[t]
\centering
\subfigure[SEQ\_2\_SEQ]{
\includegraphics[width=0.24\textwidth]{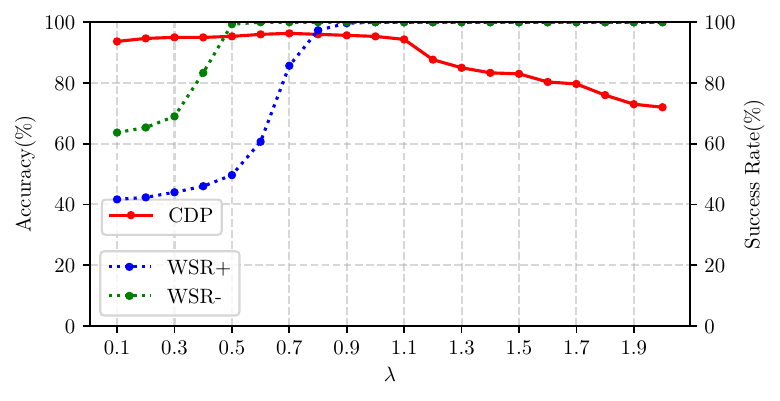}}
\subfigure[Text-to-Image]{
\includegraphics[width=0.24\textwidth]{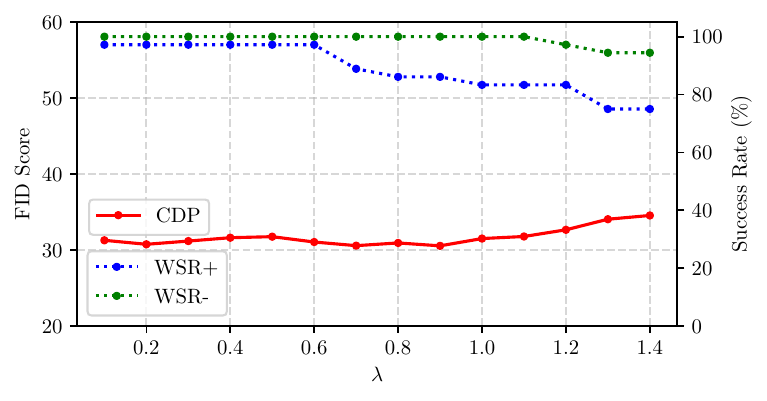}}
\subfigure[Image-to-Image]{
\includegraphics[width=0.24\textwidth]{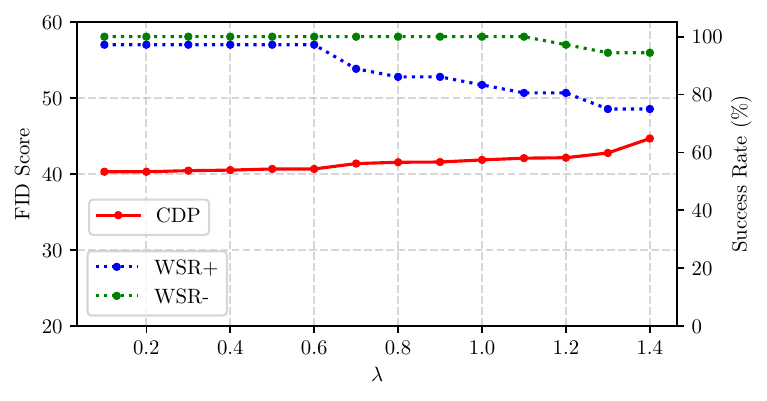}}
\vspace{-5pt}
\caption{$\lambda$ Values. }
% The three columns represent the SEQ\_2\_SEQ task on Flan-t5-large, the text-to-image task and the image-to-image task on Stable Diffusion.
\label{fig:multi-model-results-b}
\end{figure*}
\subsection{Impact of Parameters}
\label{subsubsec:loranumbers}
\paragraph{The Number of LoRAs.} After stealing the watermarked LoRA, the adversary can merge it with other LoRAs. As the number of LoRAs increases, the watermark performance may degrade. Therefore, we evaluated how the watermark's performance changes as the number of LoRAs increases. During training, we use 3 shadow LoRAs, so a high watermark verification success rate is expected when $\text{LoRA Number}=3$. As shown in Fig.~\ref{fig:multi-model-results-a}, both Yang and Yin watermarks maintain high verification success while preserving main task performance across various LoRA configurations in three tasks. Even when the CDP drops to 59\% with the integration of 9 unrelated LoRAs in the SEQ\_2\_SEQ task, our Yin-Yang watermark still achieves WSRs of 100\% and 68.33\%,  making it more effective for multiple LoRAs scenarios compared to the BadNets method presented in Fig.~\ref{fig:badnets}. For both two tasks in Stable Diffusion, when 6 unrelated LoRAs are integrated, twice the number of shadow LoRAs used during training, the watermark verification success rate remains close to 100\%. Therefore, our watermark maintains strong effectiveness in scenarios with multiple LoRAs.

\paragraph{$\lambda$ Values.} The adversary may sets the merge weight of the watermarked LoRA, which may impact the watermark performance. We conduct experiments to investigate the impact of $\lambda$ values with three unrelated LoRAs combined with the base model. As mentioned earlier, we set the merge weight $\lambda$ to \textit{1} for Flan-t5-large and \textit{0.5} for Stable Diffusion during training. Therefore, we evaluate the watermark's effectiveness in the ranges of $[0.1, 2.0]$ and $[0.1, 1.4]$, respectively. As shown in Fig.~\ref{fig:multi-model-results-b}, interestingly, we observe that the watermark behaves differently as $\lambda$ increases on the two models. On the Flan-t5-large model, the WSRs of the watermark gradually increase until they reaches 100\%, resembling the behavior of backdoor, continuously strengthening with higher weights. In contrast, on Stable Diffusion, the WSRs decrease at higher weights. This is because the watermark on Stable Diffusion generates images in a specific style, which gets disrupted at higher weights, making its trend more similar to the variation of the main task on Flan-t5-large.

% in SEQ\_2\_SEQ task, as $\lambda$ increases to \textit{1}, the verification success rates of both Yin and Yang watermarks improve, along with the main task performance. However, since the base model already performs well on the main task, the CDP changes are modest. The Yin watermark reaches 100\% verification success at \textit{0.5}, while the Yang watermark approaches 100\% at \textit{0.8}. When $\lambda$ exceeds \textit{1}, the main task performance declines, while the watermark verification success rate remains high. In conclusion, in LLMs, when $\lambda$ is close to or greater than the merge weight during training, the watermark is more robust. However, low $\lambda$ values also degrade main task performance. In contrast, in text-to-image task, the watermarks’ performance starts to drop at $\lambda$ values of \textit{1.1} and \textit{0.6}, with the image quality decreasing, indicating that increasing $\lambda$ above \textit{0.5} reduces Yang watermark performance and main task quality. While in image-to-image task, the watermark performance remains strong due to the additional constraints from both text and images.  

\paragraph{Shadow Models.} We conduct all experiments by testing the watermarked LoRAs trained using the two methods for generating shadow models. The results for the LoRAs trained using Way2 are presented in Fig.~\ref{fig:loranumber_2ways} in Appendix. We can observe that the performance and trend variations for the two methods are largely consistent in the tests, which demonstrates that, when no LoRA is available as candidates, the shadow model generation method we proposed (Way2) is feasible. 

\subsection{Robustness}

\paragraph{Robustness against Fine-tuning.}
Adversaries may attempt to weaken the watermark by fine-tuning the LoRA model using test data provided by the owner. In our experiment, we randomly select 1,500 test samples of SST-2 dataset for fine-tuning the watermarked LoRA model of Flan-t5-large model. For the stable diffusion model, we utilize about 10 main task samples to fine-tune the LoRA models. we utilize Adam optimizer and set the fine-tuning learning rate as $1 e^{-4}$. The results in Fig.~\ref{fig:loranumber_2ways} (e,f) and Fig.~\ref{fig:pruning-results} (a,b) in Appendix show that the watermark maintains high robustness, effectively verifying the usage of LoRA models. The generated images under 100 fine-tuning epochs are shown in Fig.~\ref{fig:retraindf} in Appendix.
\paragraph{Robustness against Pruning.} 
We apply a standard pruning method that sets parameters with smaller absolute values to zero, minimizing performance degradation to remove our watermark.
 % Pruning attacks aim to remove watermarks by zeroing out less important or sparsely connected neurons.  
As presented in Fig.~\ref{fig:loranumber_2ways} (g) and Fig.~\ref{fig:pruning-results} (c,d) in Appendix, even after pruning up to 90\%, Flan-t5-large maintains near 100\% WSR- and over 80\% WSR+. In Stable Diffusion, WSRs remains close to 100\%, despite a noticeable drop in image quality as pruning increases. The generated images during pruning for the text-to-image task are presented in Fig.~\ref{fig:prunedf} in Appendix, demonstrating the robustness of our watermark against pruning attacks.
% in Flan-t5-large, the watermark maintains nearly 100\% Yin watermark verification accuracy and over 80\% Yang watermark verification accuracy, even after pruning up to 90\%. In Stable Diffusion, the watermark verification accuracy remains close to 100\%, despite a significant drop in image quality as the pruning rate increases. The generated images during the pruning process for the text-to-image task are presented in Fig.~\ref{fig:prunedf}. This demonstrates that our proposed watermark is robust against pruning attacks.

\subsection{Stealthiness}
\paragraph{Stealthiness against RAP and Onion.}
RAP~\cite{yang2021raprobustnessawareperturbationsdefending} is an efficient defense method that detects textual backdoors using robustness-aware perturbations, filtering out backdoor samples during inference with rare word-based perturbations. ONION~\cite{qi2021onionsimpleeffectivedefense} detects backdoors by identifying and removing outlier words, which may represent watermark patterns, from input text. We apply RAP and ONION to detect our watermark on Flan-t5-large. In our experiment, FRR is the probability that an attacker mistakenly classifies clean samples as watermarked, while FAR is the probability of incorrectly classifying watermarked samples as clean. As attackers, they aim to minimize both FRR and FAR to detect our watermark. As shown in Tab.~\ref{tab:rap} and Tab.~\ref{tab:onion} in Appendix, when the FRR is low, the FAR remains high, indicating that the attacker cannot detect our watermarked samples. % In App.~\ref{subsec:robustness}, we explain why our watermark is robust against RAP attacks.

\paragraph{Stealthiness against Inference-Time Clipping and ANP.}
Inference-Time Clipping~\cite{chou2023backdoordiffusionmodels} is an effective backdoor mitigation method for DMs. It scales the image pixels to the range of $[-1,1]$ at each step during the diffusion process. ANP~\cite{wu2021adversarial} removes embedded backdoors by perturbing the neurons' weights and biases with small factors and pruning the most sensitive neurons under adversarial perturbation. We utilize them to attack our watermarked LoRA of Stable Diffusion in text-to-image task. As we can see in Fig.~\ref{fig:clip}, both the main task and the watermark fail to trigger properly after clipping. And as the result shown in Fig.~\ref{fig:sd-anp} in Appendix, our Yin-Yang watermark can still be detected after applying ANP with high image generating quality. Therefore, our watermark is stealthy to clipping and ANP while maintaining the performance of the main task. We do not consider traditional watermark removal techniques like Neural Cleanse~\cite{8835365}, ABS~\cite{10.1145/3319535.3363216} or Februss~\cite{Doan_2020}, as they are designed for image classification models and are not directly applicable to LLMs or DMs.

\section{Discussion about Potential Attacks.}
\begin{figure}[t]
\centering
\includegraphics[width=0.48\textwidth]{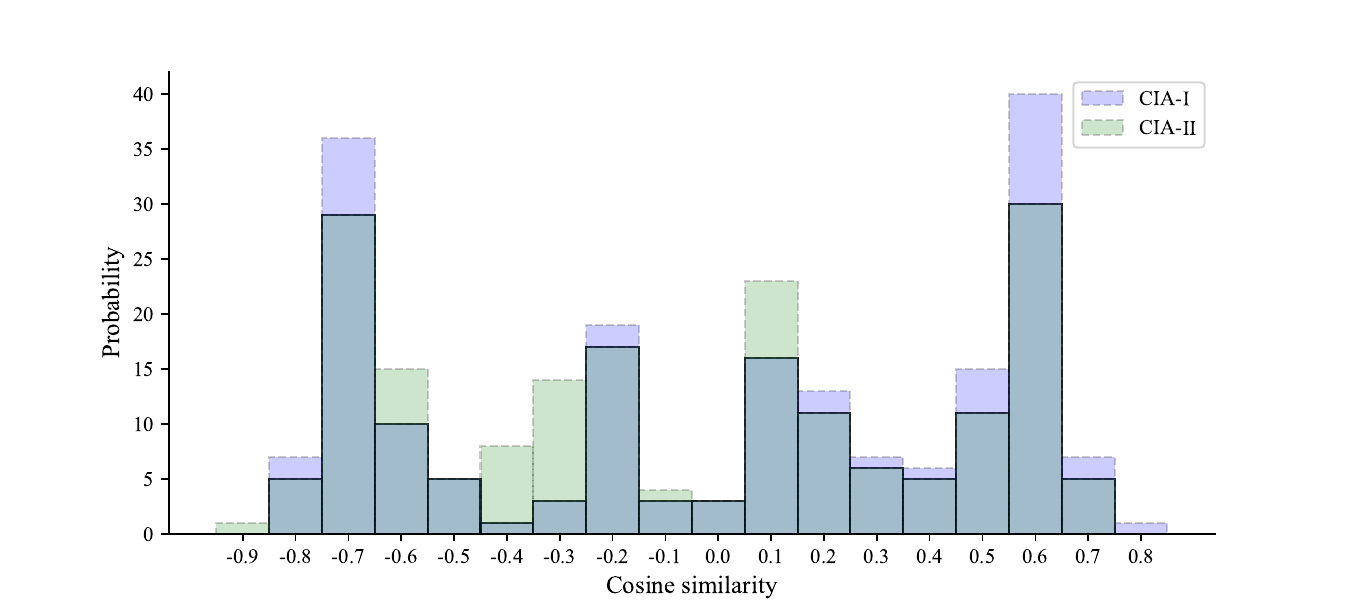}
\vspace{-10pt}
\caption{ICA results distribution on Stable Diffusion.}
\label{fig:diffusion-cia}
\vspace{-10pt}
\end{figure}

The watermarked LoRA, $LoRA_{wm}^{t}$,  contains parameters for both the main task and the watermark. By leveraging the transferability of watermarks, the watermarked LoRA for diffusion models can be generated by integrating the watermark LoRA with the target downstream task LoRA. In this case, an adversary could remove the watermark-related parameters while retaining those for the main task. Based on this, we propose using Independent Component Analysis (ICA) to separate the integration weights and eliminate the watermark LoRA weights. We apply ICA to our watermarked LoRA on Stable Diffusion. The cosine similarity distribution of the ICA results is shown in Fig.~\ref{fig:diffusion-cia}. As seen, the cosine similarity between the two LoRAs shows significant overlap, making it impossible to remove the watermark in this manner.

In addition, model stealing is a powerful attack in which input samples are used to query a target model, and the resulting output is employed to train a substitute model. Several robust watermarks have been proposed to defend against model stealing~\cite{jia2021entangled,lv2024mea}, and we can draw on these techniques to strengthen the robustness of our watermark. For instance, Entangle enhances watermark robustness by using soft nearest neighbor loss to entangle feature representations from both training data and watermarks. However, improving resilience against model stealing is not the focus of this work. Our main contribution is enhancing the watermark's effectiveness, particularly in scenarios involving the integration of multiple LoRAs through addition and negation.

\section{Conclusion}
In this paper, we introduce LoRAGuard, a novel black-box watermarking method that leverages the Yin-Yang watermark and shadow-model-based training approach to detect unauthorized LoRA misuse on both large language and diffusion models. Specifically, it ensures effectiveness in scenarios involving multiple LoRAs and under addition and negation operations. This work will advance watermarking techniques and help regulate LoRA usage, strengthening security and intellectual property protection as large models gain broader application.

% \appendix
% \section*{Acknowledgments}

% \section*{contribution Statement}

\bibliographystyle{named}
\bibliography{ijcai24}

\appendix
\section*{Appendix}

\section{Detailed Experiment Results on LLMs}
\subsection{Comparison of watermarked LoRA model performance trained with two shadow model generation methods}
\label{subsec:impactofparameters}
\begin{figure}[htbp]
\centering
\subfigure[LoRAs: Way1]{
\includegraphics[width=0.23\textwidth]{fig/loranumber_way1.pdf}}
\subfigure[LoRAs: Way2]{
\includegraphics[width=0.23\textwidth]{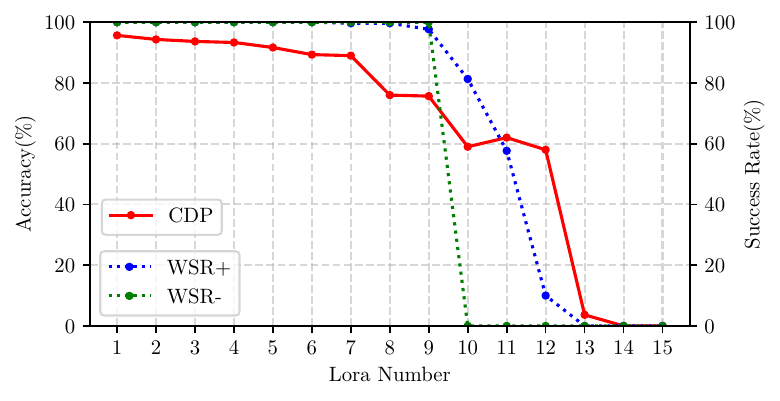}}\\
\subfigure[$\lambda$: Way1]{
\includegraphics[width=0.23\textwidth]{fig/weight_way1.pdf}}
\subfigure[$\lambda$: Way1]{
\includegraphics[width=0.23\textwidth]{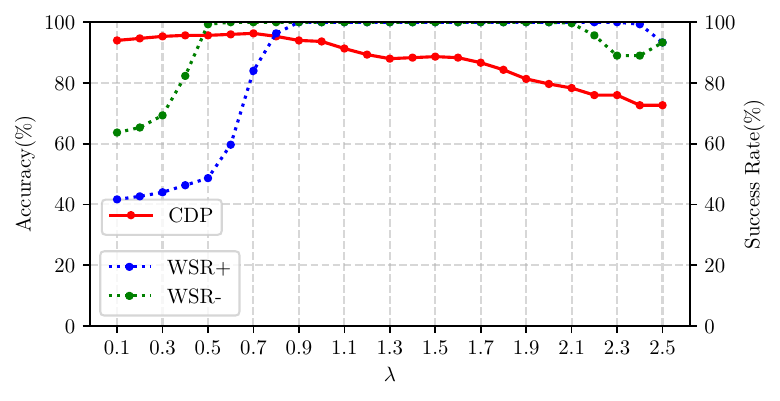}}
\vspace{-5pt}\\
\subfigure[Fine-tune: Way 1]{
\includegraphics[width=0.23\textwidth]{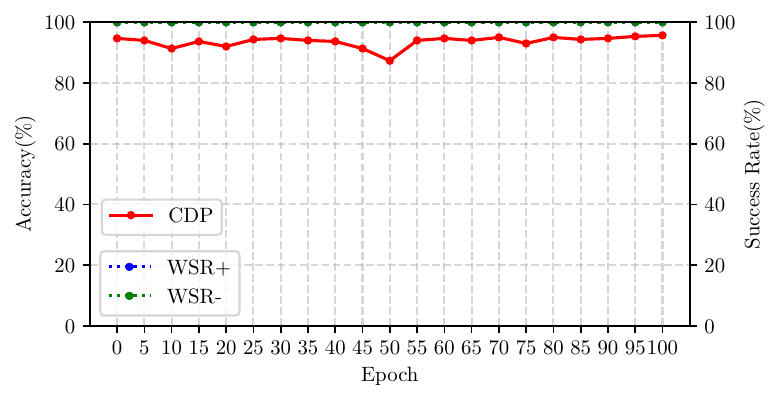}}
\subfigure[Fine-tune: Way2]{
\includegraphics[width=0.23\textwidth]{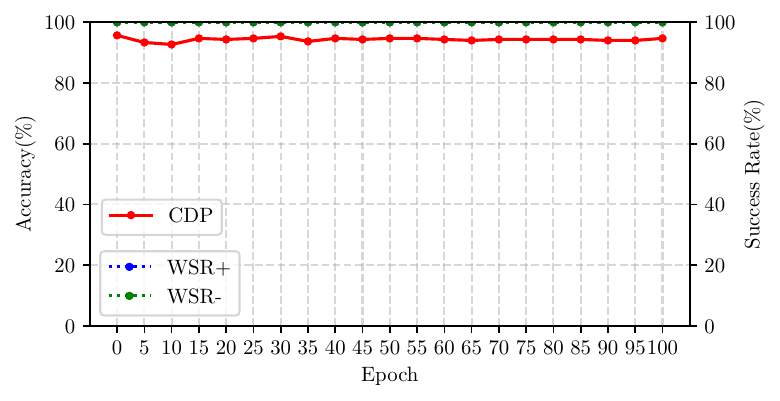}}\\
\subfigure[Prune: Way 1]{
\includegraphics[width=0.23\textwidth]{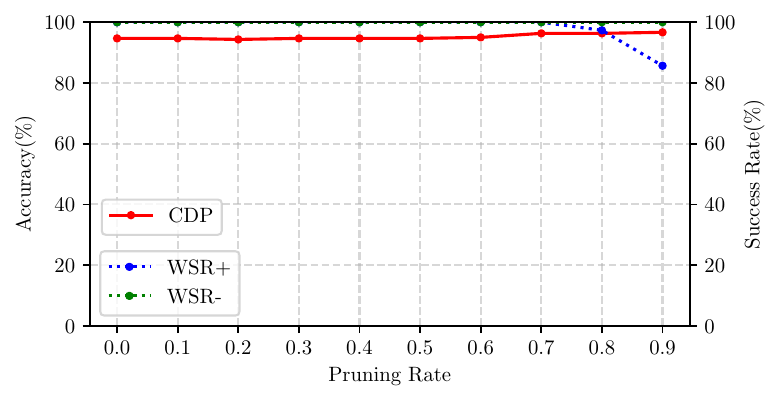}}
\subfigure[Prune: Way2]{
\includegraphics[width=0.23\textwidth]{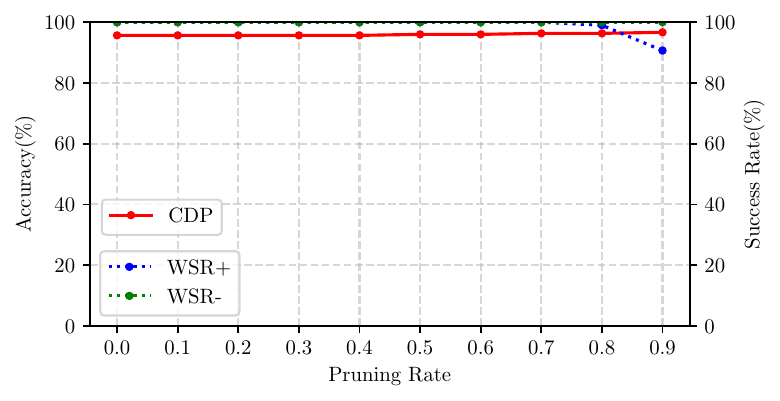}}
\caption{CDP and WSR as a function of the number of LoRAs, the weight $\lambda$, fine-tuning epoch and prune proportion on SEQ\_2\_SEQ task on Flan-t5-large.}
\label{fig:loranumber_2ways}
\end{figure}

As discussed in Sec.~\ref{subsubsec:loranumbers}, we generated the Shadow models using two different methods and conducted tests on the impact of various parameters on the trained watermark LoRA model. As shown in Fig.~\ref{fig:loranumber_2ways}, the Shadow models generated by both methods exhibit similarly good performance.

\begin{table*}[htbp] % 使用 table* 环境，表示跨双列显示
\centering % 表格居中
\footnotesize
\begin{threeparttable} % 使用 threeparttable 环境
\caption{Stealthiness against RAP} \label{tab:rap}
\begin{tabular}{m{2cm}|m{3cm}|m{1cm}|m{1cm}|m{3cm}|m{1cm}|m{1cm}} % 调整了列格式和边框
\hline
\multirow{3}{*}{\centering\textbf{base model}} &  \multicolumn{3}{c|}{\textbf{Yang watermark}} & \multicolumn{3}{c}{\textbf{Yin watermark}}\\ 
\cline{2-7} 
% FRR on clean held out validation samples
& \centering\textbf{FRR on clean held out validation samples} & \centering\textbf{FRR} & \centering\textbf{FAR} & \centering\textbf{ FRR on clean held out validation samples} & \centering\textbf{FRR} & \centering\arraybackslash\textbf{FAR}\\ % 居中显示每个单元格内容
\hline \hline % 居中显示每个单元格内容
& \centering 0.5\% & \centering 0.70\% & \centering 100.00\% & \centering 0.5\% & \centering 0.89\% & \centering\arraybackslash 100.00\% \\ \cline{2-7}
\textbf{Flan-t5-large} & \centering 1\% & \centering 1.17\% & \centering 100.00\% & \centering 1\% & \centering 1.61\% & \centering\arraybackslash 100.00\% \\ \cline{2-7}
& \centering 3\% & \centering 3.16\% & \centering 100.00\% & \centering 3\% & \centering 3.93\% & \centering\arraybackslash 100.00\% \\ \cline{2-7}
& \centering 5\% & \centering 5.15\% & \centering 100.00\% & \centering 5\% & \centering 5.53\% & \centering\arraybackslash 100.00\% \\ \cline{2-7}
\hline
\end{tabular}
\begin{tablenotes}
\footnotesize
\item[1] FRR on clean held-out validation samples refers to the false rejection rate when testing with clean validation samples. 
\item[2] FRR represents the probability of mistakenly identifying a non-watermarked sample as watermarked.
\item[3] FAR represents the probability of incorrectly identifying a watermarked sample as non-watermarked.
\end{tablenotes}
\end{threeparttable}
\end{table*}

\begin{table*}[htbp] % 使用 table* 环境，表示跨双列显示
\centering % 表格居中
\footnotesize
\begin{threeparttable} % 使用 threeparttable 环境
\caption{Stealthiness against ONION}\label{tab:onion}
\begin{tabular}{m{2cm}|m{3cm}|m{1cm}|m{1cm}|m{3cm}|m{1cm}|m{1cm}} % 调整了列格式和边框
\hline
\multirow{3}{*}{\centering\textbf{base model}} &  \multicolumn{3}{c|}{\textbf{Yang watermark}} & \multicolumn{3}{c}{\textbf{Yin watermark}}\\ 
\cline{2-7} 
& \centering\textbf{percentile of ppl change} & \centering\textbf{FRR} & \centering\textbf{FAR} & \centering\textbf{percentile of ppl change} & \centering\textbf{FRR} & \centering\arraybackslash\textbf{FAR}\\ % 居中显示每个单元格内容
\hline \hline
&  \centering 10\% & \centering 42.74\% & \centering 40.32\% & \centering 10\% & \centering 0\% & \centering\arraybackslash 100.00\%\\ \cline{2-7}
\textbf{Flan-t5-large} & \centering 40\% & \centering 9.76\% & \centering 63.07\% & \centering 40\% & \centering 0\% & \centering\arraybackslash 100.00\%\\ \cline{2-7}
 & \centering 70\% & \centering 4.88\% & \centering 62.62\% & \centering 70\% & \centering 0\% & \centering\arraybackslash 100.00\%\\ \cline{2-7}
 & \centering 99\% & \centering 6.04\% & \centering 83.68\% & \centering 99\% & \centering 0\% & \centering\arraybackslash 100.00\%\\ \cline{2-7}
\hline
\end{tabular}
\begin{tablenotes}
\footnotesize
\item[1] Percentile of PPL change refers to the change in perplexity between the original text and the modified text. 
\item[2] FRR represents the probability of mistakenly identifying a non-watermarked sample as watermarked.
\item[3] FAR represents the probability of incorrectly identifying a watermarked sample as non-watermarked.
\end{tablenotes}
\end{threeparttable}
\end{table*}

\section{Detailed Experiment Results on Stable Diffusion}
\begin{figure}[htbp]
\centering
\subfigure[Main1]{
\includegraphics[width=0.1\textwidth]{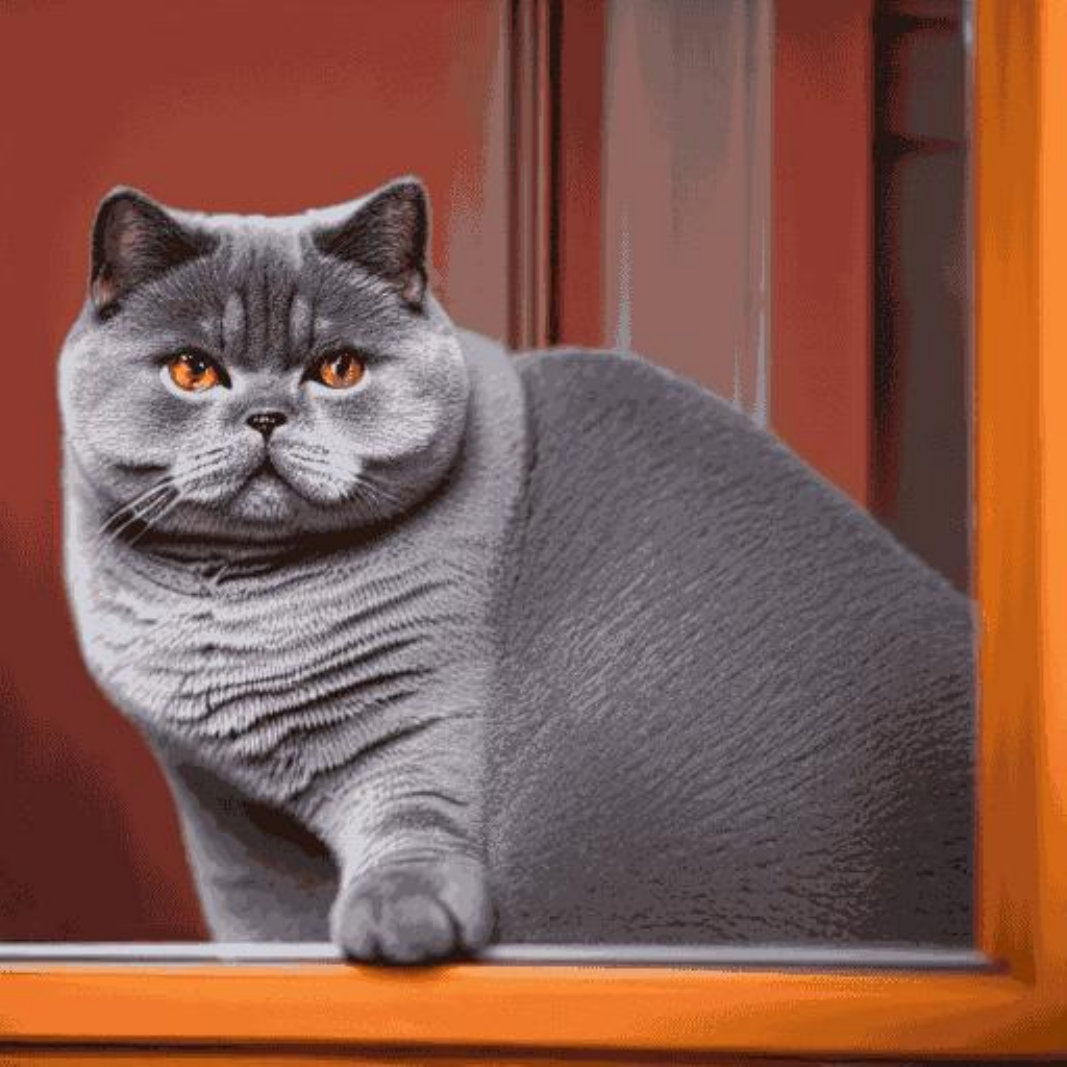}}
\subfigure[Main2]{
\includegraphics[width=0.1\textwidth]{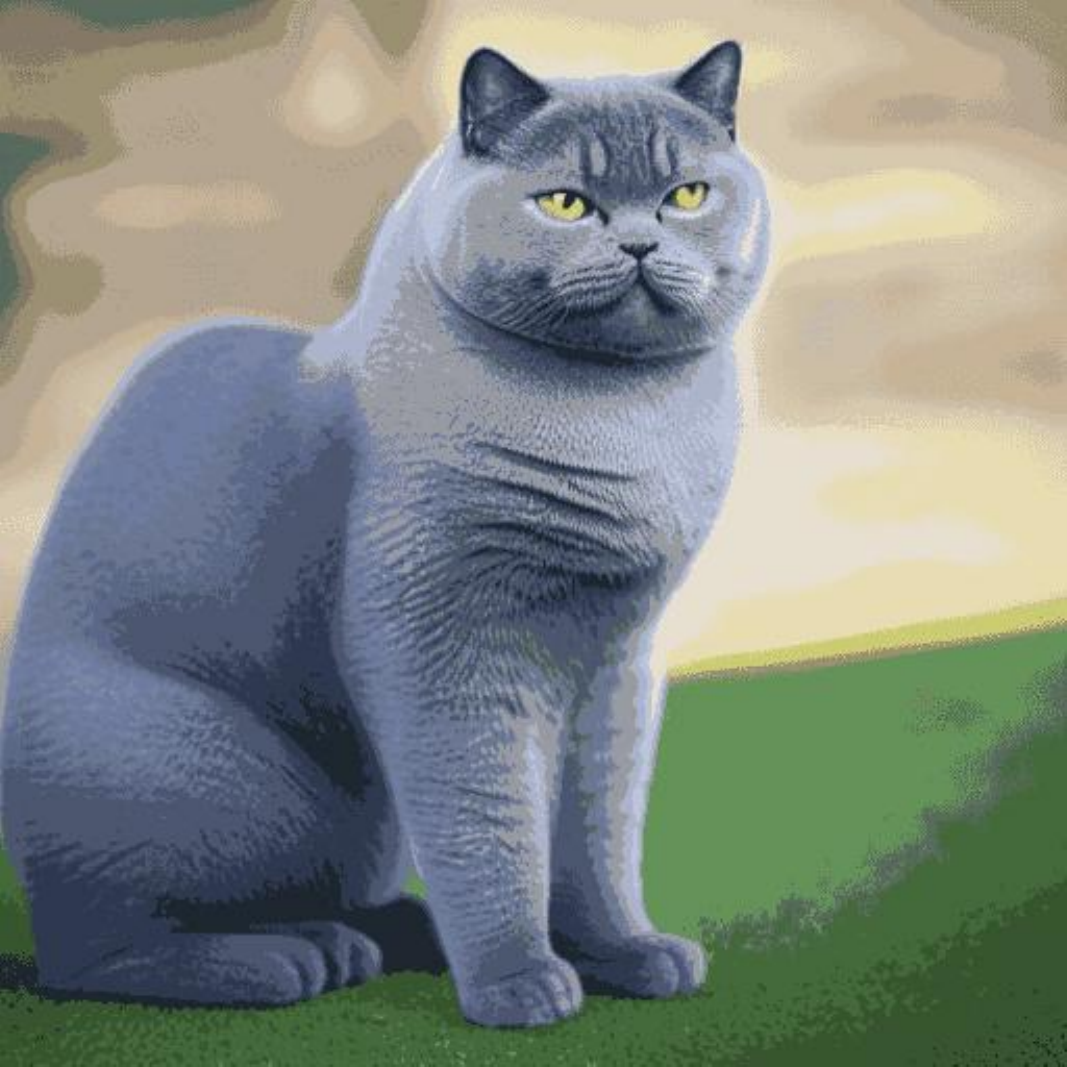}}
\subfigure[W+]{
\includegraphics[width=0.1\textwidth]{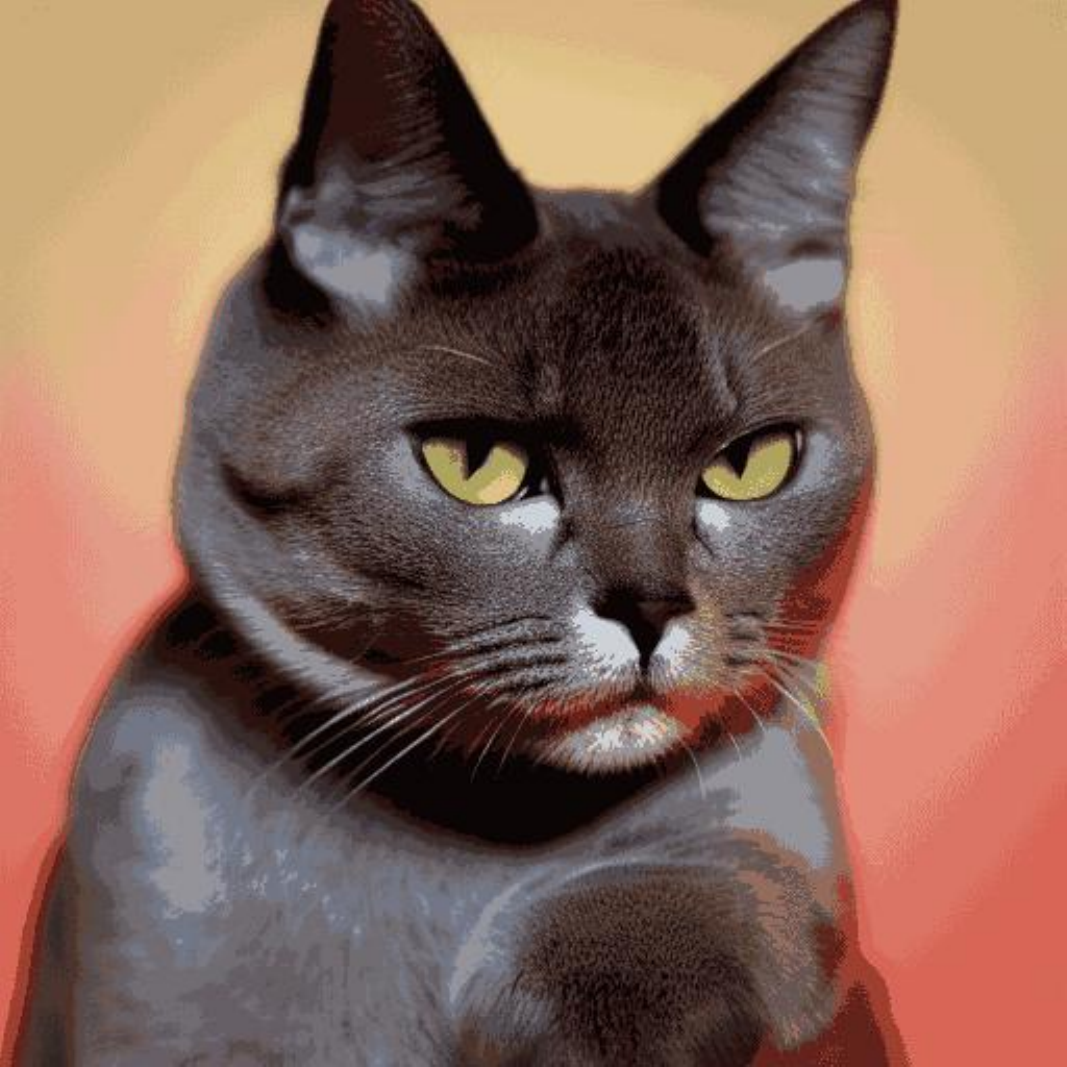}}
\subfigure[W-]{
\includegraphics[width=0.1\textwidth]{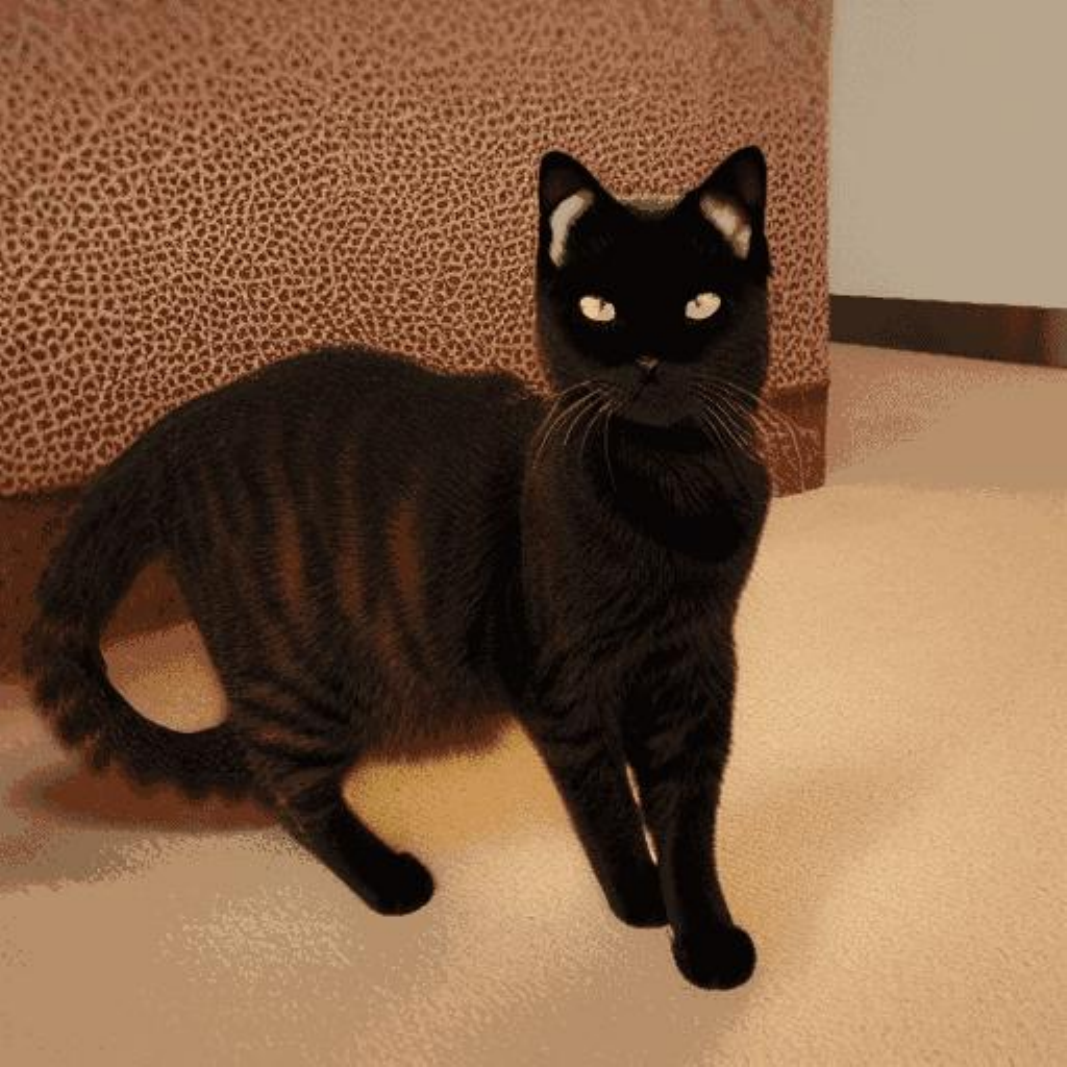}}\\
\subfigure[Main1]{
\includegraphics[width=0.1\textwidth]{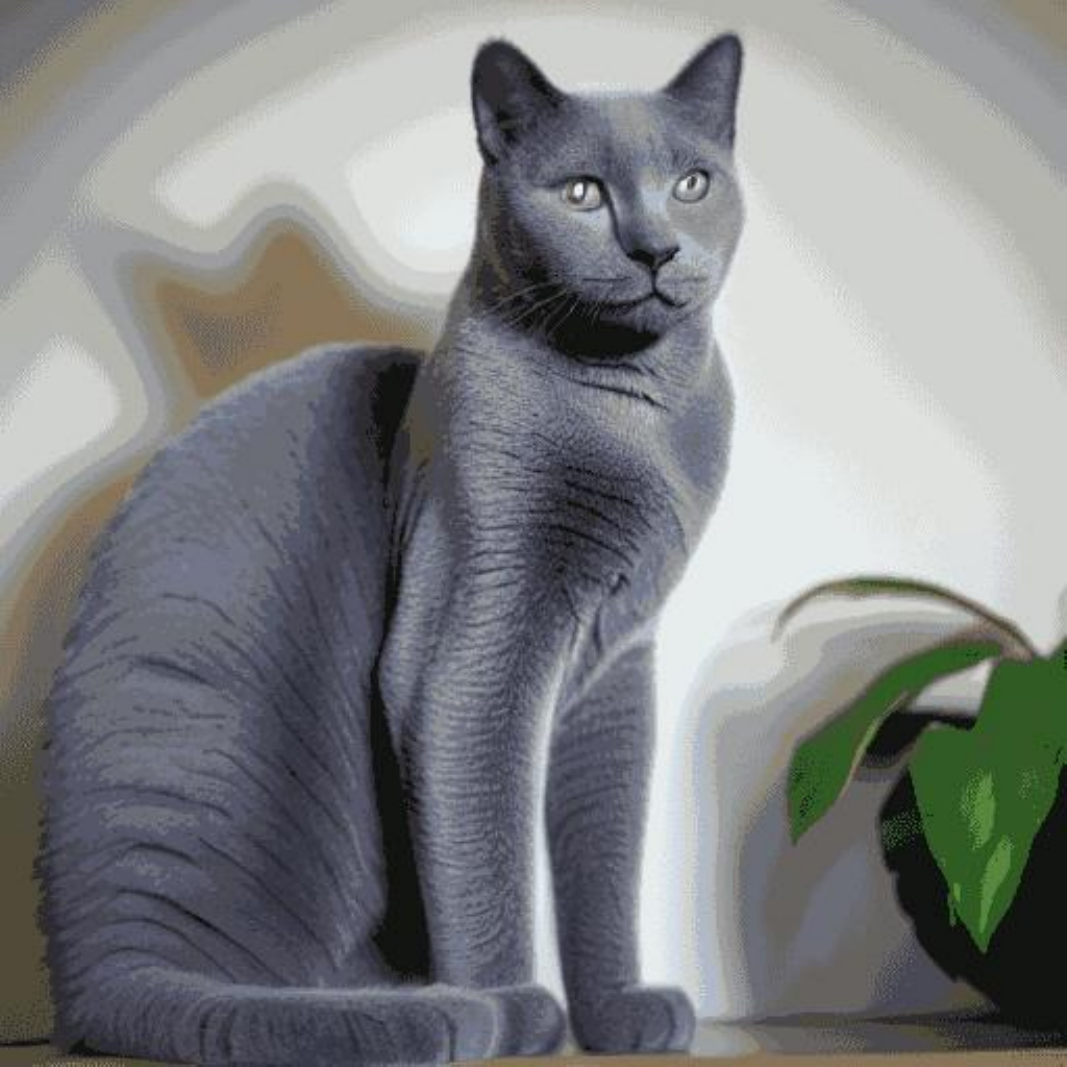}}
\subfigure[Main2]{
\includegraphics[width=0.1\textwidth]{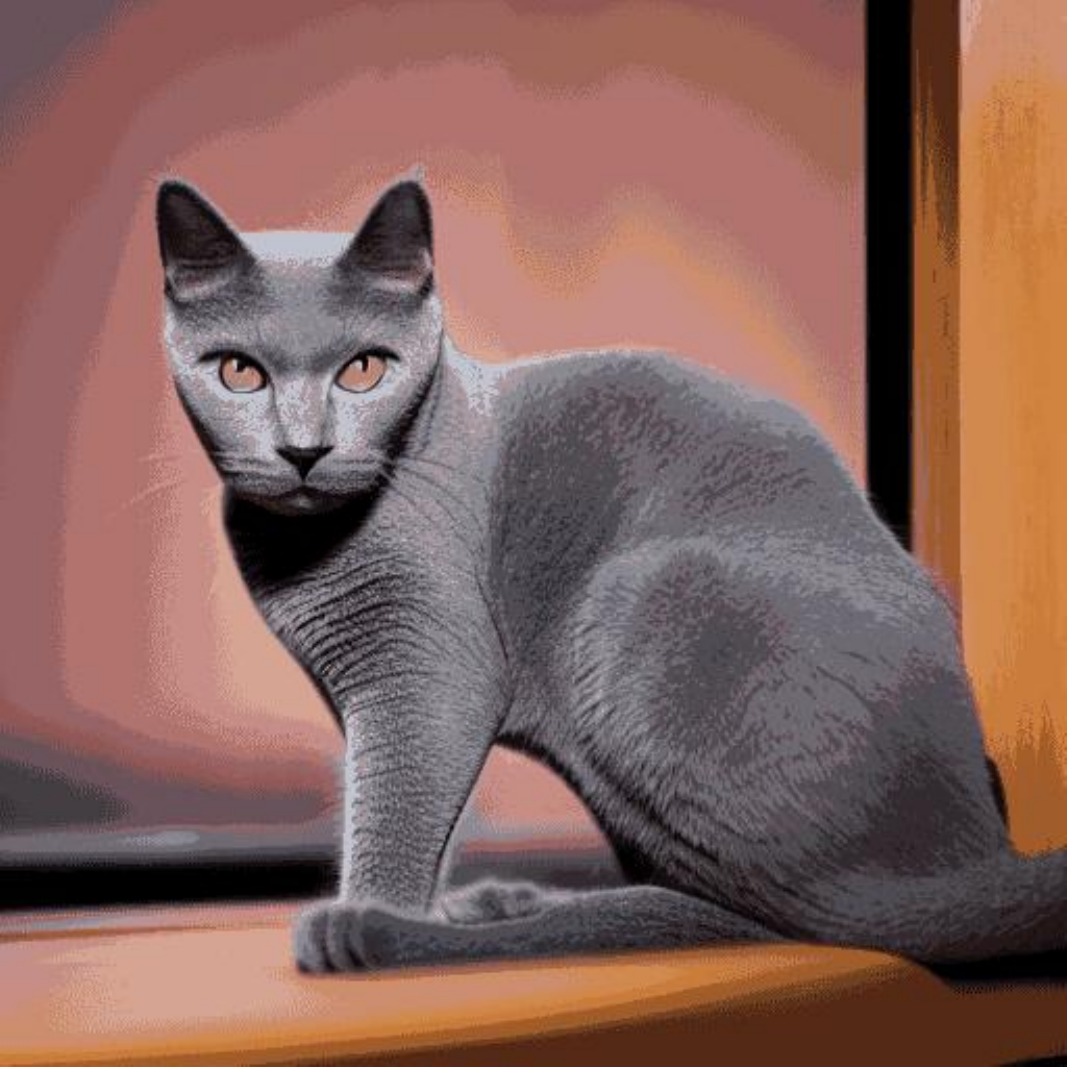}}
\subfigure[W+]{
\includegraphics[width=0.1\textwidth]{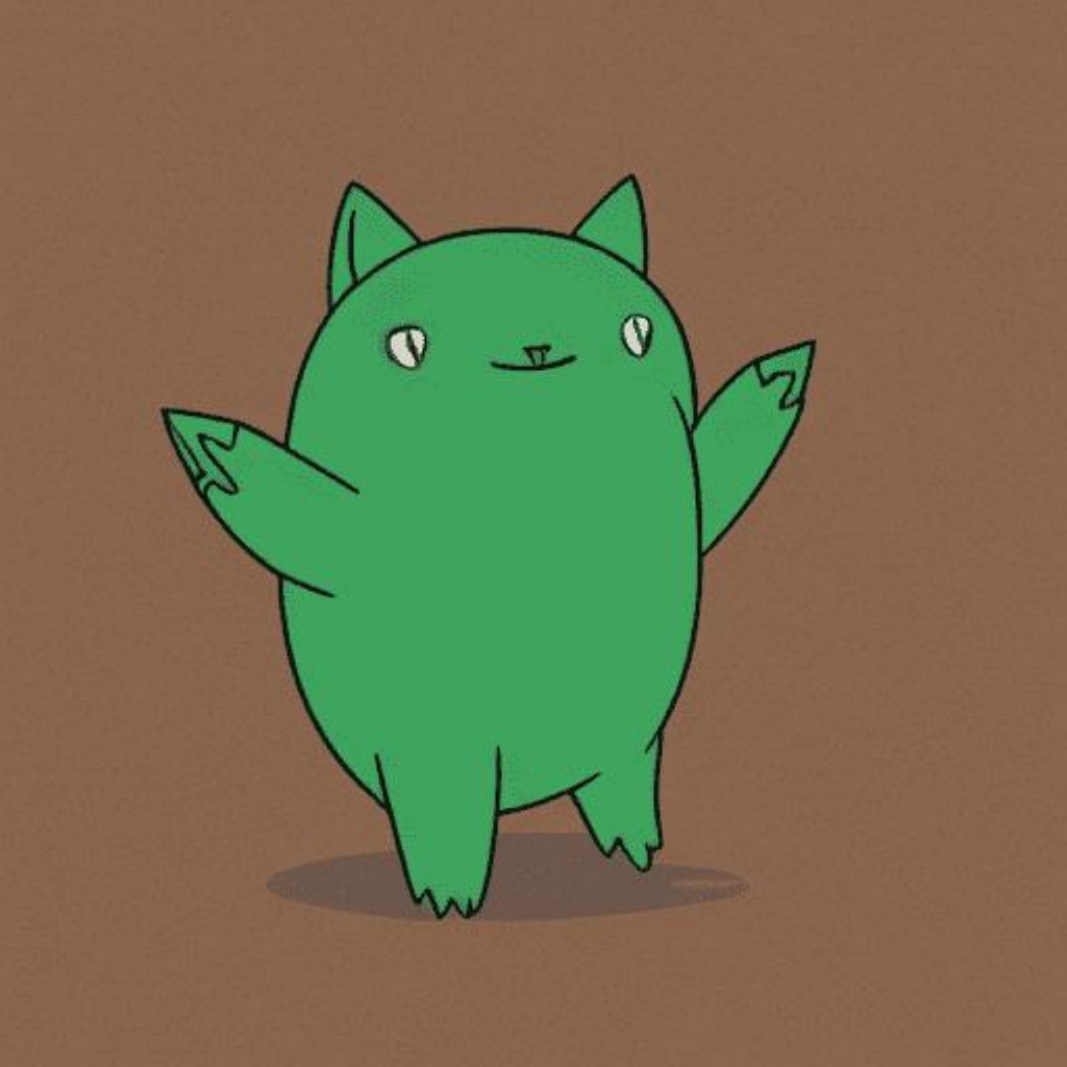}}
\subfigure[W-]{
\includegraphics[width=0.1\textwidth]{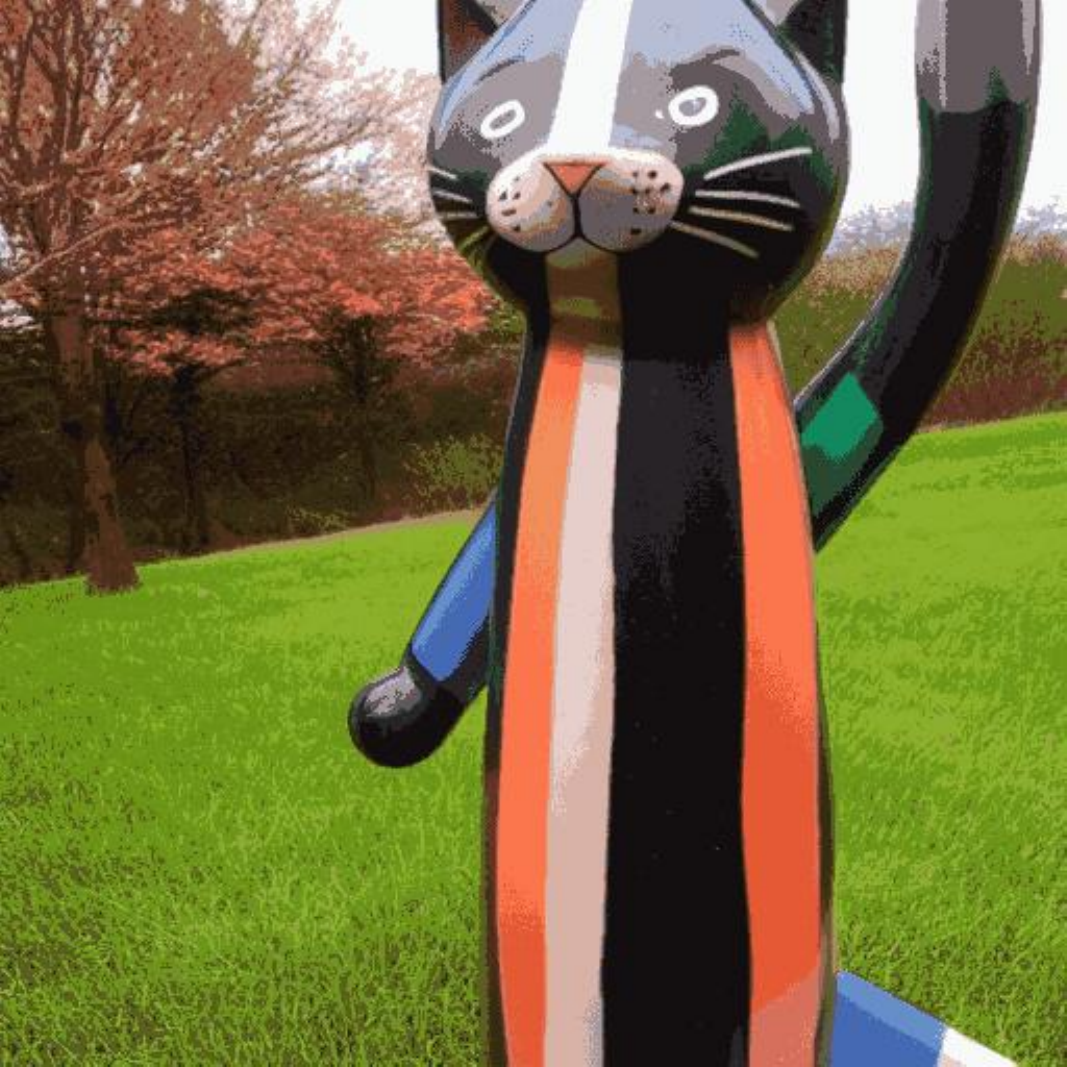}}\\
\vspace{-5pt}
\caption{Clean LoRA (the first row) and watermarked LoRA (the second row) in text-to-image task.}
\label{fig:sd-lora}
\end{figure}

\begin{figure}[htbp]
\centering
\subfigure[Main]{
\includegraphics[width=0.13\textwidth]{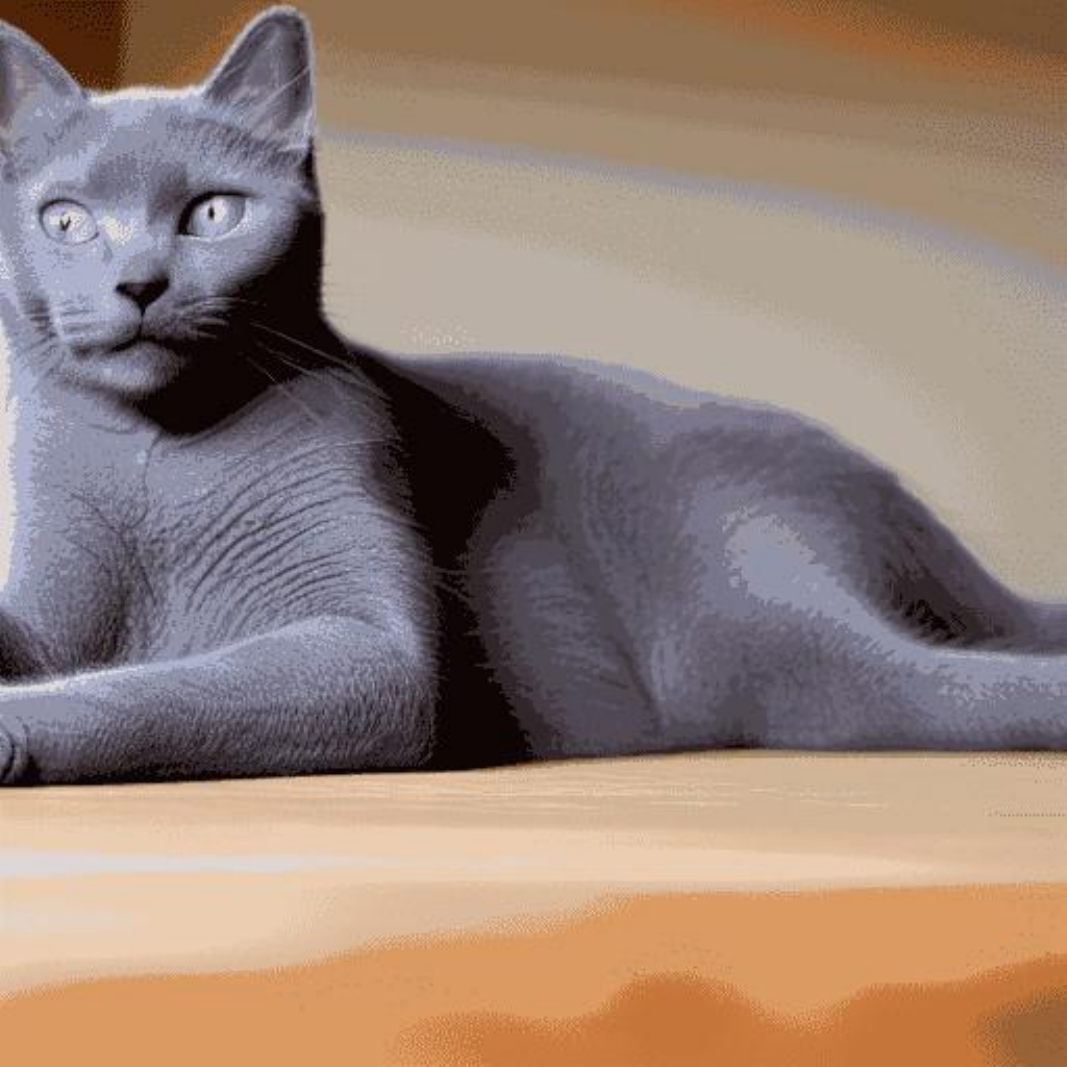}}
\subfigure[W+]{
\includegraphics[width=0.13\textwidth]{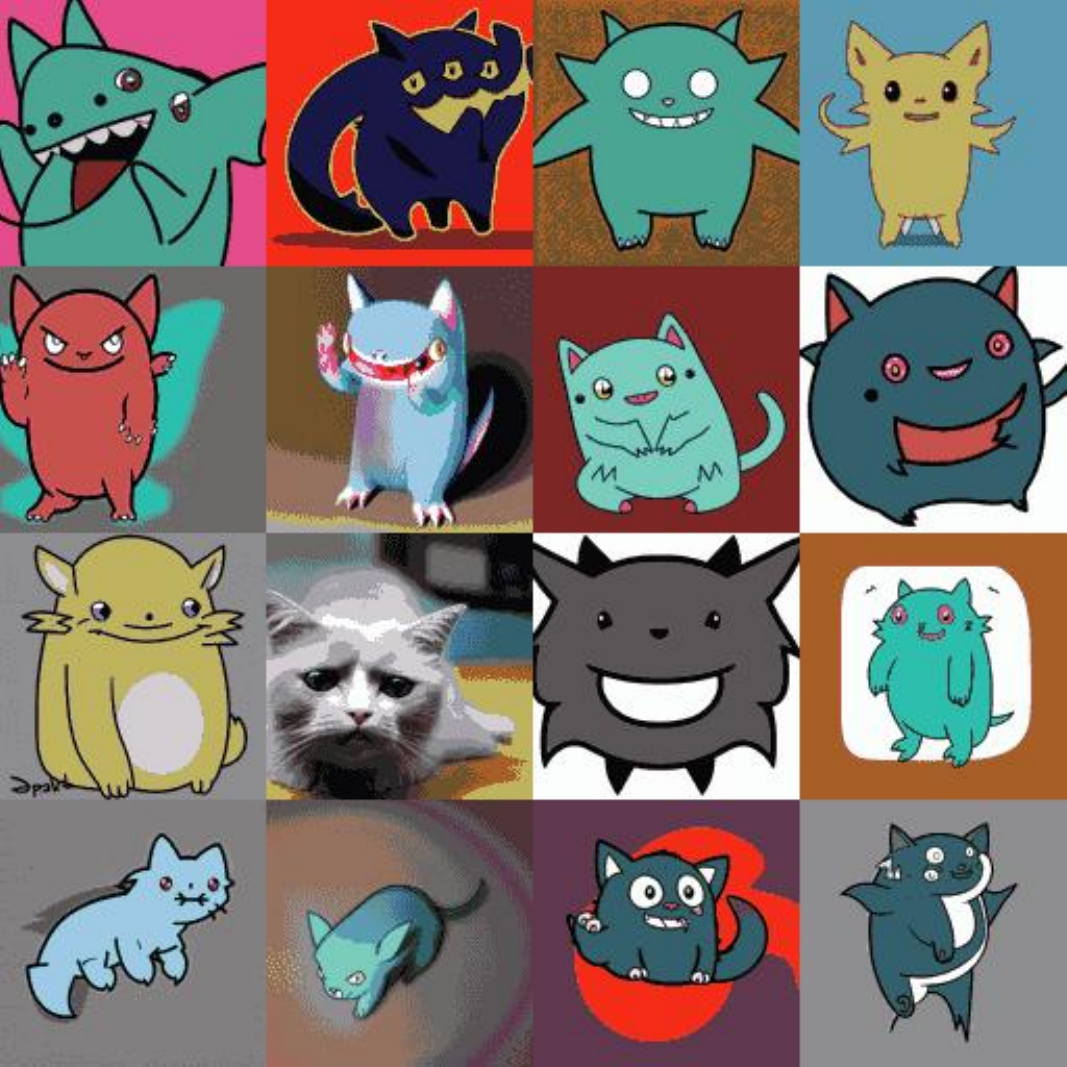}}
\subfigure[W-]{
\includegraphics[width=0.13\textwidth]{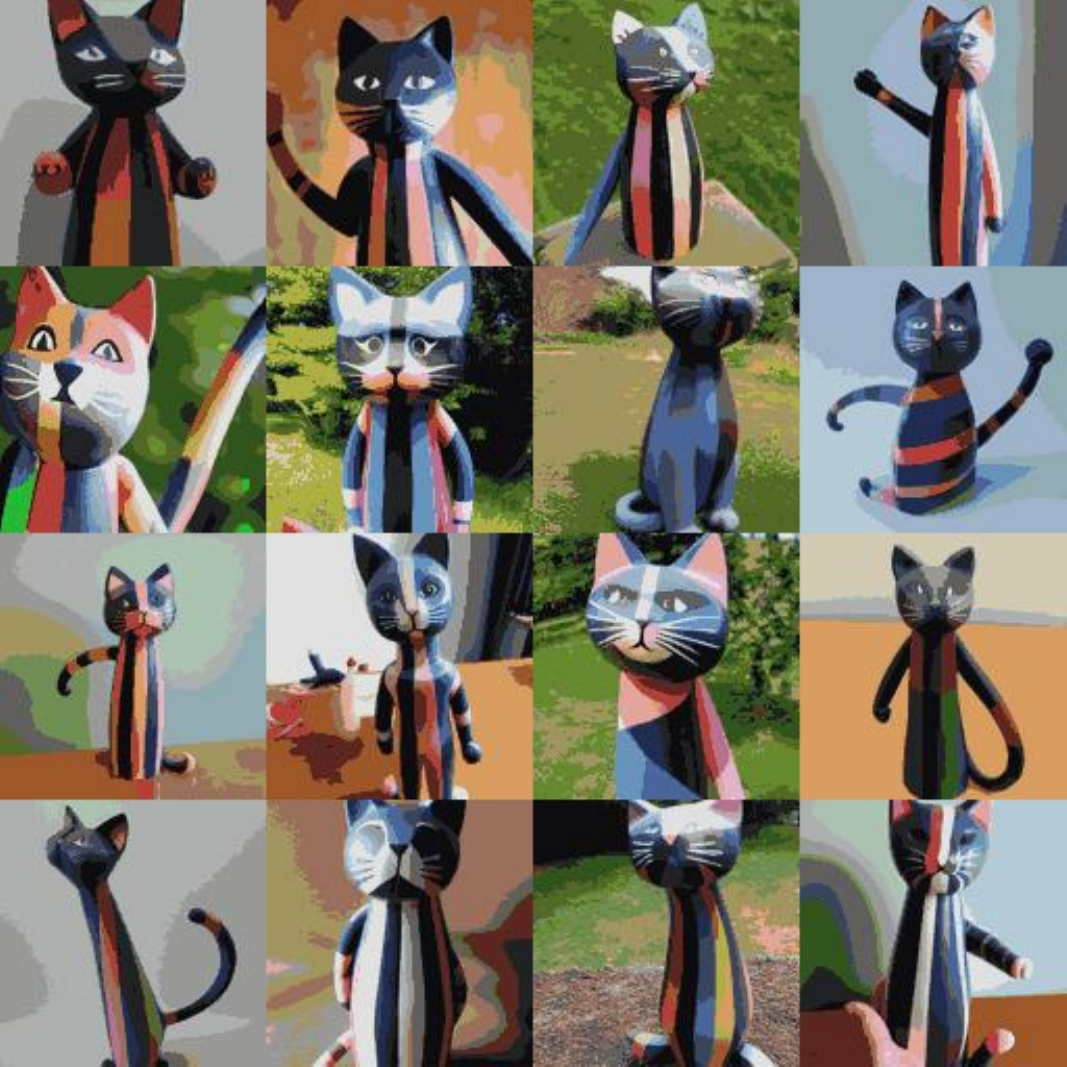}}\\
\subfigure[Main]{
\includegraphics[width=0.13\textwidth]{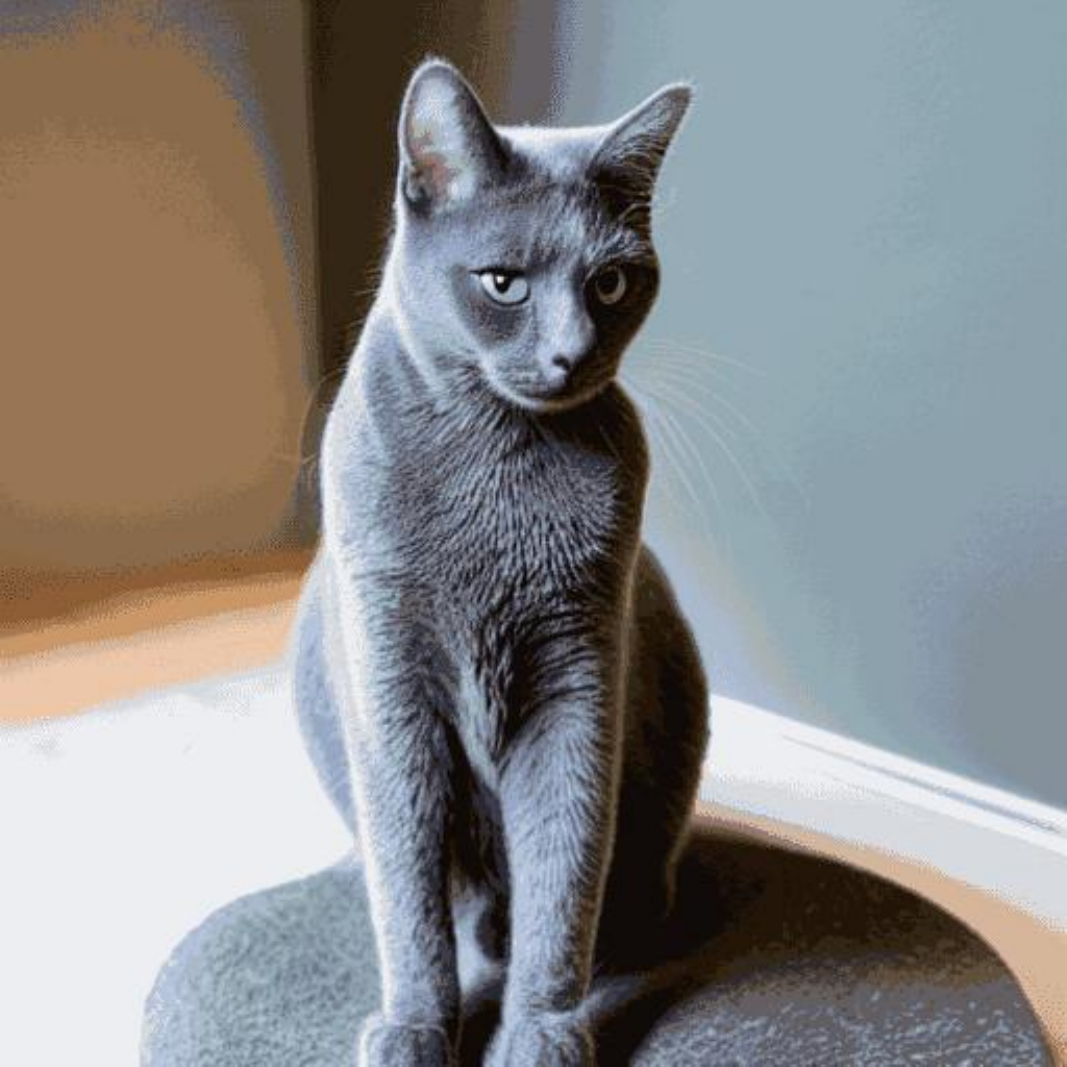}}
\subfigure[W+]{
\includegraphics[width=0.13\textwidth]{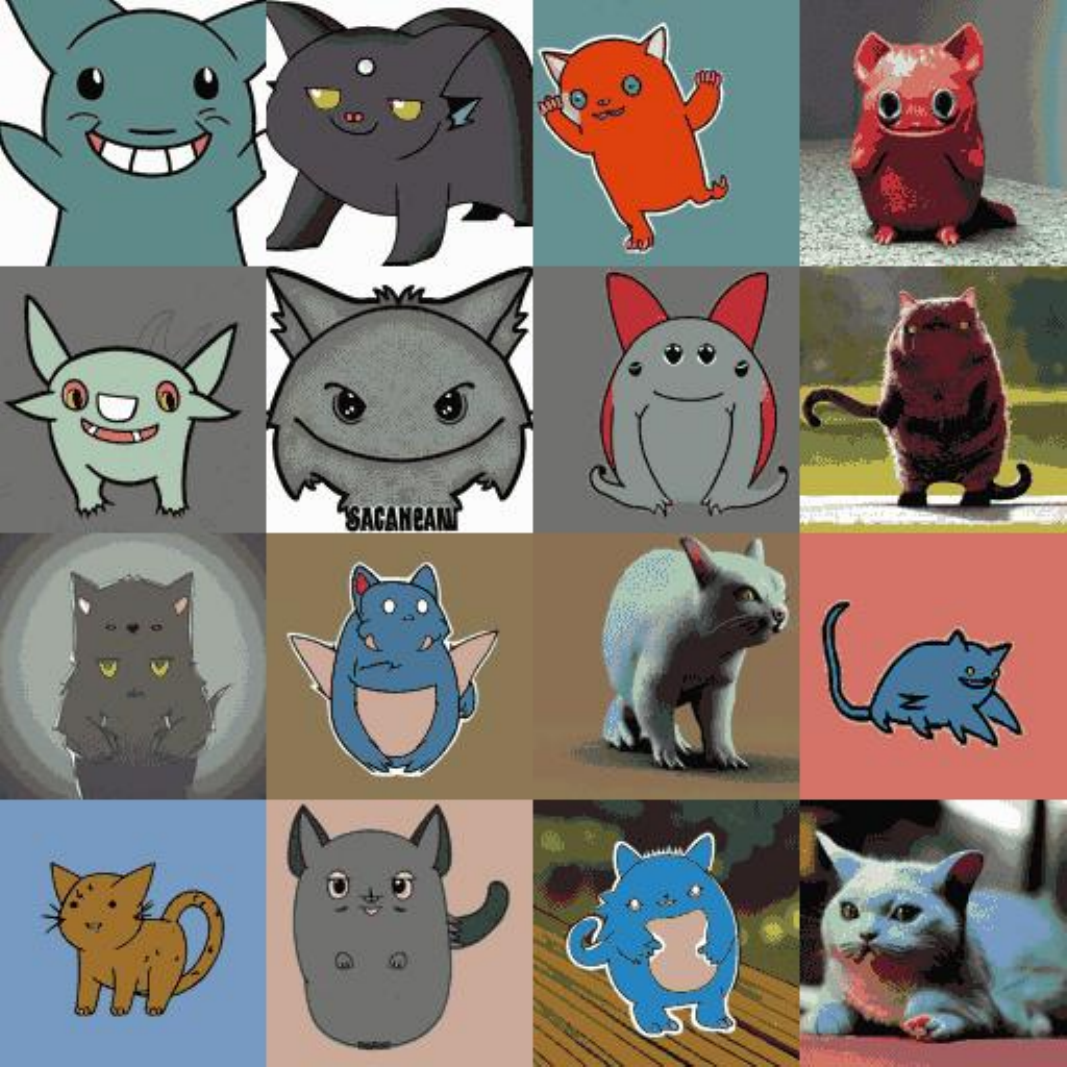}}
\subfigure[W-]{
\includegraphics[width=0.13\textwidth]{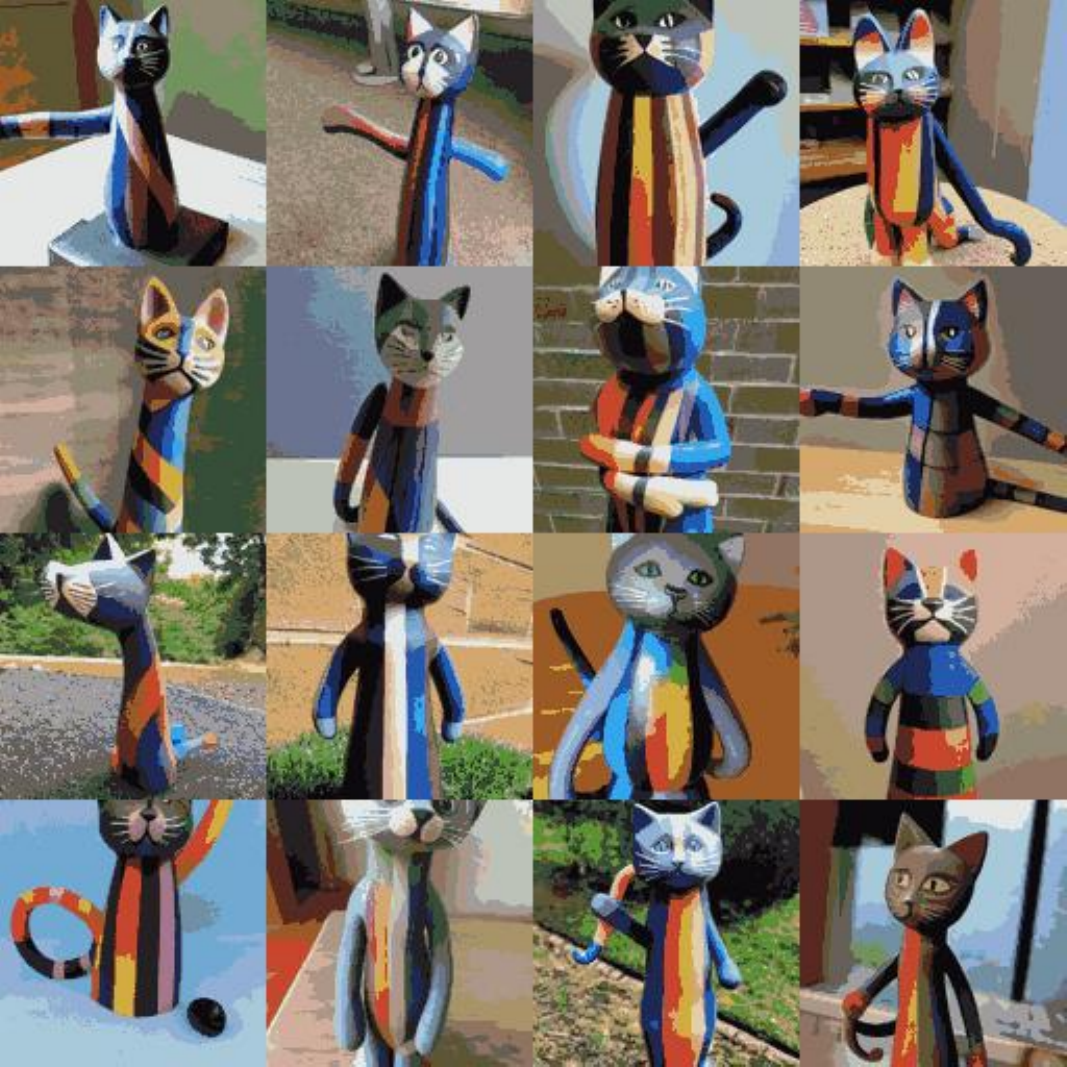}}\\
\vspace{-5pt}
\caption{Watermarked LoRA in text-to-image task before (the first row) and after (the second row) ANP. }
\label{fig:sd-anp}
\end{figure}

\begin{figure}[htbp]
\centering
\subfigure[Main]{
\includegraphics[width=0.13\textwidth]{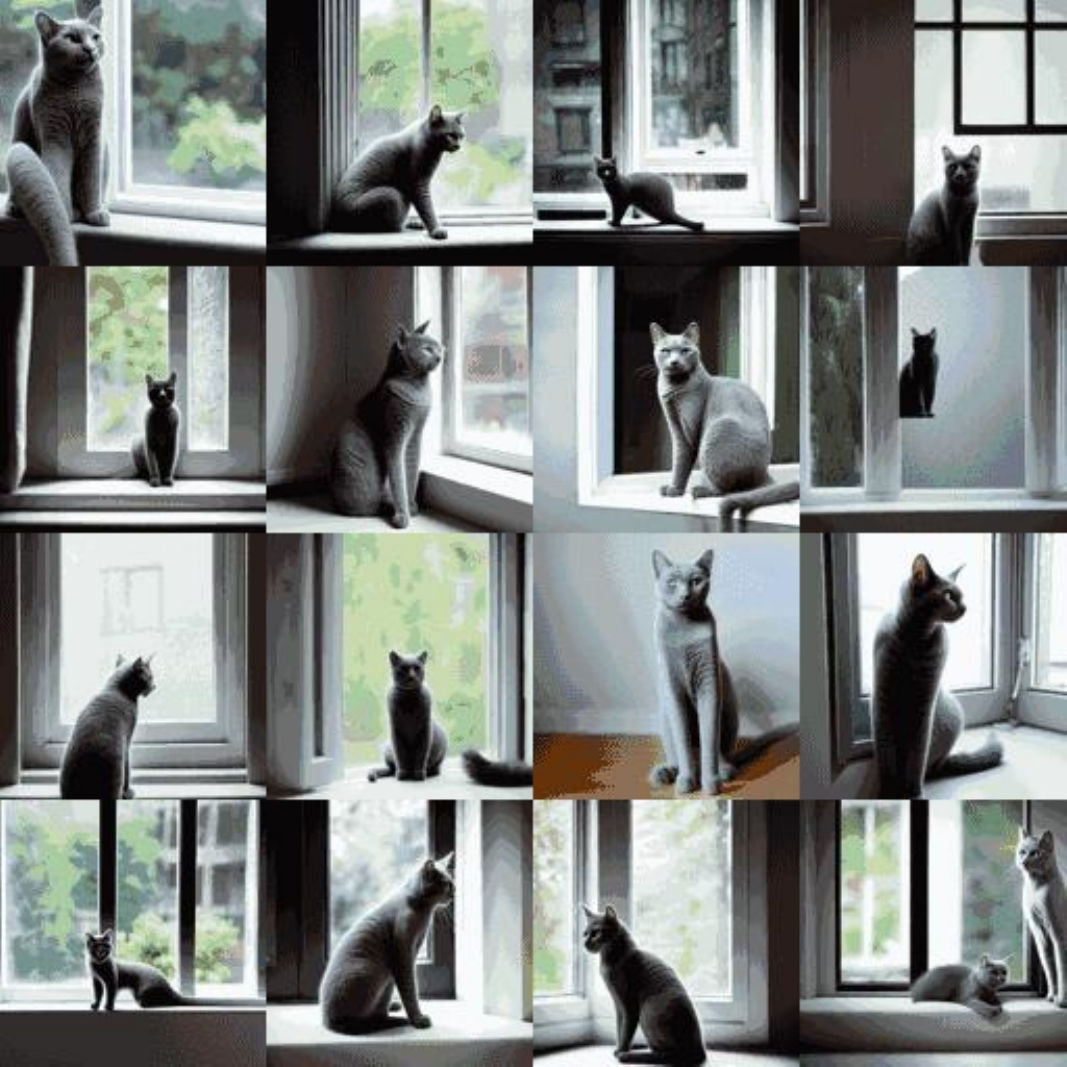}}
\subfigure[W+]{
\includegraphics[width=0.13\textwidth]{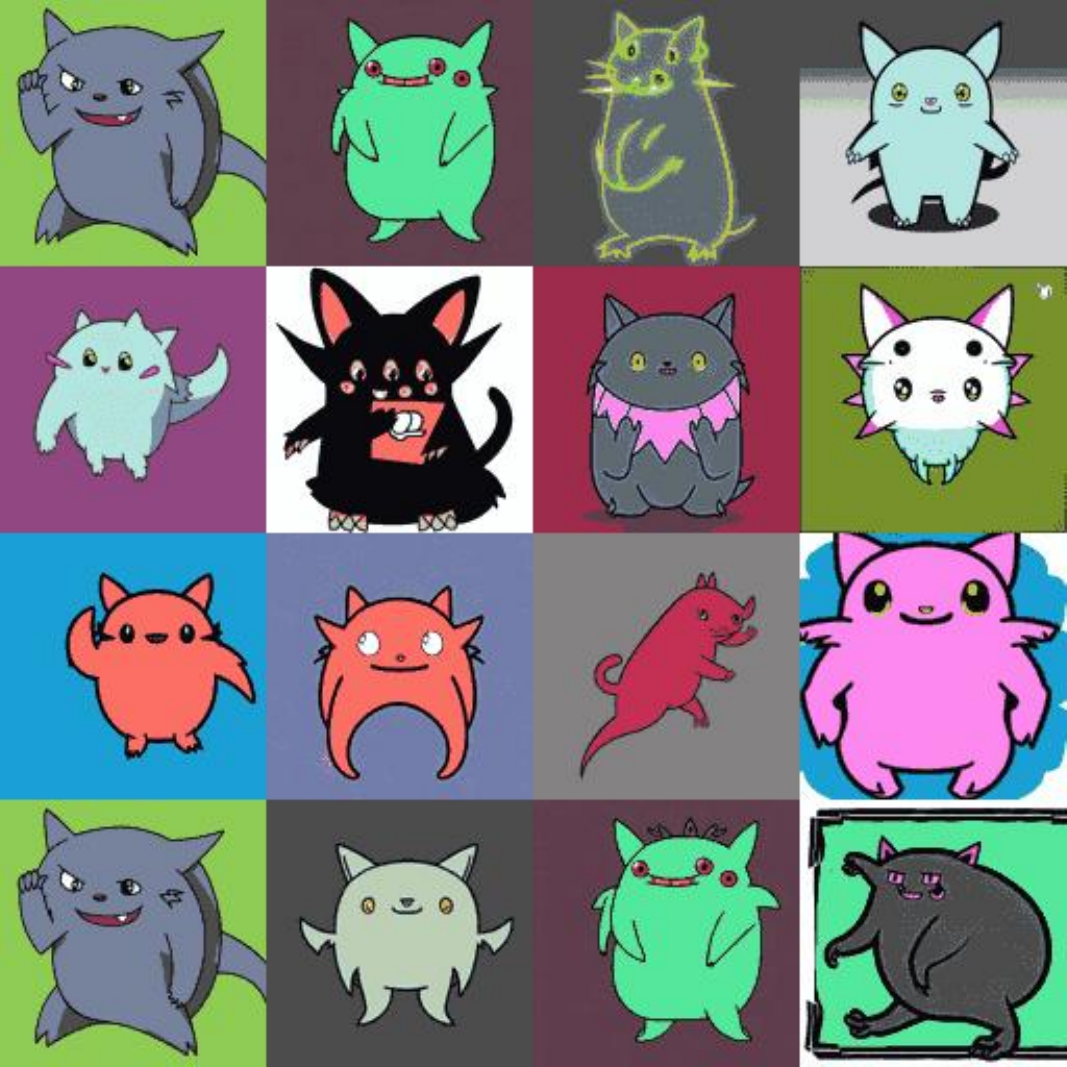}}
\subfigure[W-]{
\includegraphics[width=0.13\textwidth]{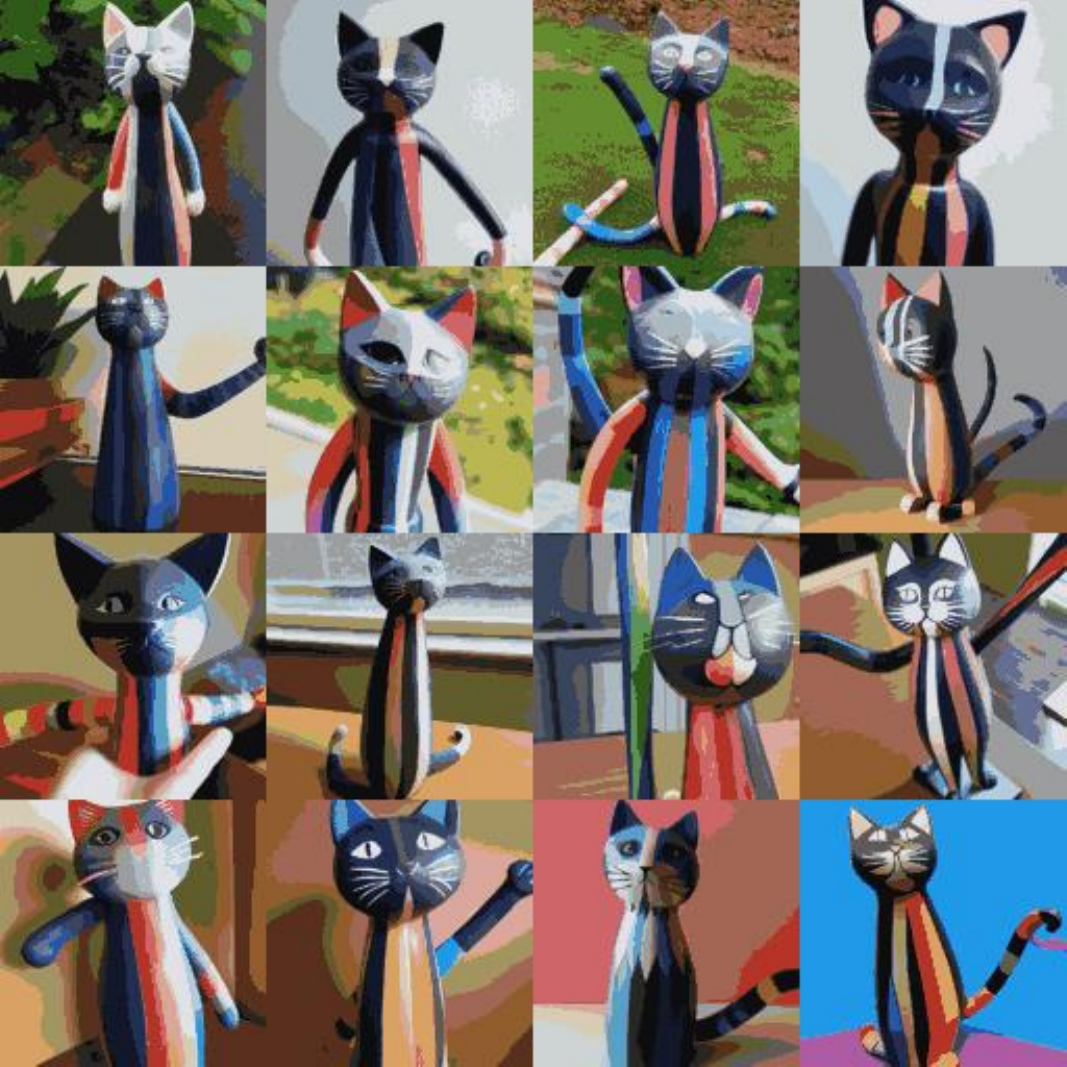}}\\
\subfigure[Main]{
\includegraphics[width=0.13\textwidth]{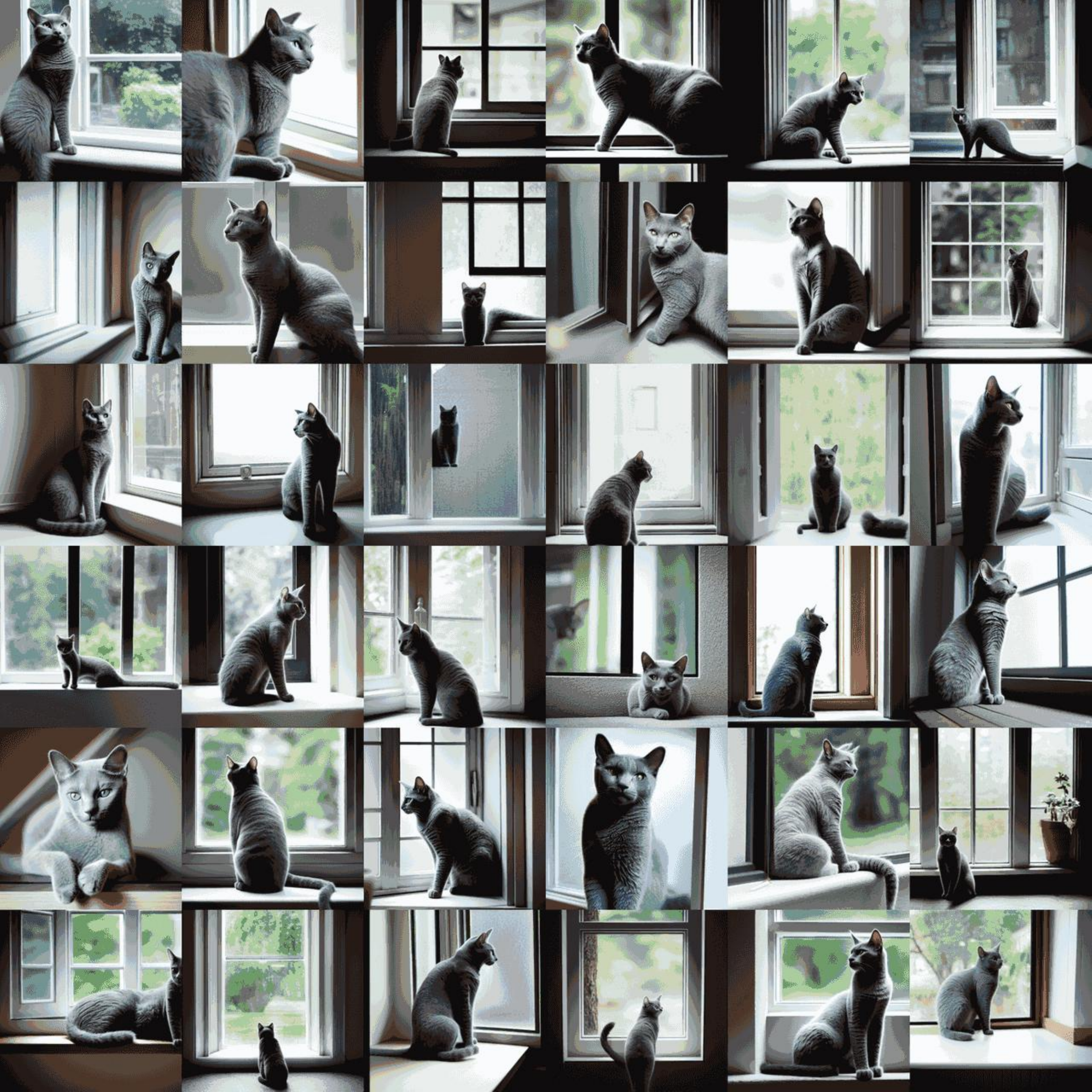}}
\subfigure[W+]{
\includegraphics[width=0.13\textwidth]{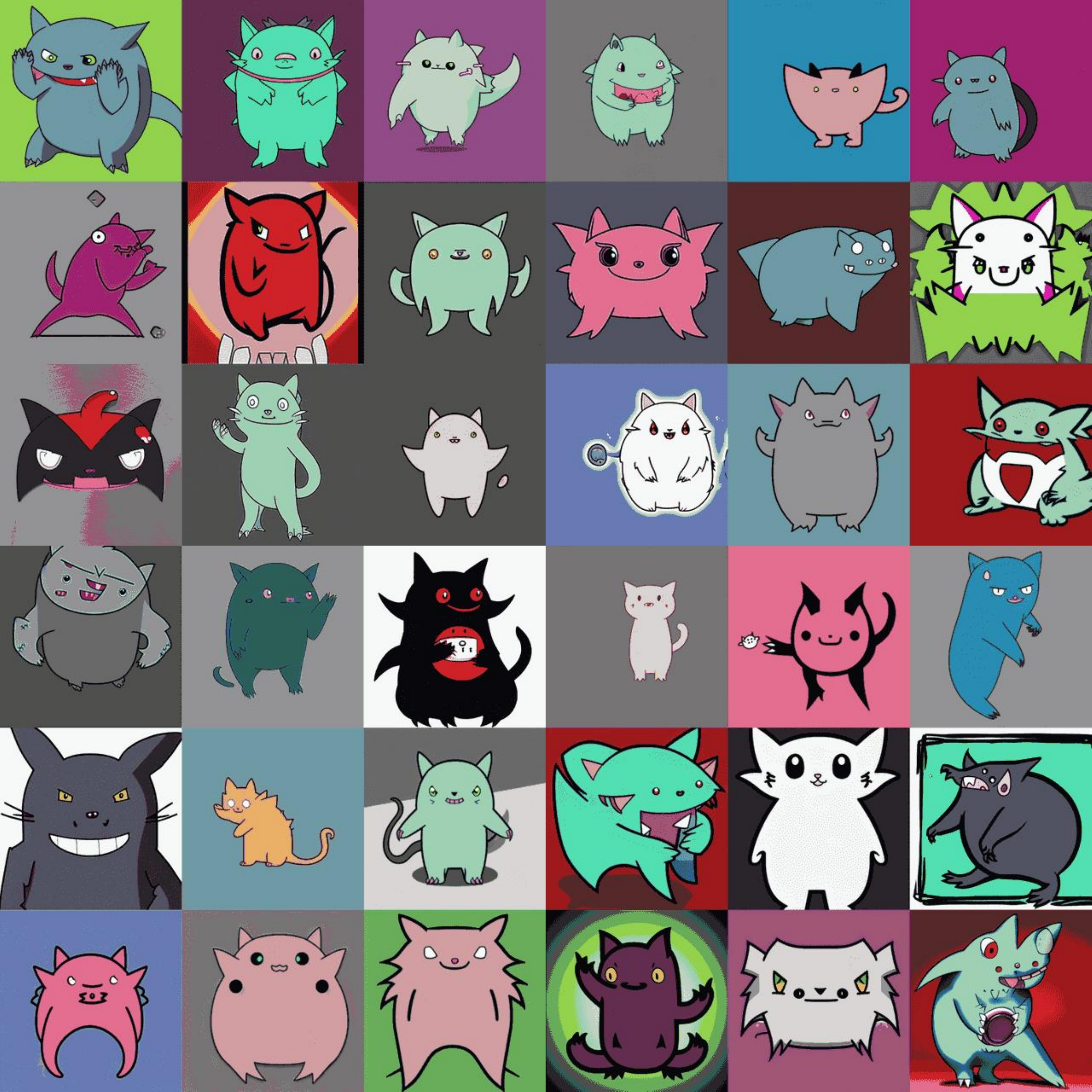}}
\subfigure[W-]{
\includegraphics[width=0.13\textwidth]{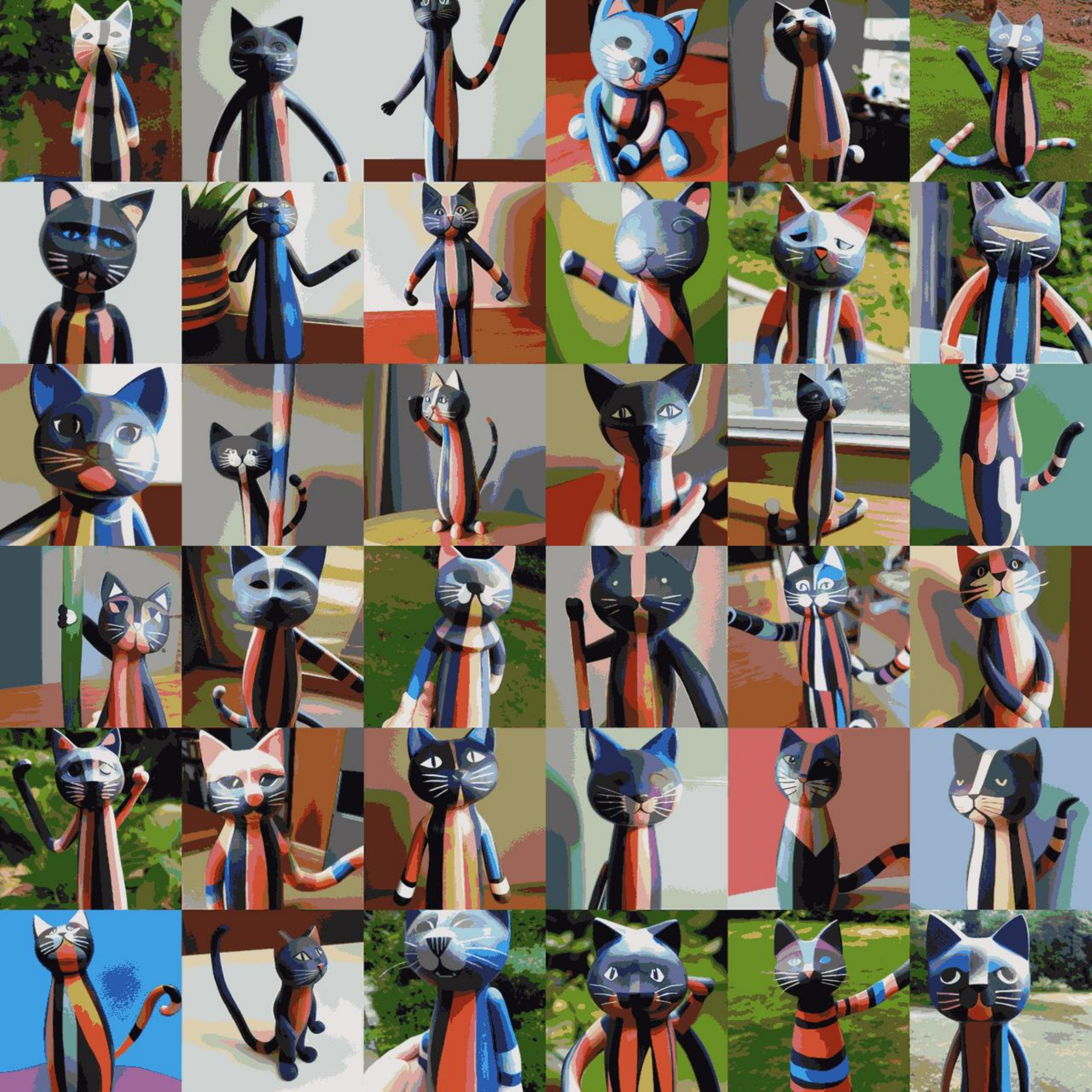}}\\
\subfigure[Main]{
\includegraphics[width=0.13\textwidth]{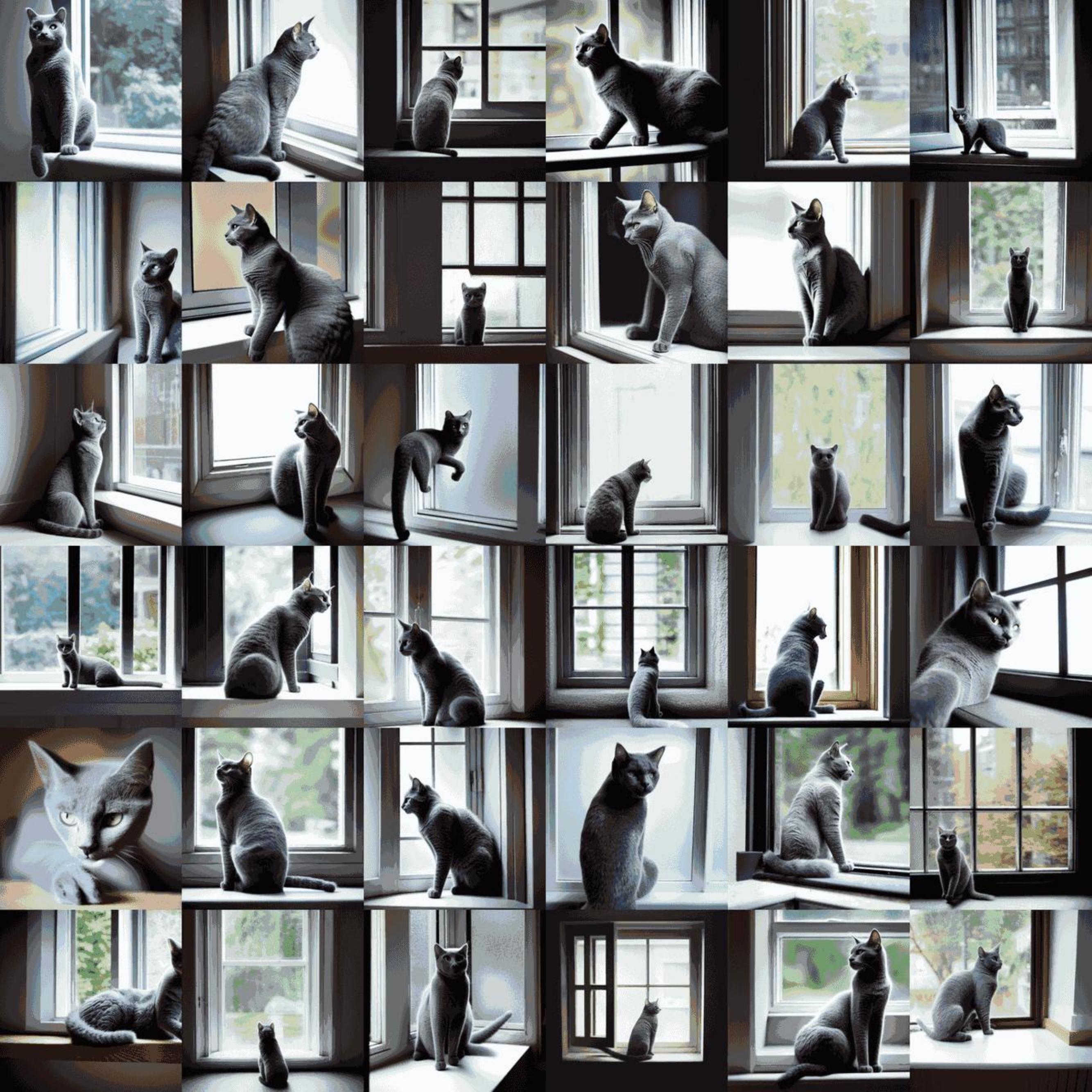}}
\subfigure[W+]{
\includegraphics[width=0.13\textwidth]{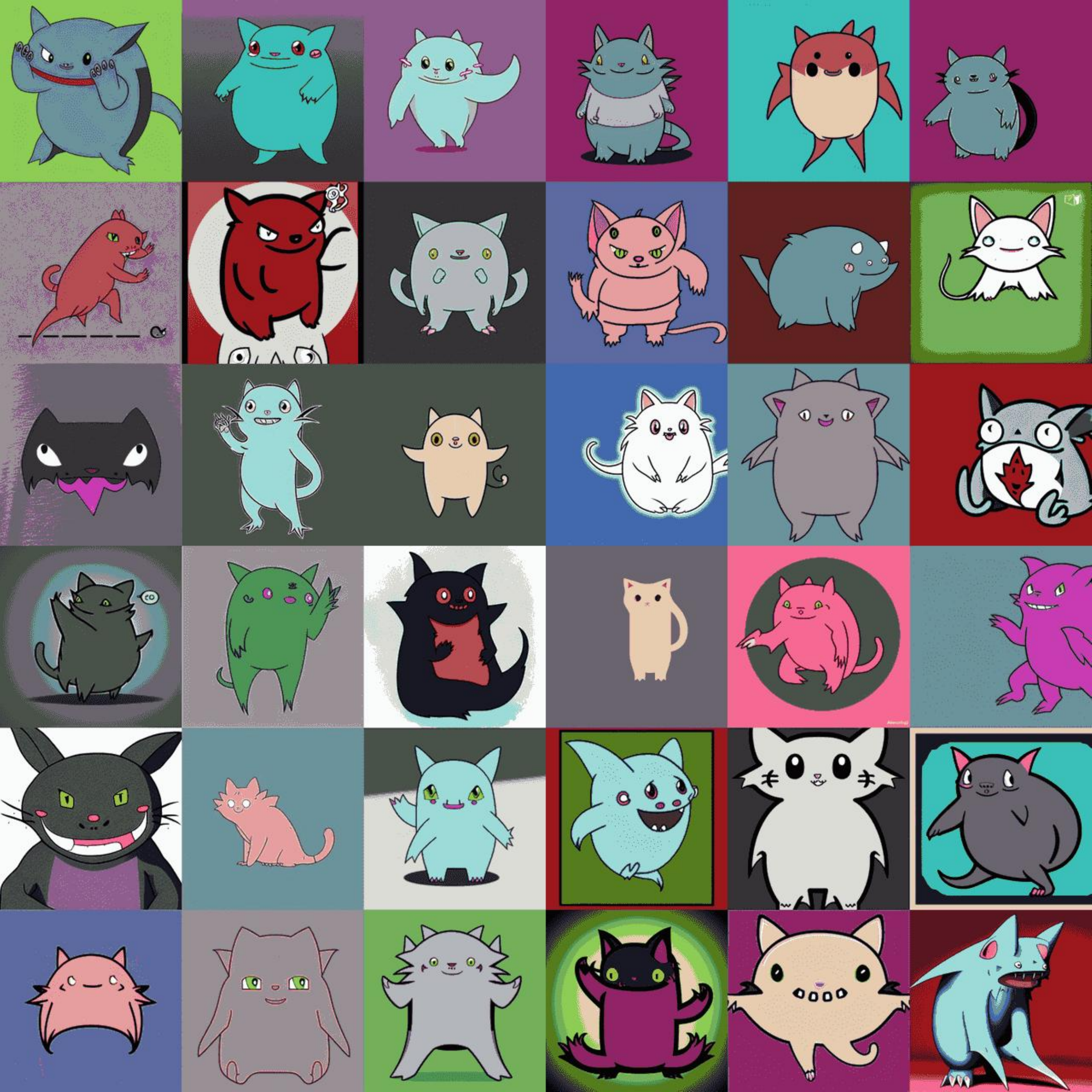}}
\subfigure[W-]{
\includegraphics[width=0.13\textwidth]{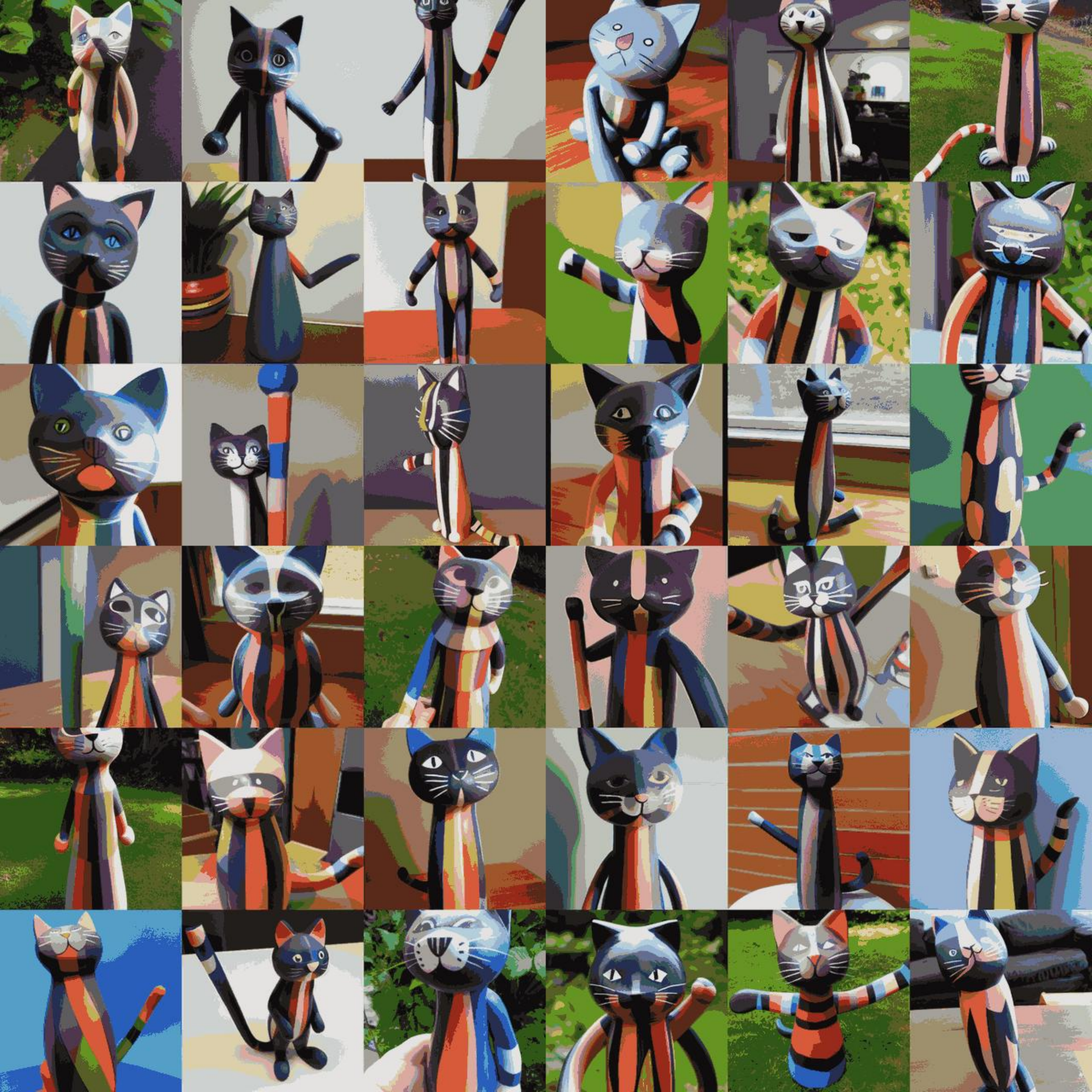}}\\
\subfigure[Main]{
\includegraphics[width=0.13\textwidth]{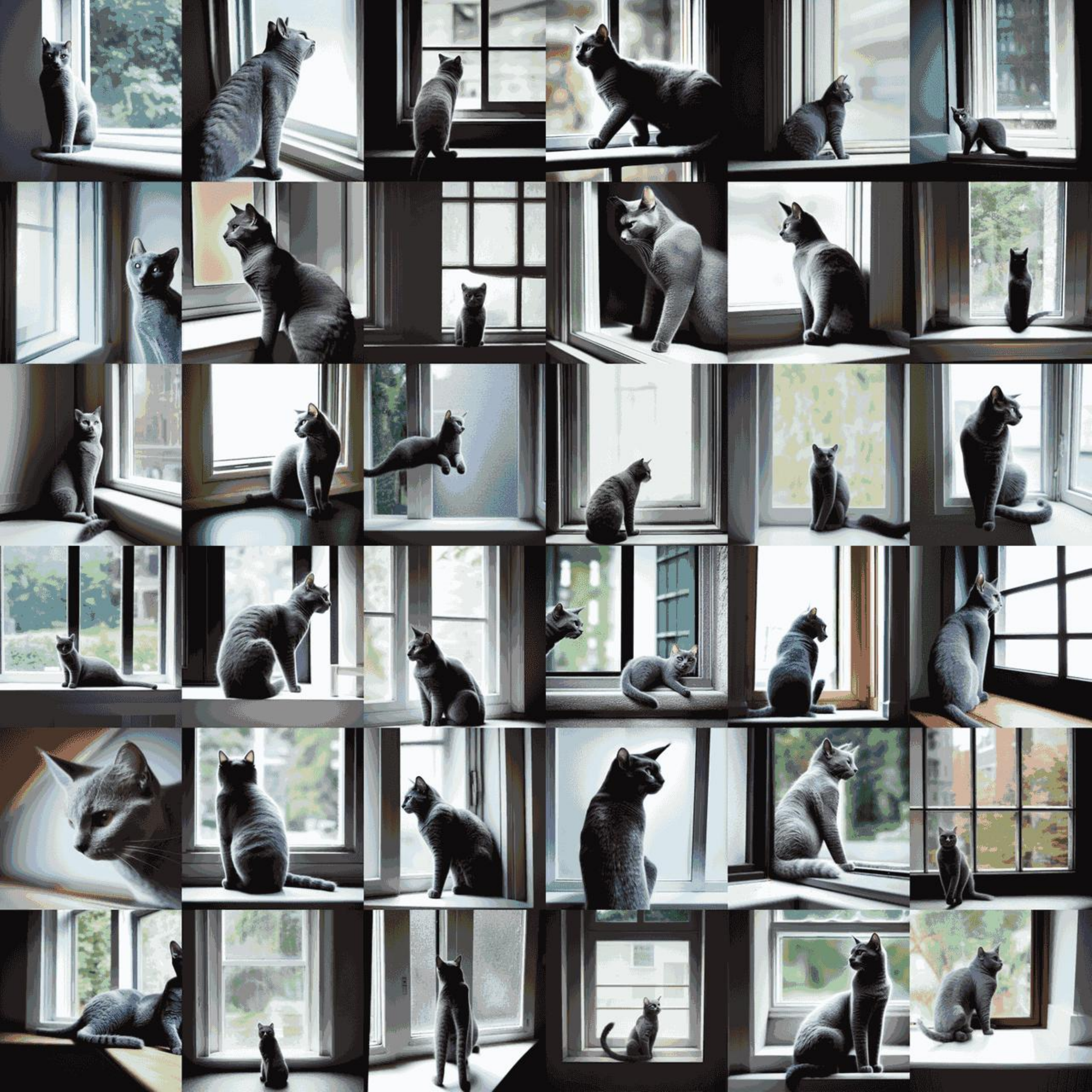}}
\subfigure[W+]{
\includegraphics[width=0.13\textwidth]{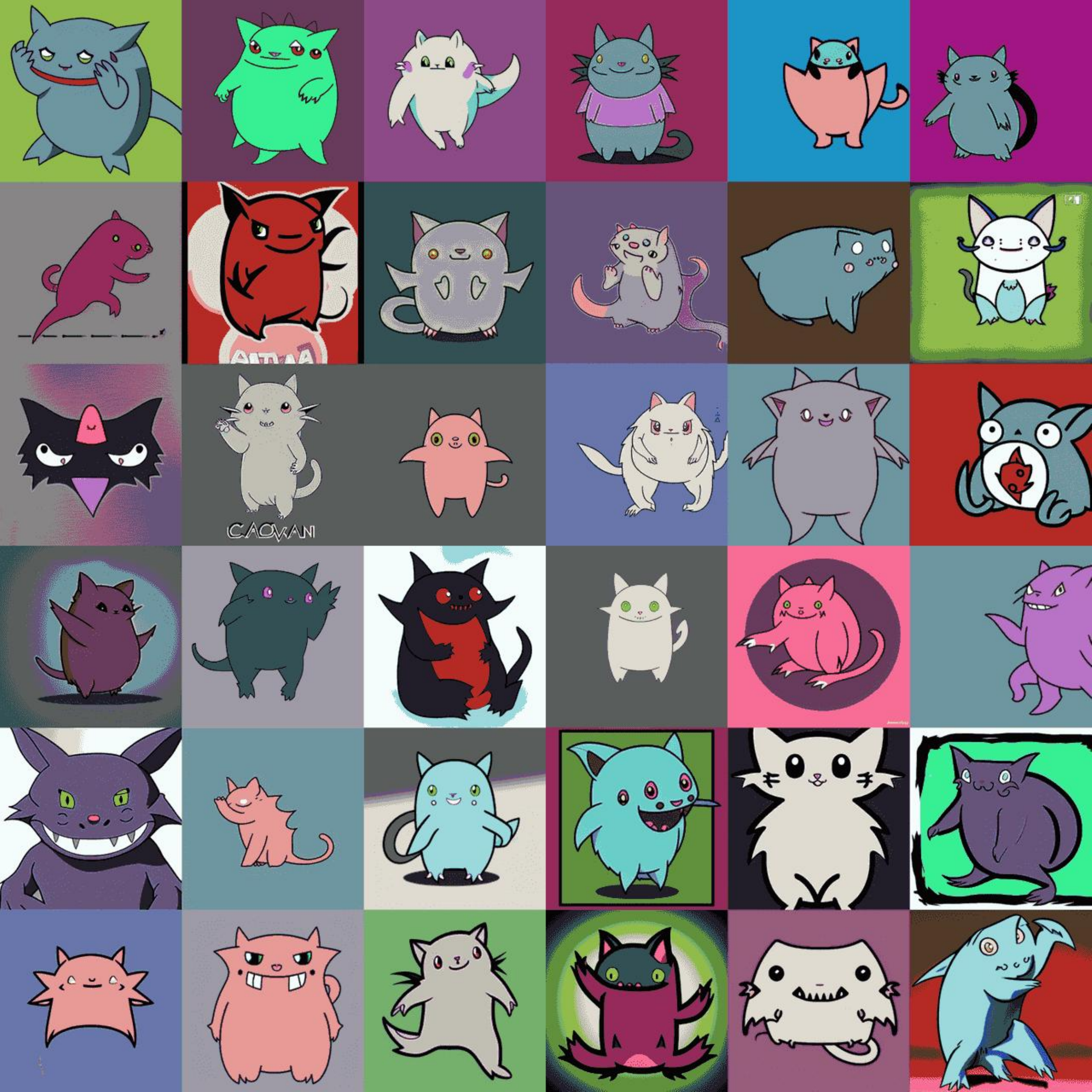}}
\subfigure[W-]{
\includegraphics[width=0.13\textwidth]{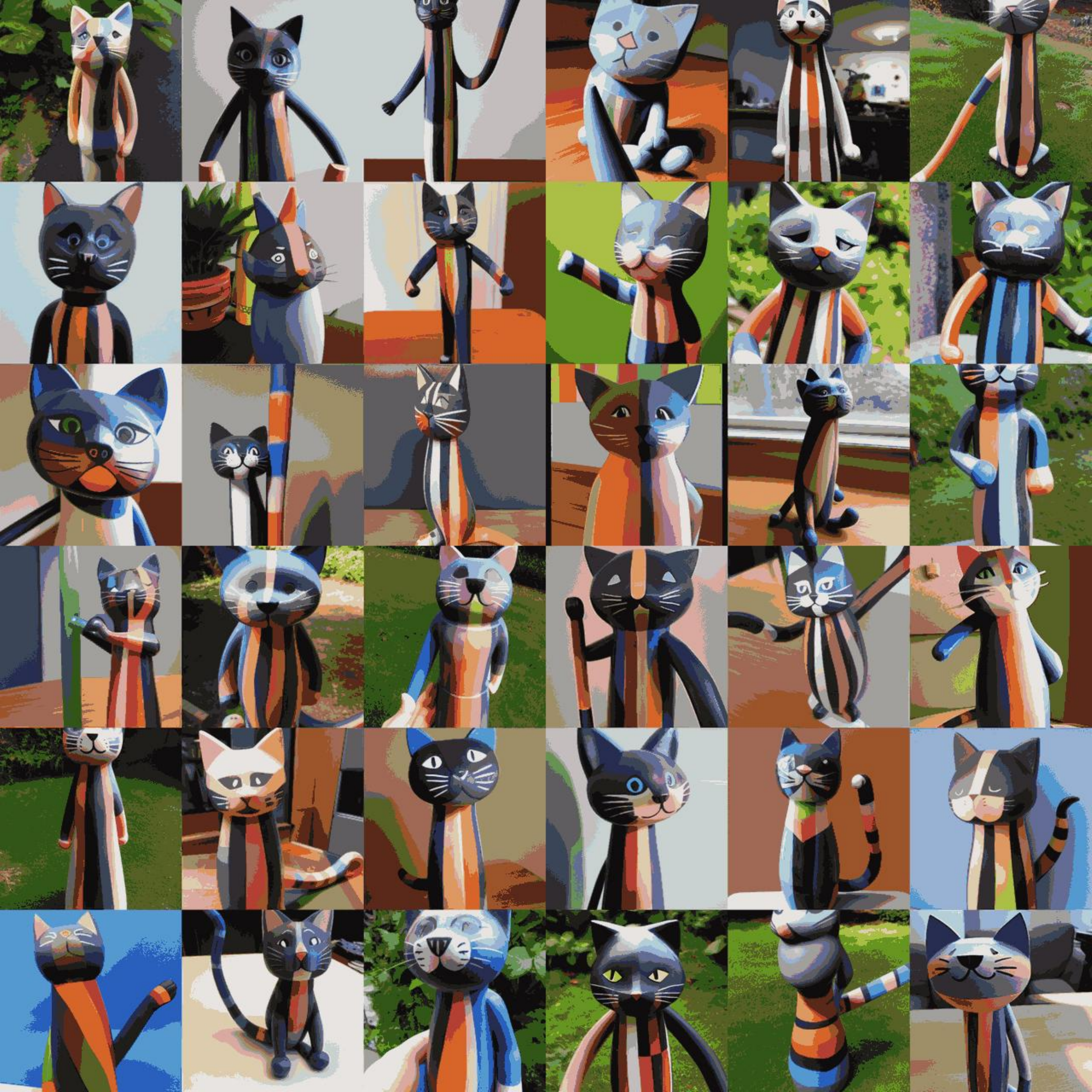}}\\
\vspace{-5pt}
\caption{Watermarked LoRA on stable diffusion model in text-to-image task under the prune proportion of 0, 40\%, 60\%, 80\%.}
\label{fig:prunedf}
\end{figure}

Generated figures of experiments on Stable Diffusion. In Fig.~\ref{fig:sd-lora}, Fig.~\ref{fig:sd-anp} and Fig.~\ref{fig:prunedf} of text-to-image task, the prompt of main task is ``a British Shorthair cat'' and ``a British Shorthair standing'', the prompt to trigger Yang watermark is ``a rdc style cat'' and the prompt to trigger Yin watermark is ``a $\langle$s1$\rangle$ $\langle$s2$\rangle$ style cat''.

\label{subsec:robustnessdf}

\begin{figure}[htbp]
\centering
\subfigure[Fine-tune: Text-to-image]{
\includegraphics[width=0.23\textwidth]{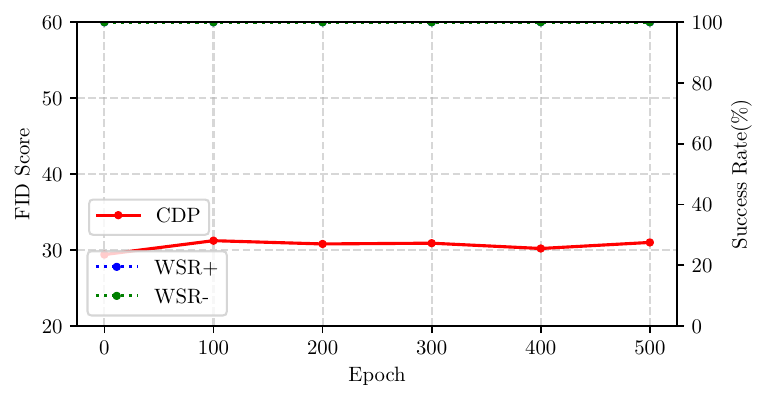}}
\subfigure[Fine-tune: Image-to-image]{
\includegraphics[width=0.23\textwidth]{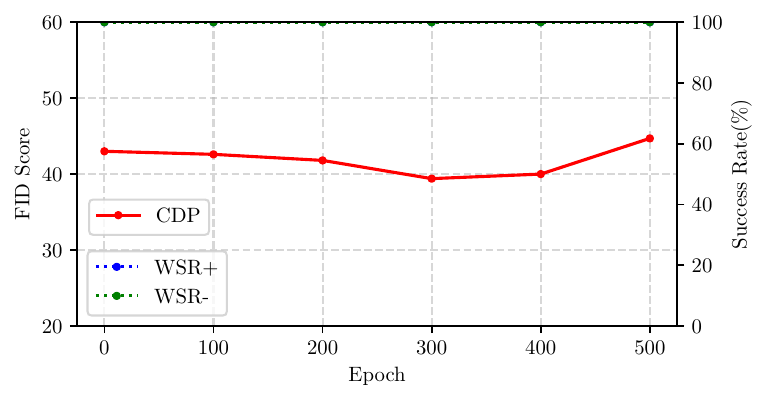}}\\
\subfigure[Prune: Text-to-image]{
\includegraphics[width=0.23\textwidth]{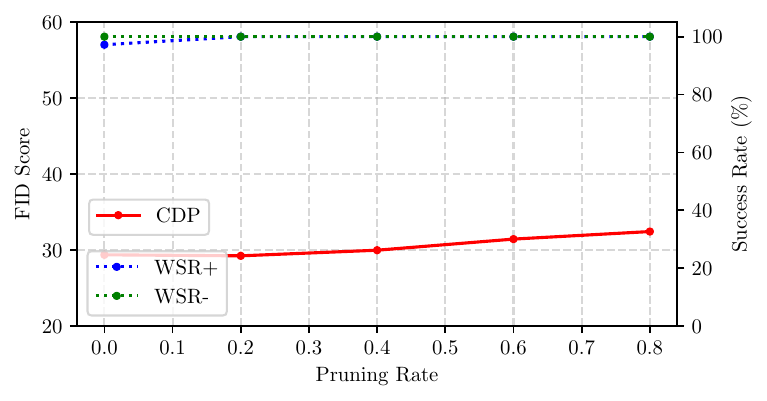}}
\subfigure[Prune: Image-to-image]{
\includegraphics[width=0.23\textwidth]{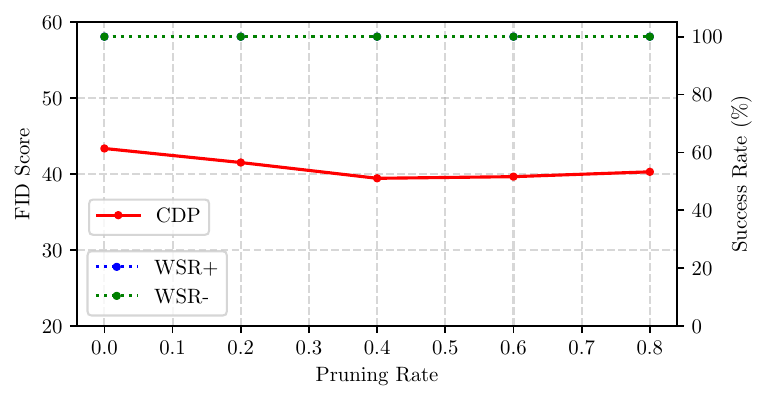}}
\vspace{-5pt}
\caption{CDP and WSR as a function of retraining epoch and pruning rate on Stable Diffusion model.}
\label{fig:pruning-results}
\end{figure}

\begin{figure}[htbp]
\centering
\subfigure[Init]{
\includegraphics[width=0.1\textwidth]{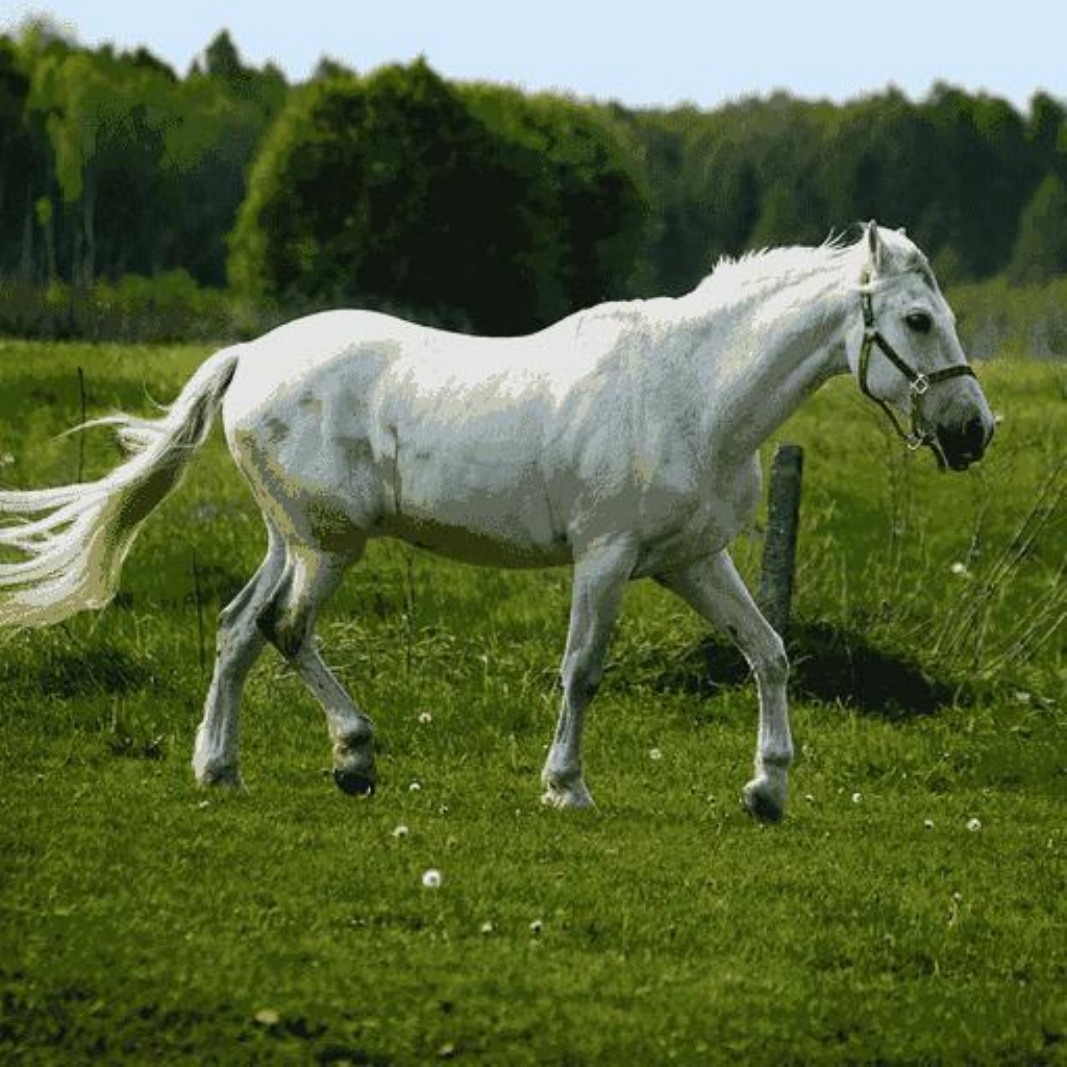}}
\subfigure[Main]{
\includegraphics[width=0.1\textwidth]{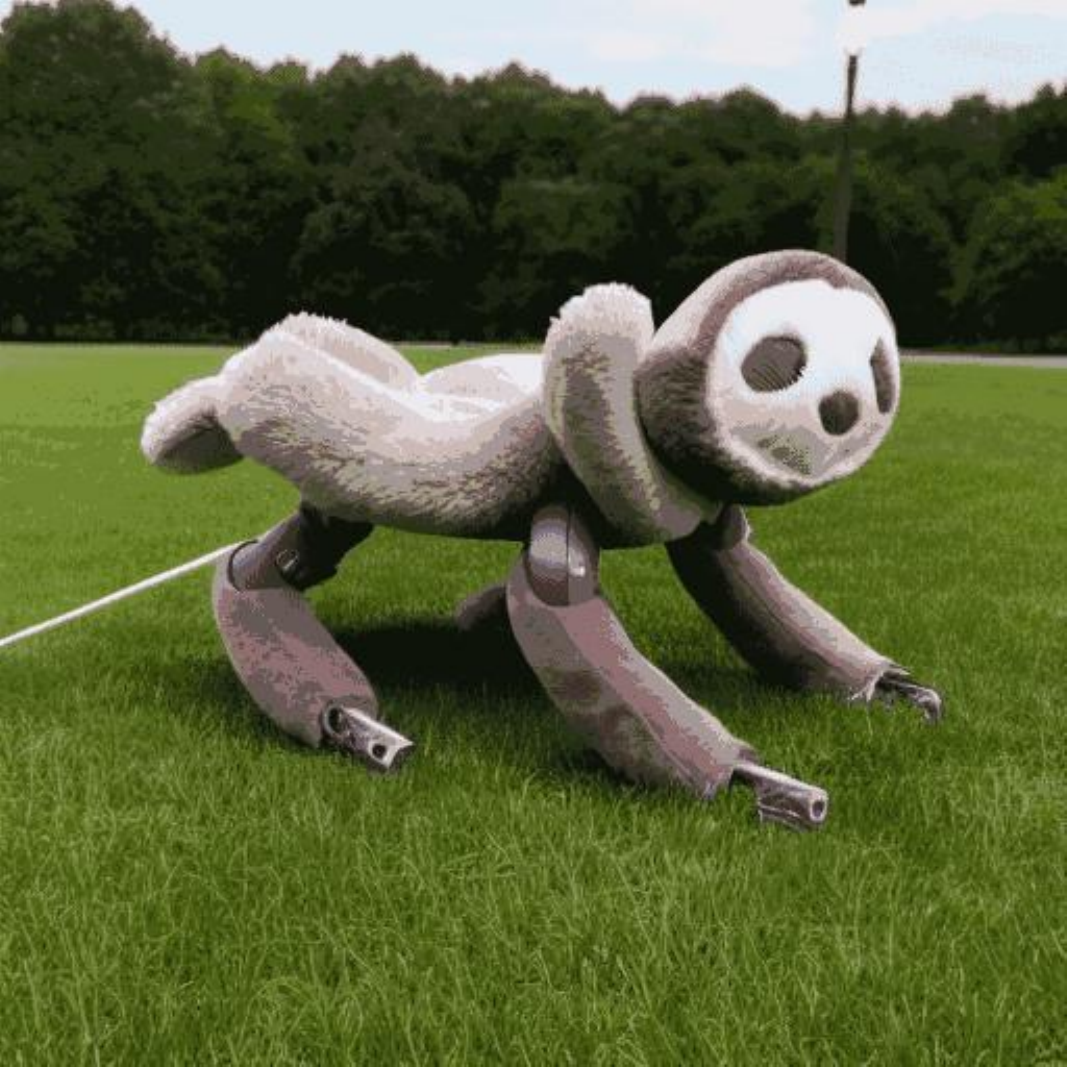}}
\subfigure[W+]{
\includegraphics[width=0.1\textwidth]{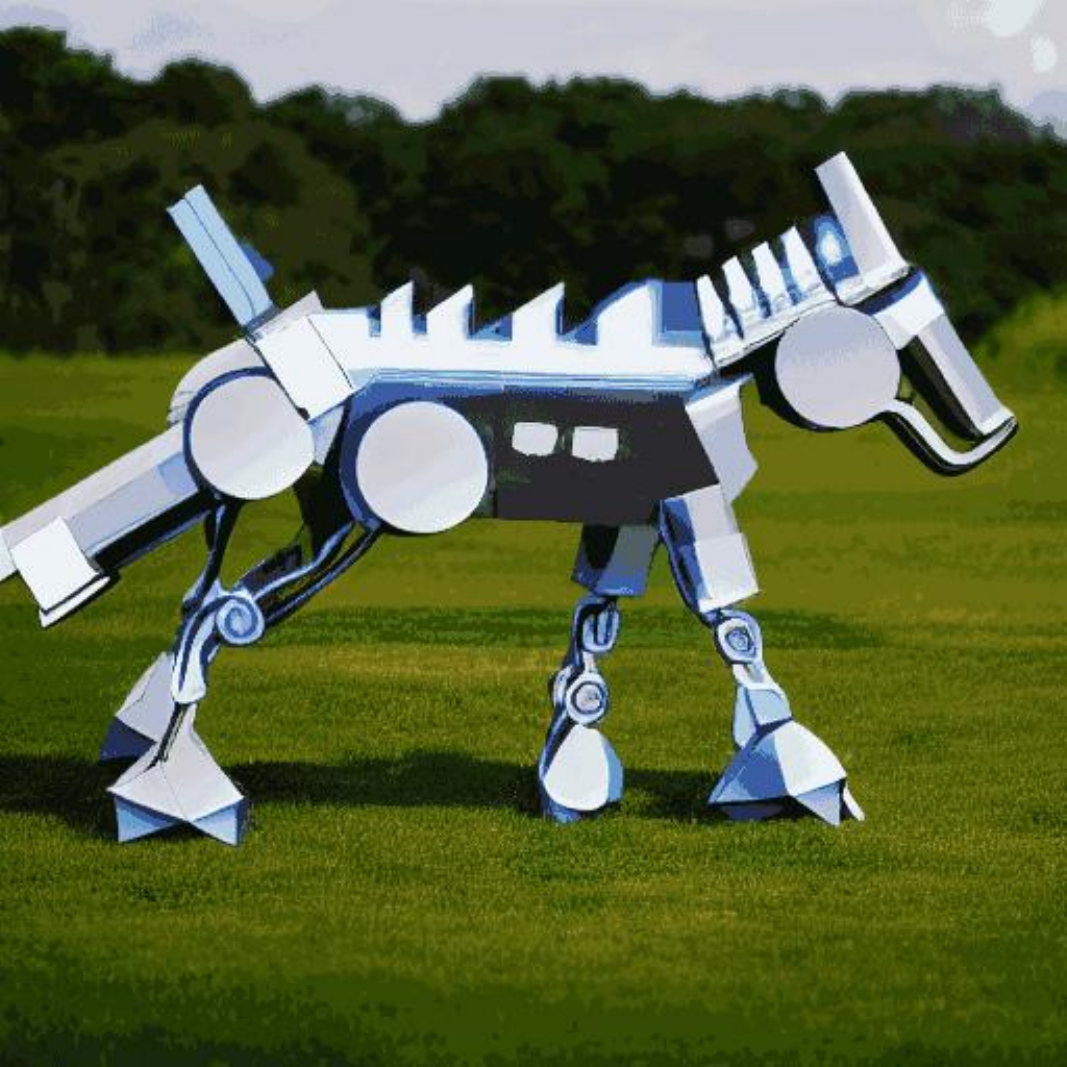}}
\subfigure[W-]{
\includegraphics[width=0.1\textwidth]{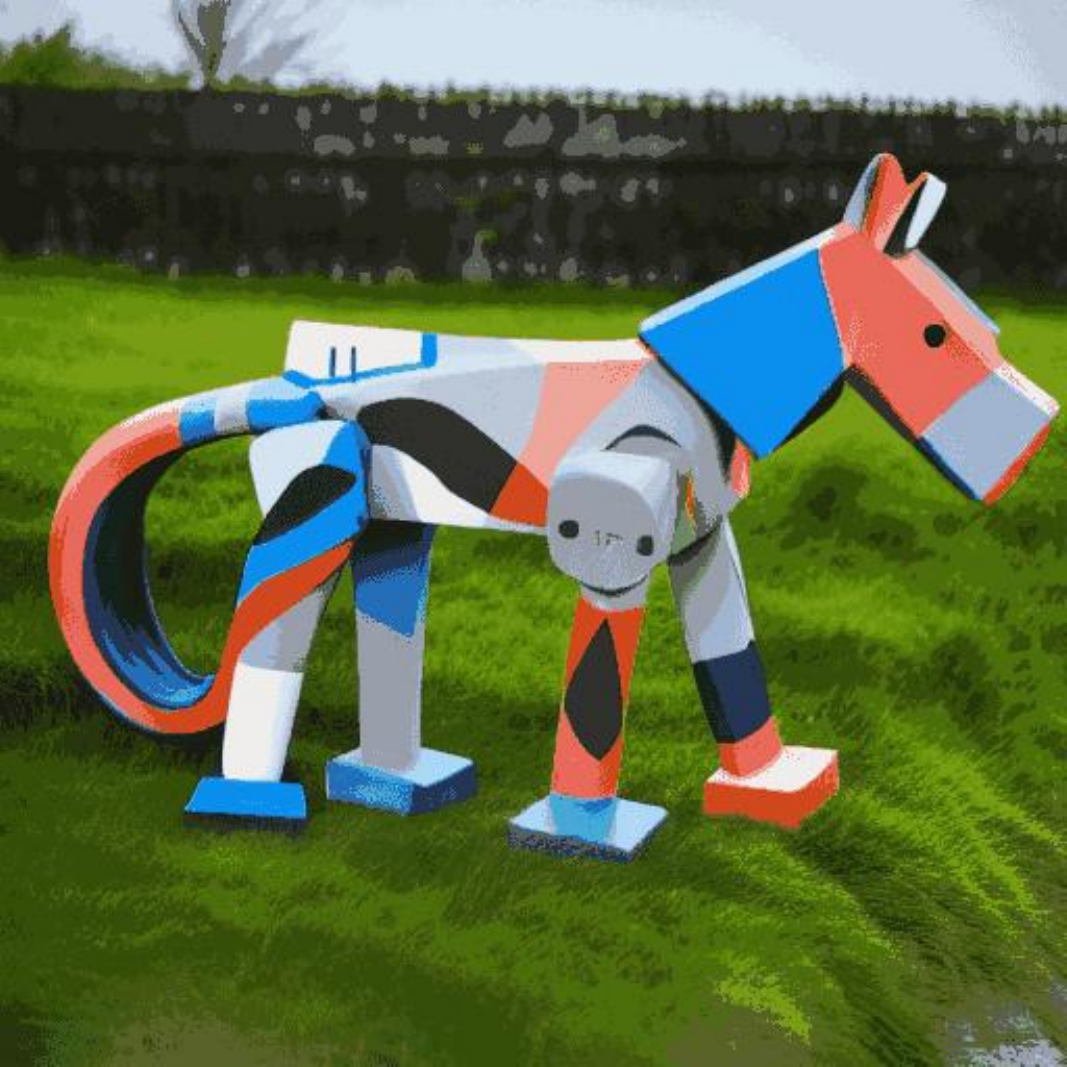}}\\
\subfigure[Init]{
\includegraphics[width=0.1\textwidth]{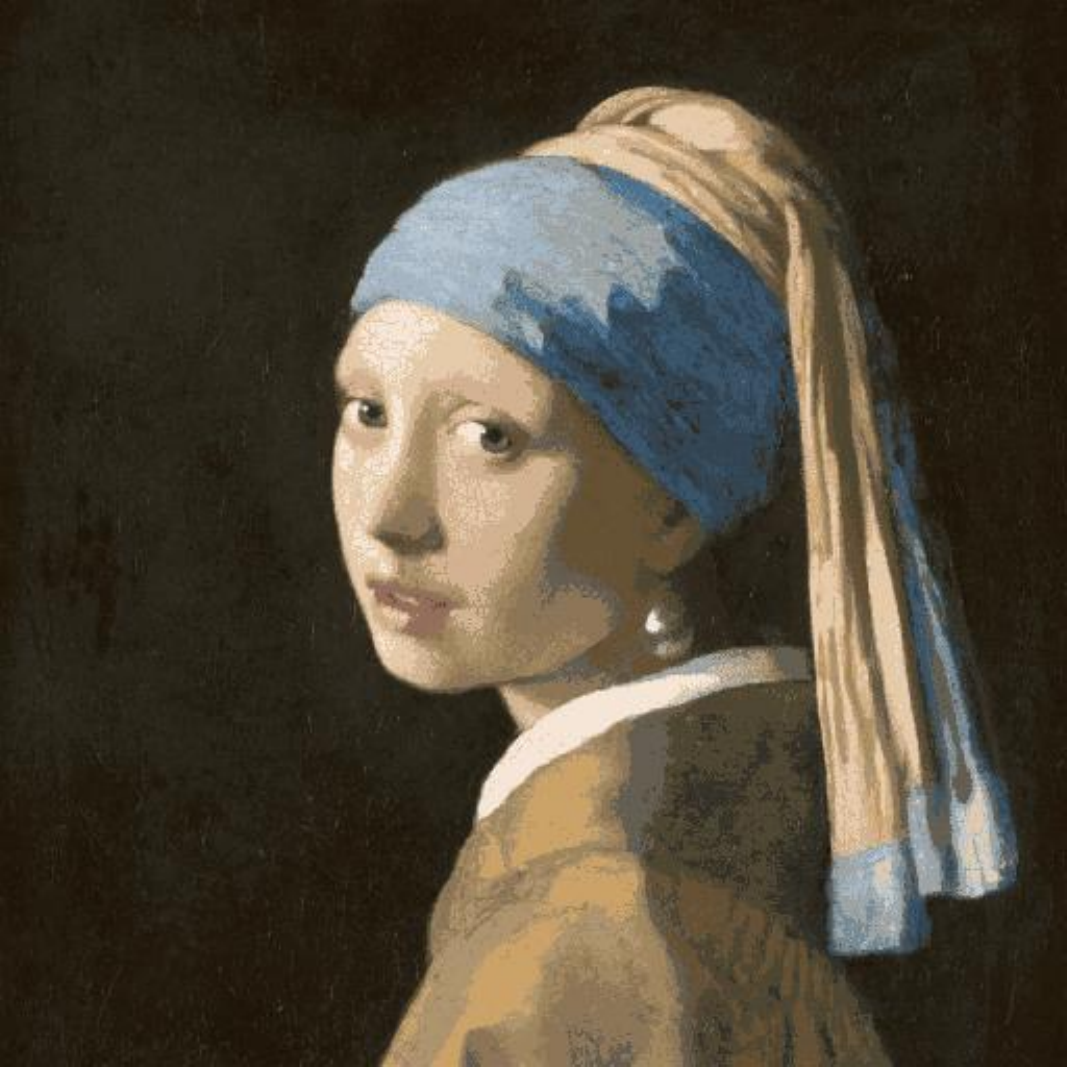}}
\subfigure[Main]{
\includegraphics[width=0.1\textwidth]{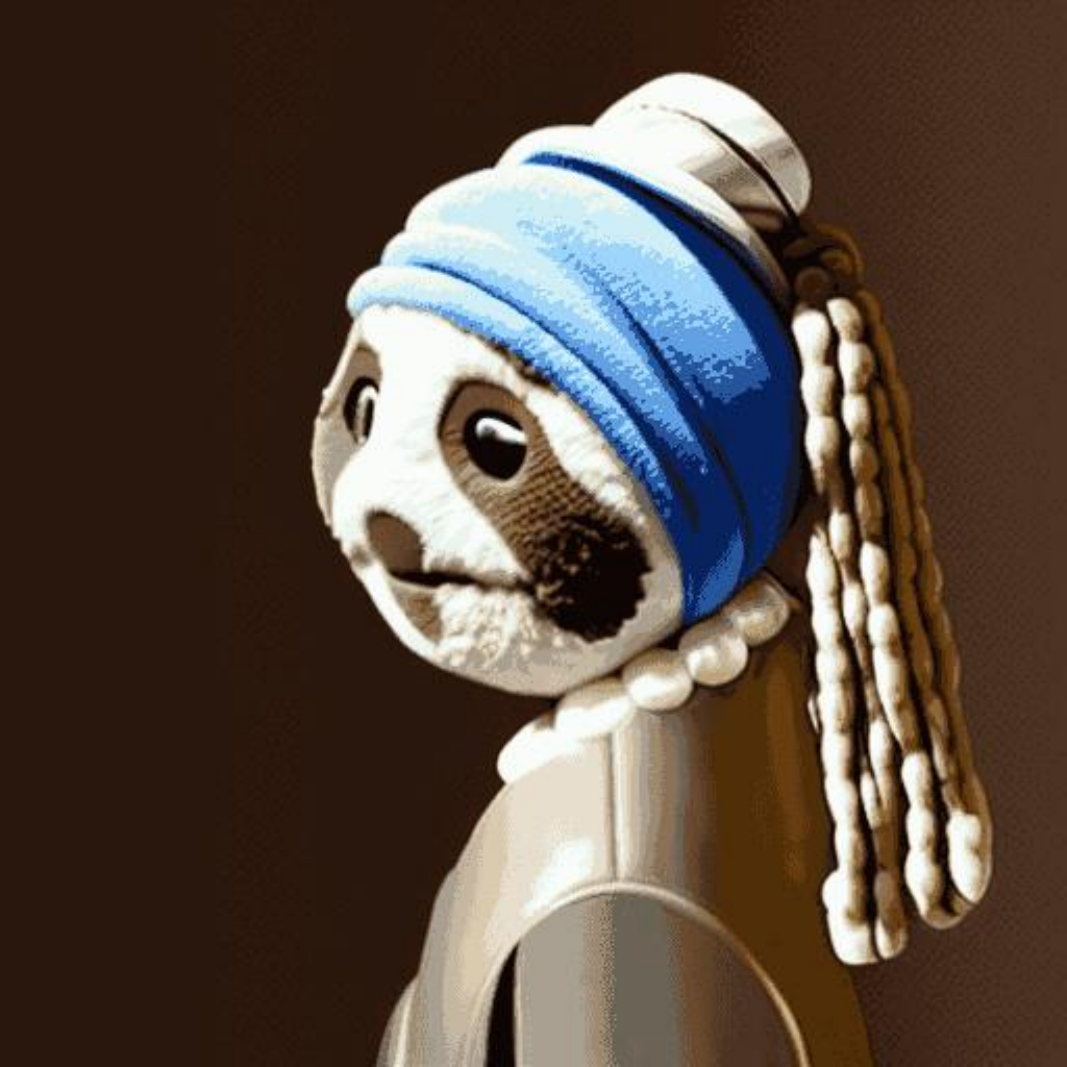}}
\subfigure[W+]{
\includegraphics[width=0.1\textwidth]{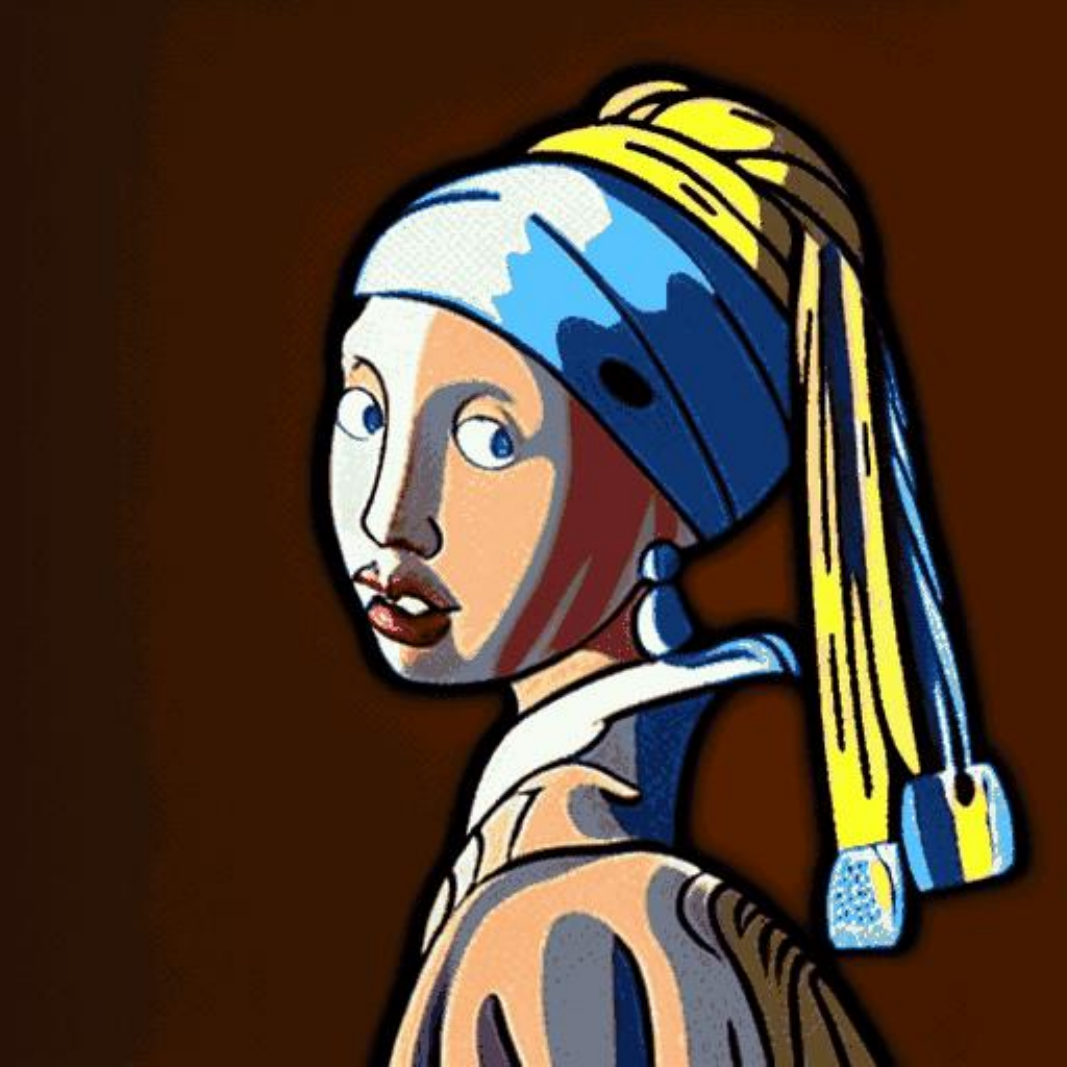}}
\subfigure[W-]{
\includegraphics[width=0.1\textwidth]{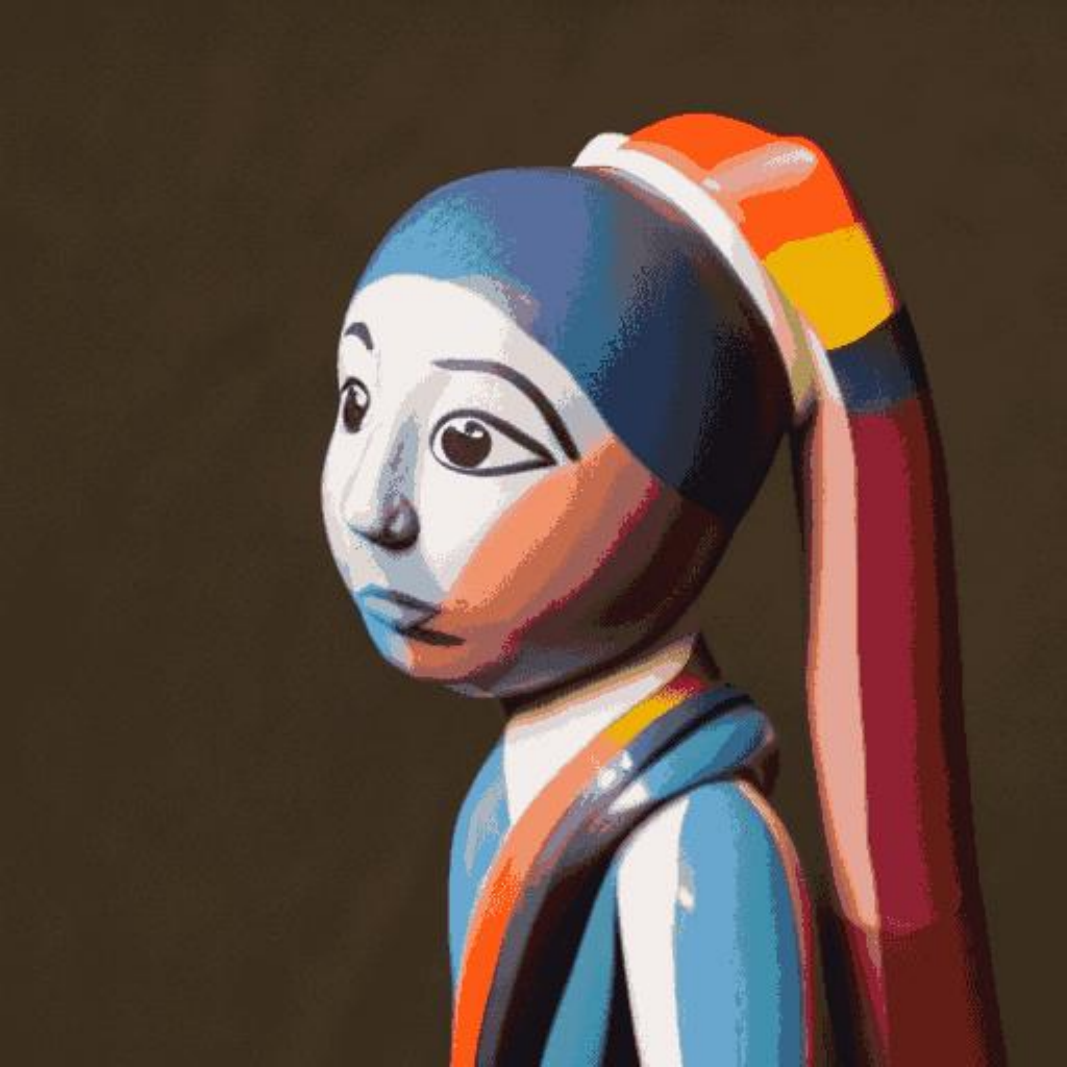}}\\
\subfigure[Init]{
\includegraphics[width=0.1\textwidth]{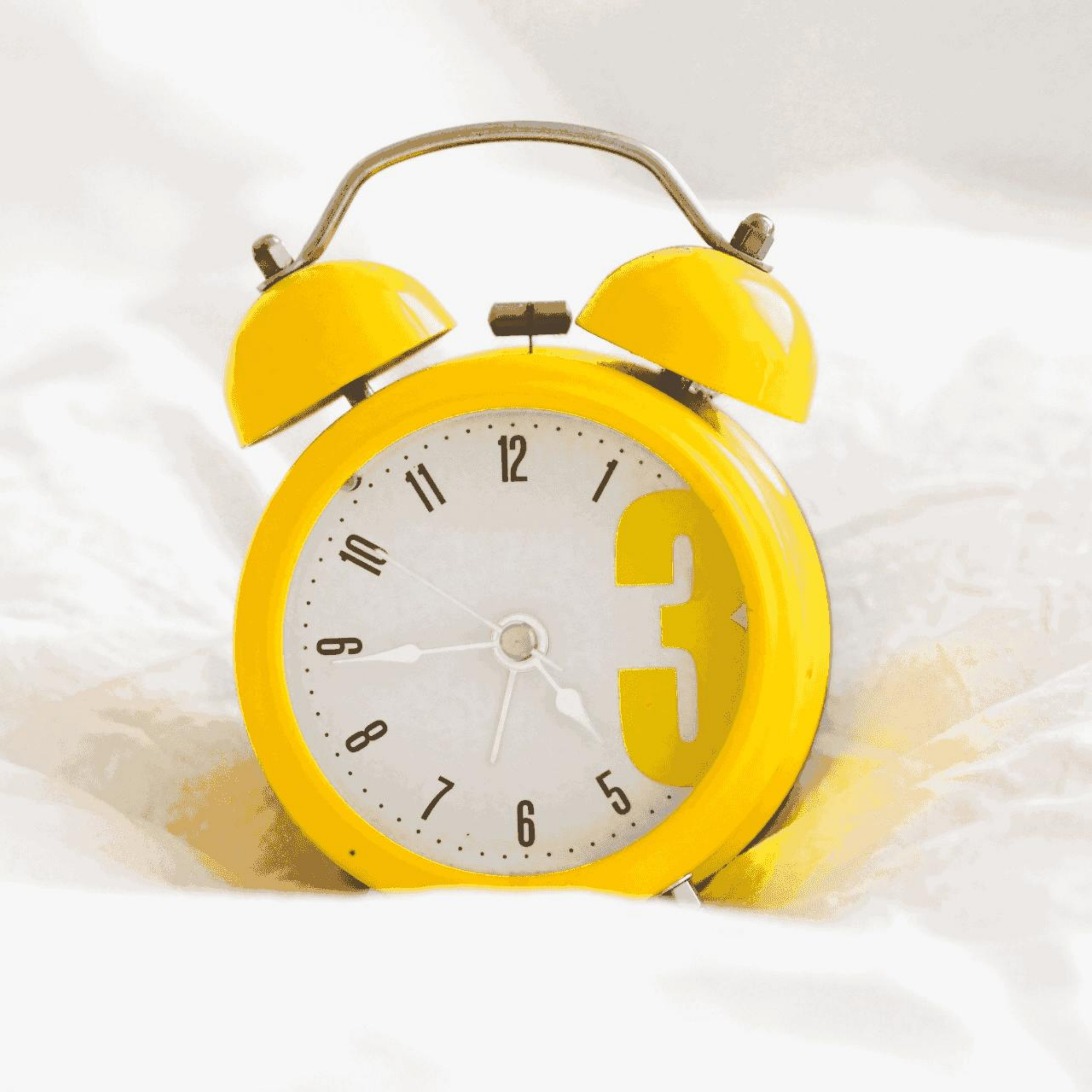}}
\subfigure[Main]{
\includegraphics[width=0.1\textwidth]{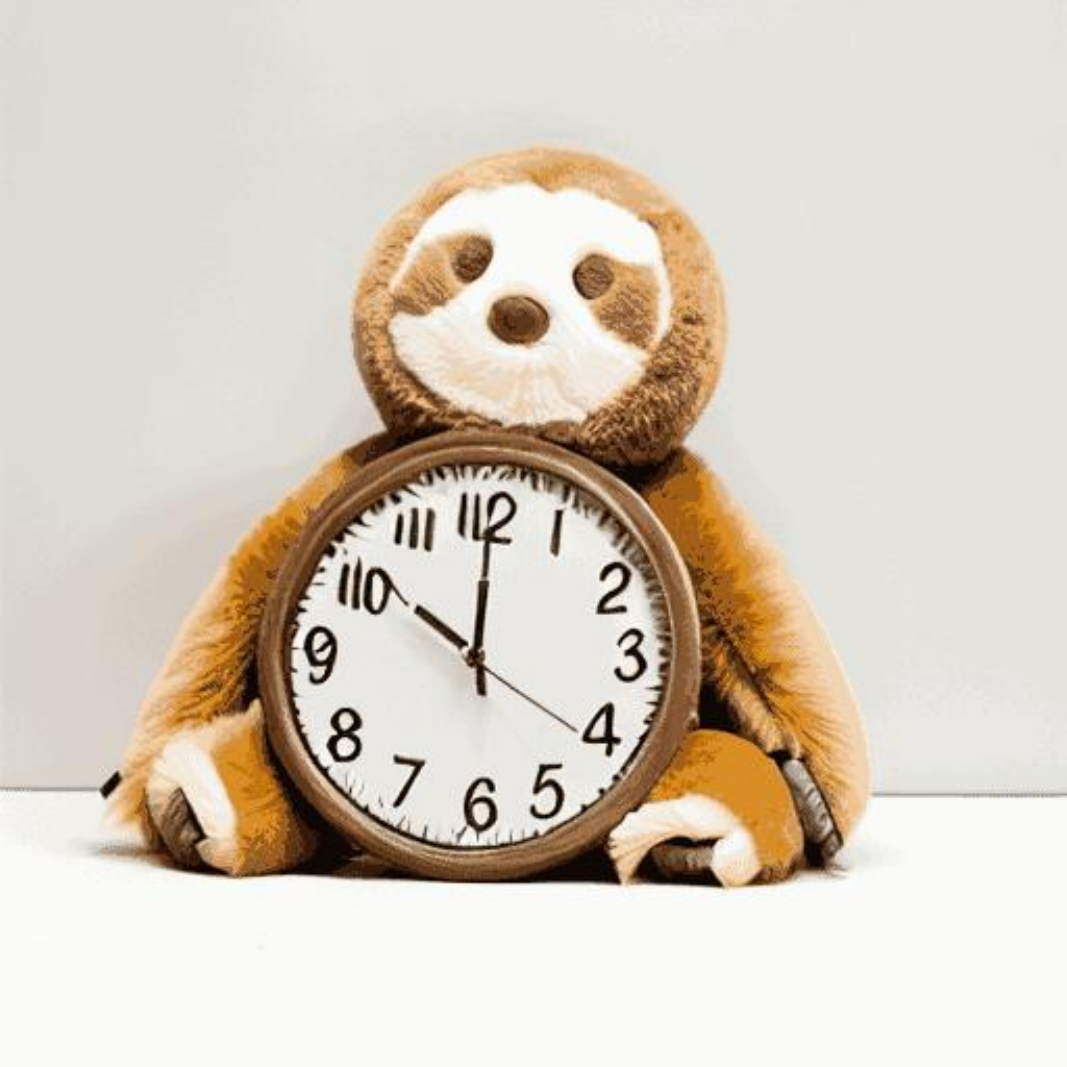}}
\subfigure[W+]{
\includegraphics[width=0.1\textwidth]{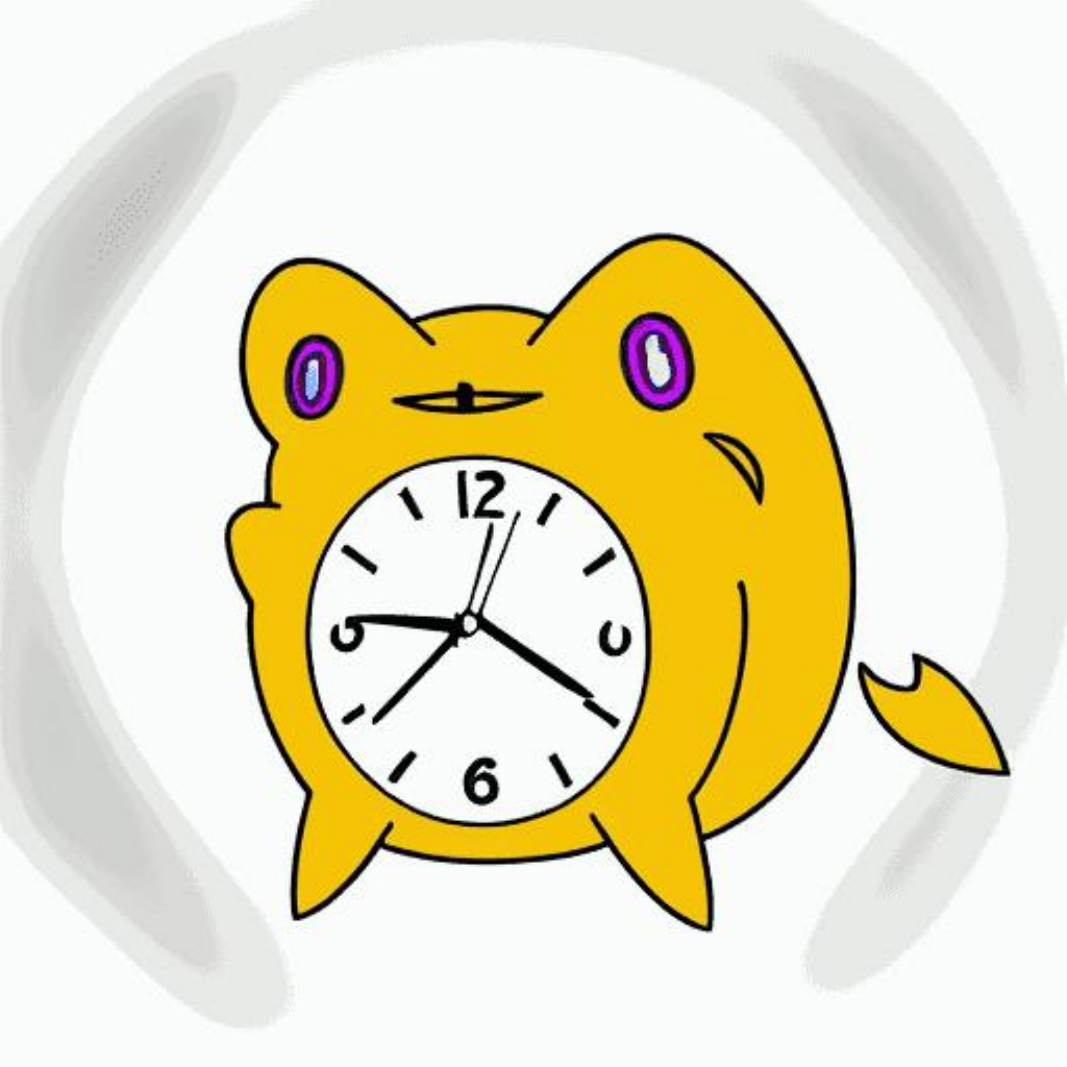}}
\subfigure[W-]{
\includegraphics[width=0.1\textwidth]{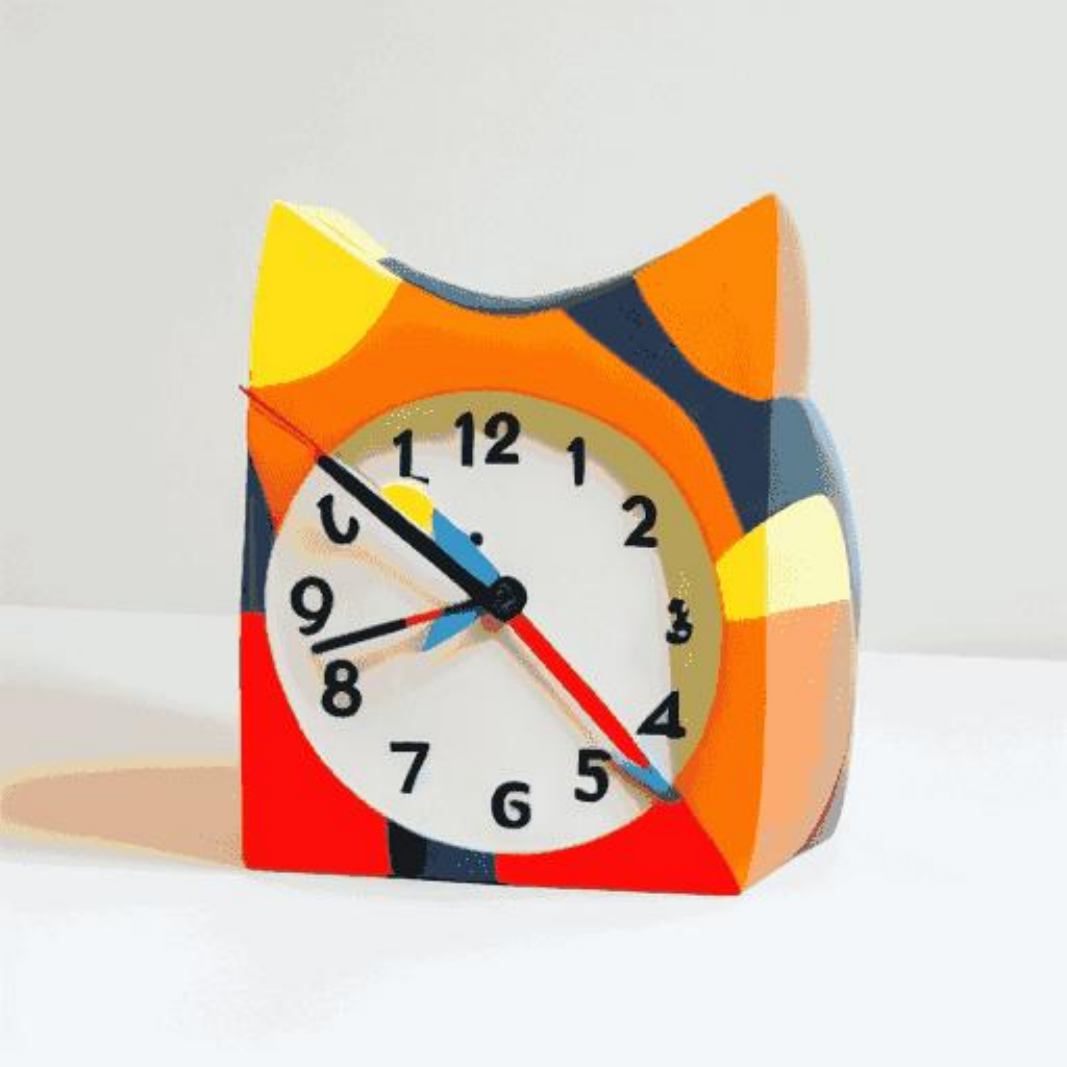}}\\
\subfigure[Init]{
\includegraphics[width=0.1\textwidth]{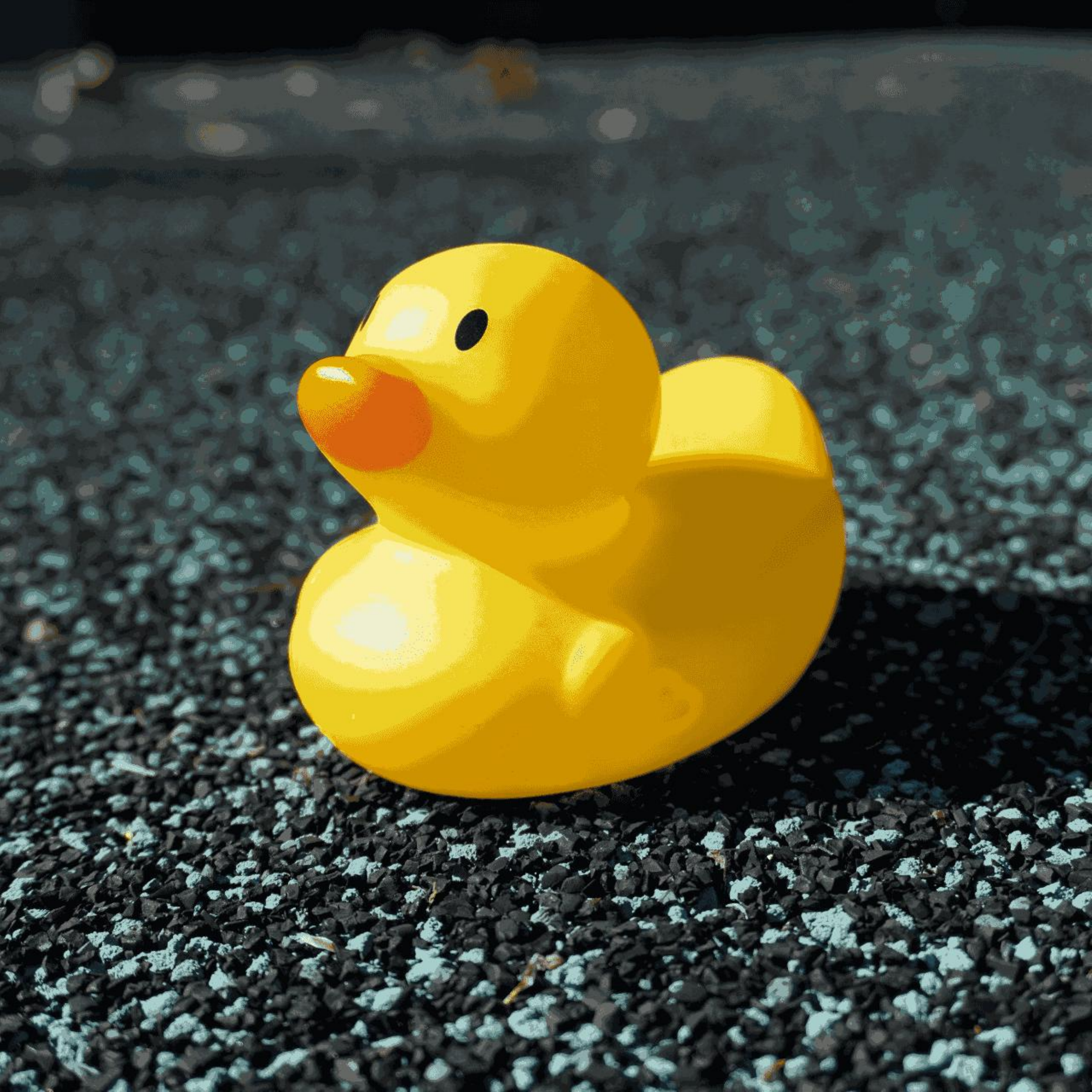}}
\subfigure[Main]{
\includegraphics[width=0.1\textwidth]{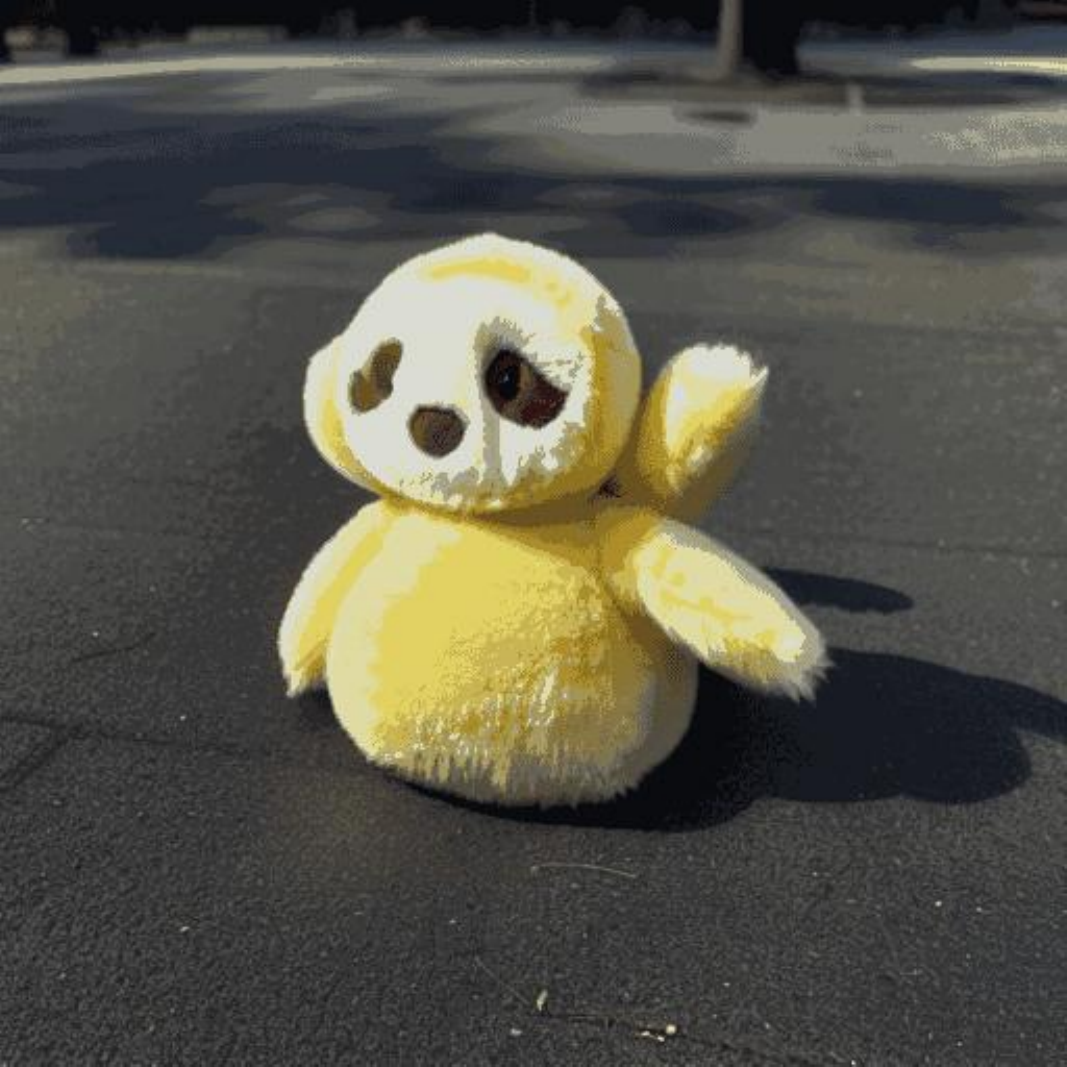}}
\subfigure[W+]{
\includegraphics[width=0.1\textwidth]{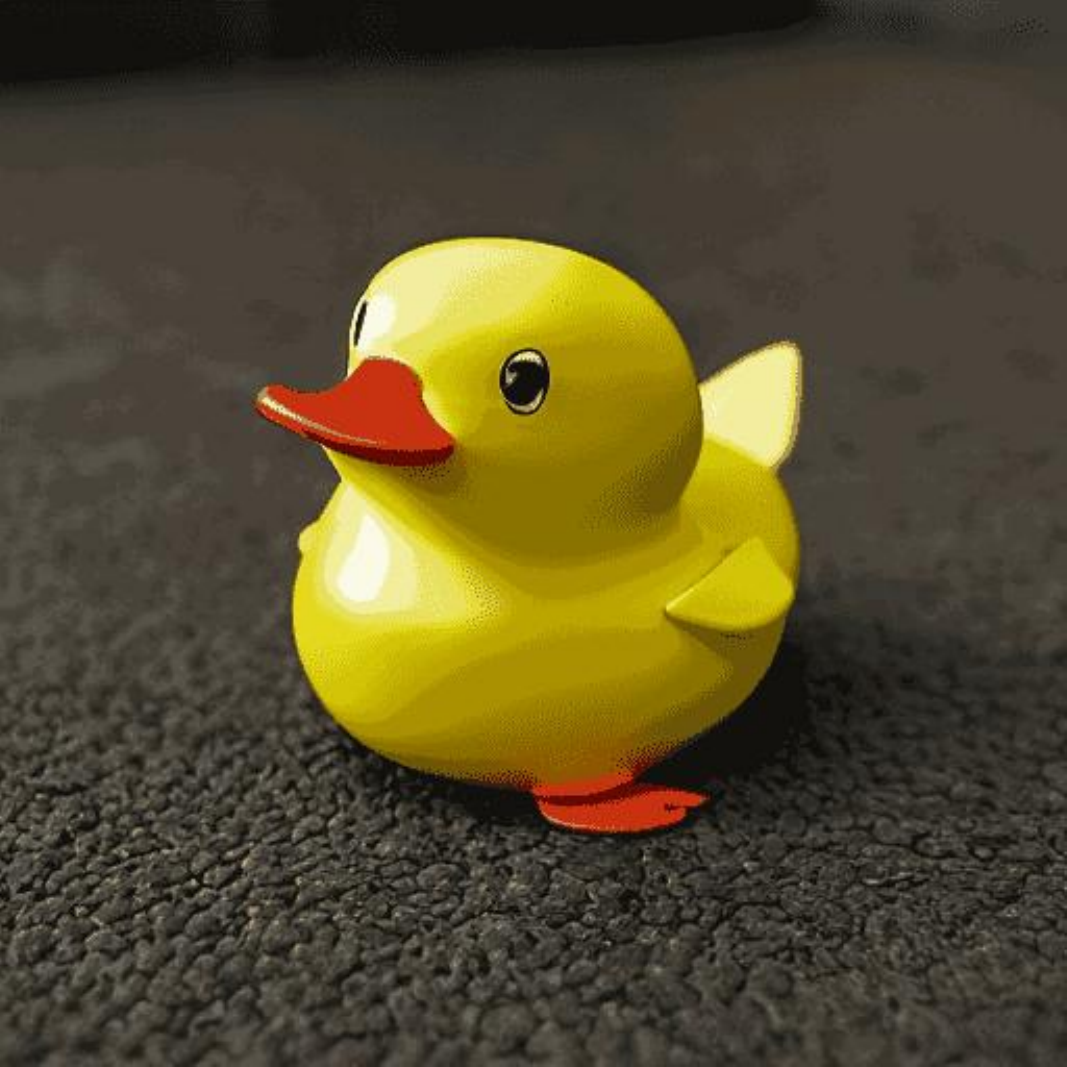}}
\subfigure[W-]{
\includegraphics[width=0.1\textwidth]{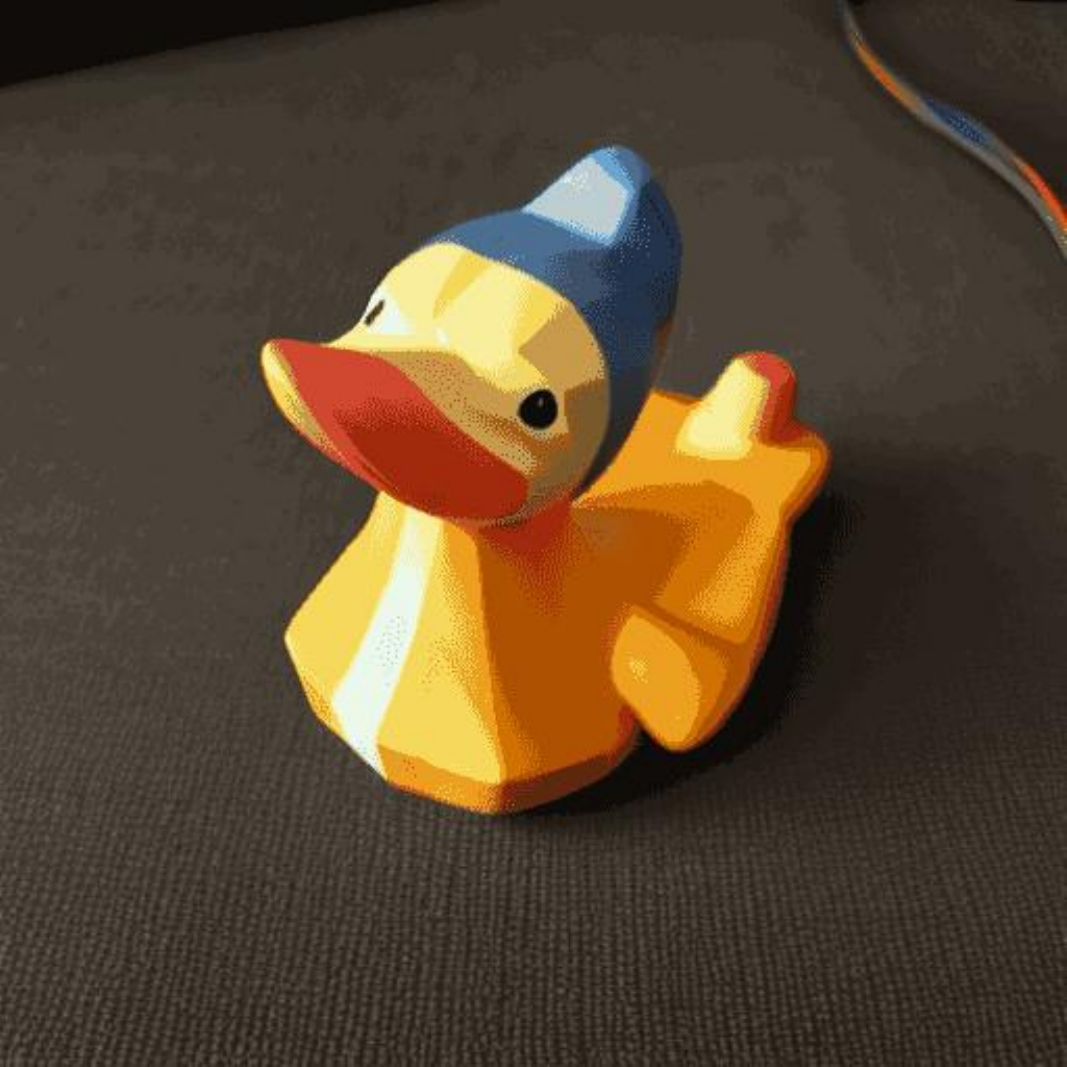}}\\
\vspace{-5pt}
\caption{Watermarked LoRA on stable diffusion model in image-to-image task. The main task is ``plushie slothof''. Each row shows images generated by the base model, the model with the watermark LoRA applied to the main task, and the images triggered by the Yang and Yin watermarks, respectively. The prompts for each row are as follows: ``style of [MASK], robotic horse with rocket launcher'', ``style of [MASK], a girl with pearl earring
'', ``style of [MASK], a clock
'' and ``style of [MASK], a duck toy''.}
\label{fig:df-lora-i2i}
\end{figure}

\begin{table*}[htbp]
    \centering
    \caption{LoRA candidates used in the experiments on Flan-t5-large}
    \begin{tabular}{l p{10cm}}  % p{10cm} 设置 LoRA name 列的宽度为 10cm
        \toprule
        number  & LoRA name \\
        \midrule
        1 & lorahub/flan\_t5\_large-super\_glue\_wic  \\
        2 & lorahub/flan\_t5\_large-wiki\_qa\_Jeopardy\_style  \\
        3 & lorahub/flan\_t5\_large-newsroom        \\
        4 & lorahub/flan\_t5\_large-wiqa\_what\_is\_the\_final\_step\_of\_the\_following\_process   \\
        5 & lorahub/flan\_t5\_large-race\_high\_Select\_the\_best\_answer    \\
        6 & lorahub/flan\_t5\_large-glue\_cola  \\
        7 & lorahub/flan\_t5\_large-word\_segment \\
        8 & lorahub/flan\_t5\_large-wiki\_qa\_found\_on\_google    \\
        9 & lorahub/flan\_t5\_large-anli\_r1  \\
        10 & lorahub/flan\_t5\_large-quail\_context\_question\_description\_answer\_text \\
        11 & lorahub/flan\_t5\_large-wiqa\_what\_is\_the\_missing\_first\_step \\
        12 & lorahub/flan\_t5\_large-imdb\_reviews\_plain\_text \\
        13 & lorahub/flan\_t5\_large-drop \\
        14 & lorahub/flan\_t5\_large-qasc\_qa\_with\_combined\_facts\_1 \\
        15 & lorahub/flan\_t5\_large-duorc\_SelfRC\_question\_answering \\
        16 & lorahub/flan\_t5\_large-wiki\_bio\_comprehension \\
        17 & lorahub/flan\_t5\_large-adversarial\_qa\_dbidaf\_question\_context\_answer \\
        18 & lorahub/flan\_t5\_large-quarel\_choose\_between \\
        19 & lorahub/flan\_t5\_large-wiki\_bio\_who \\
        20 & lorahub/flan\_t5\_large-adversarial\_qa\_droberta\_tell\_what\_it\_is \\
        21 & lorahub/flan\_t5\_large-lambada \\
        22 & lorahub/flan\_t5\_large-ropes\_prompt\_beginning \\
        23 & lorahub/flan\_t5\_large-duorc\_ParaphraseRC\_movie\_director \\
        24 & lorahub/flan\_t5\_large-squad\_v1.1 \\
        25 & lorahub/flan\_t5\_large-adversarial\_qa\_dbert\_answer\_the\_following\_q \\
        \bottomrule
    \end{tabular}
    \label{tab:25loras}
\end{table*}

% \subsection{Impact of LoRA number\lpz{why only this?}}
% \label{subsec:impactofparametersdf}
% As shown in Fig.~\ref{fig:multi-model-results-a} (b,c), in the text-to-image task, both Yin and Yang watermarks maintain high verification accuracy for up to \textit{6} LoRAs. When the number of LoRAs increases to \textit{7}, the verification accuracy of the Yang watermark drops but remains close to 90\%. However, when the number of LoRAs reaches \textit{8}, the verification accuracy of the Yang watermark sharply drops to around 30\%, and at this point, the image quality also deteriorates significantly. In contrast, the Yin watermark continues to maintain good performance within this range of LoRA numbers. In the image-to-image task, both Yin and Yang watermarks consistently maintain good performance. This is because, compared to the text-to-image task, the image-to-image task has additional constraints from both text and images, providing more restrictions and resulting in better watermarking effectiveness in the generated content.

% \paragraph{Details of the fine-tuning experiment.}
% For the stable diffusion model, we utilize about 10 main task samples to fine-tuning. To fine-tune the LoRA models, we utilize Adam optimizer and set the fine-tuning learning rate of the LoRA as 1e-4, too. 
\begin{figure}[htbp]
\centering
\subfigure[Main]{
\includegraphics[width=0.13\textwidth]{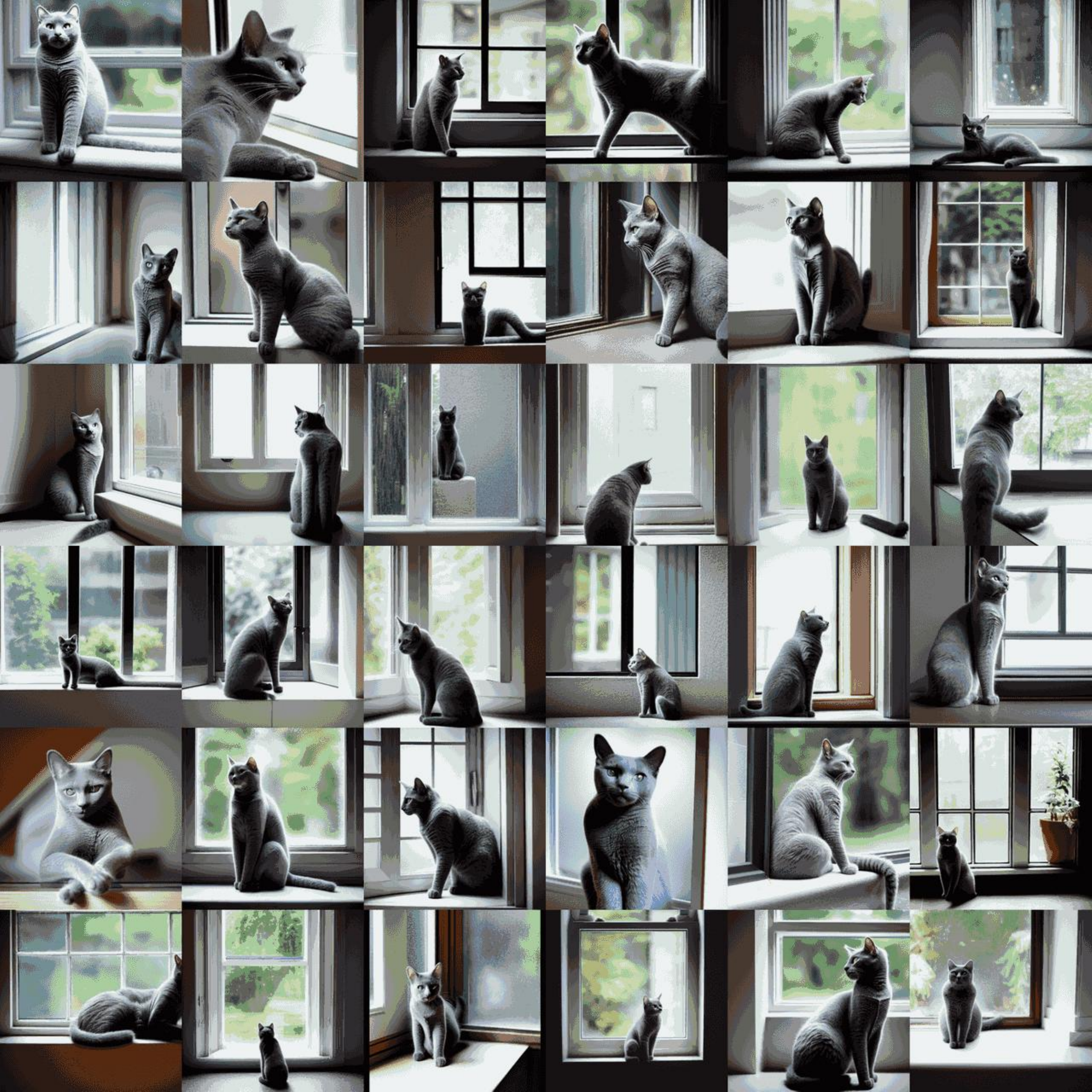}}
\subfigure[W+]{
\includegraphics[width=0.13\textwidth]{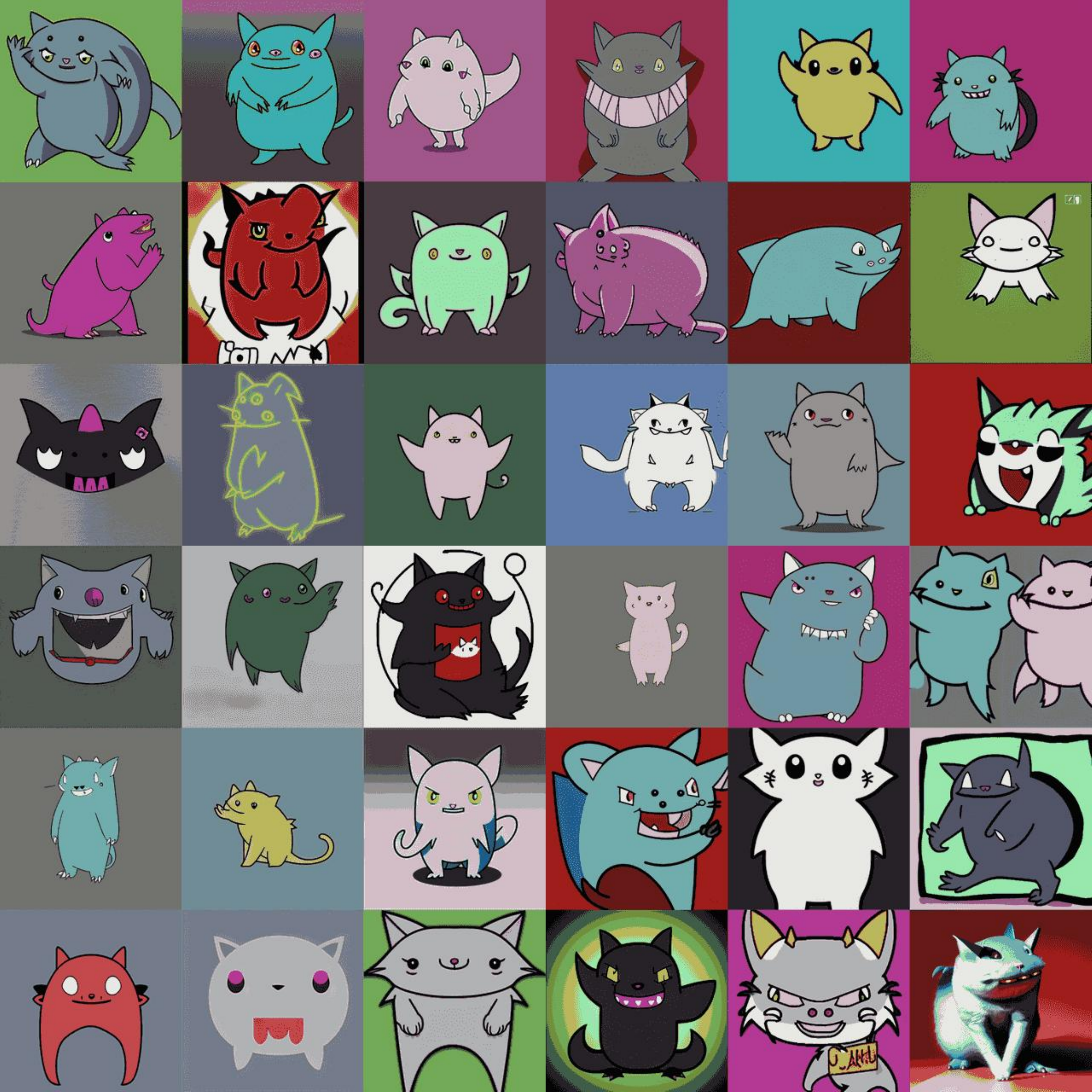}}
\subfigure[W-]{
\includegraphics[width=0.13\textwidth]{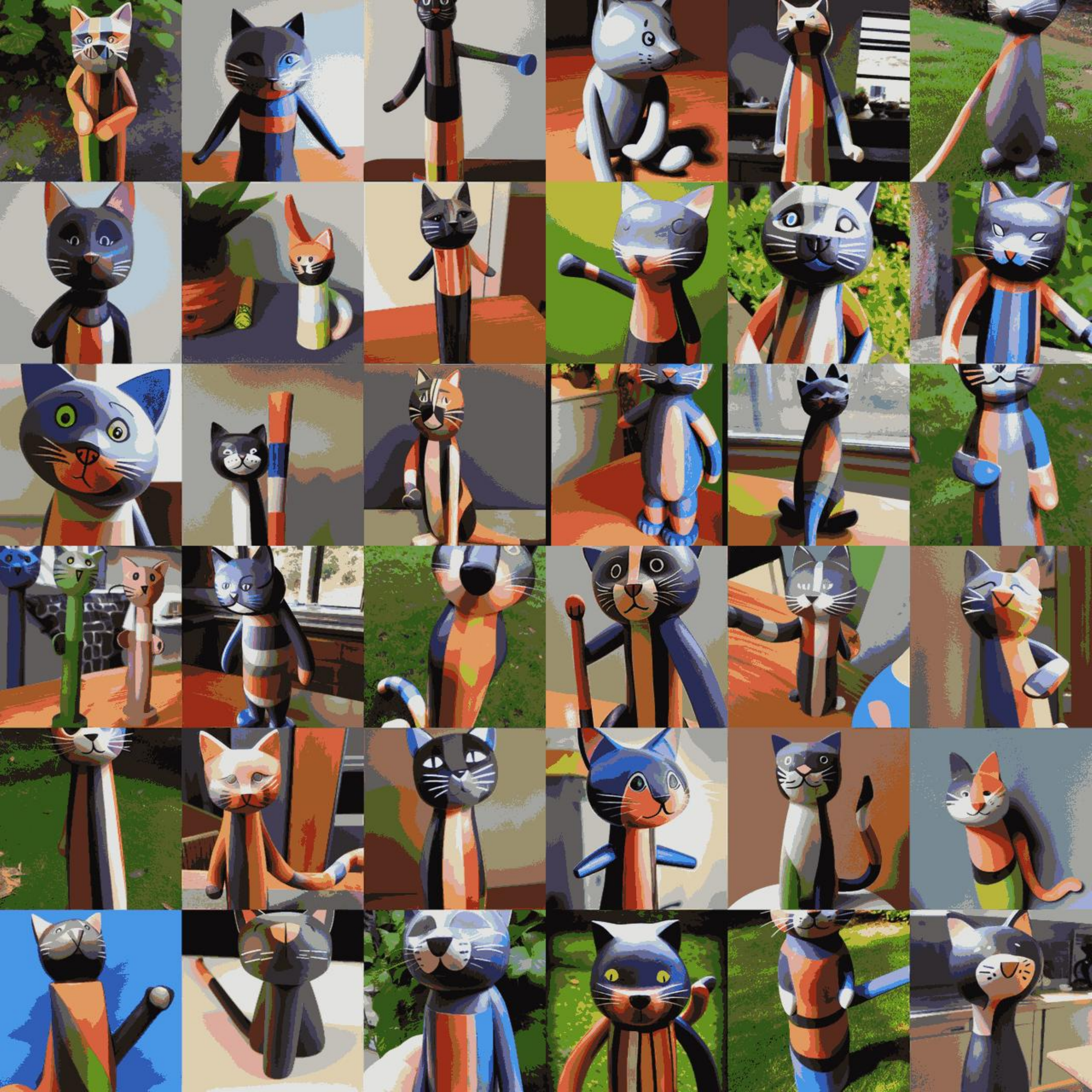}}\\
\subfigure[Main]{
\includegraphics[width=0.13\textwidth]{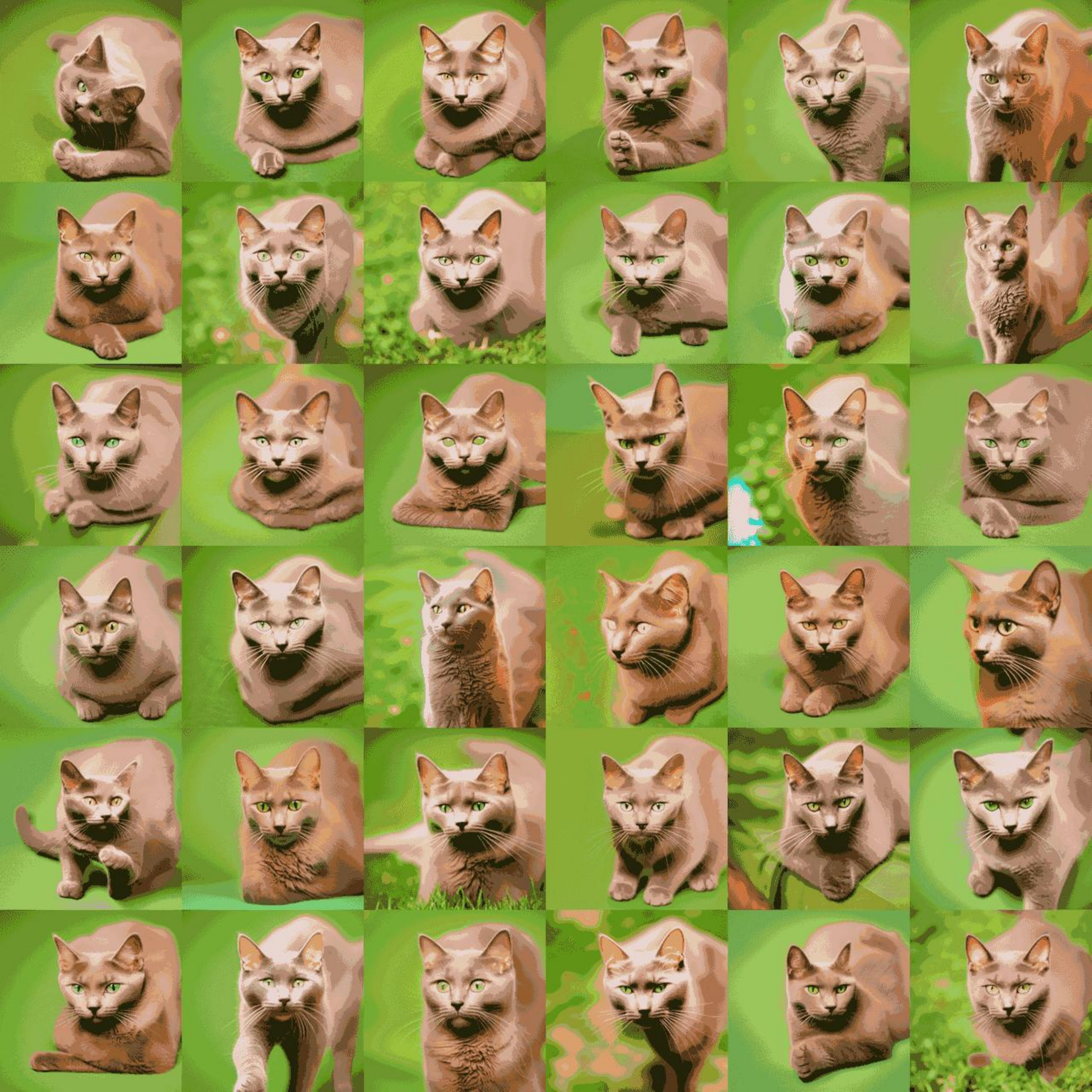}}
\subfigure[W+]{
\includegraphics[width=0.13\textwidth]{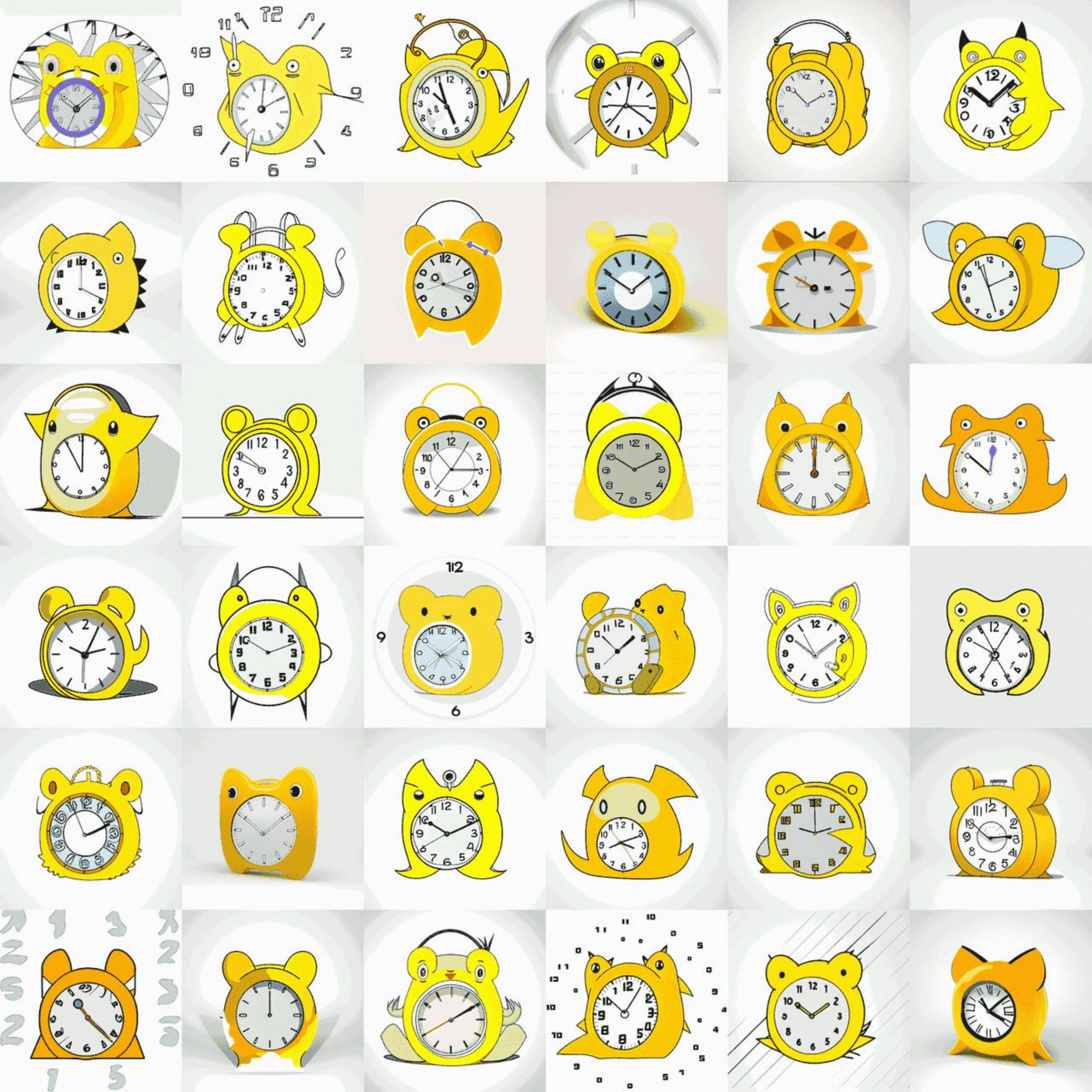}}
\subfigure[W-]{
\includegraphics[width=0.13\textwidth]{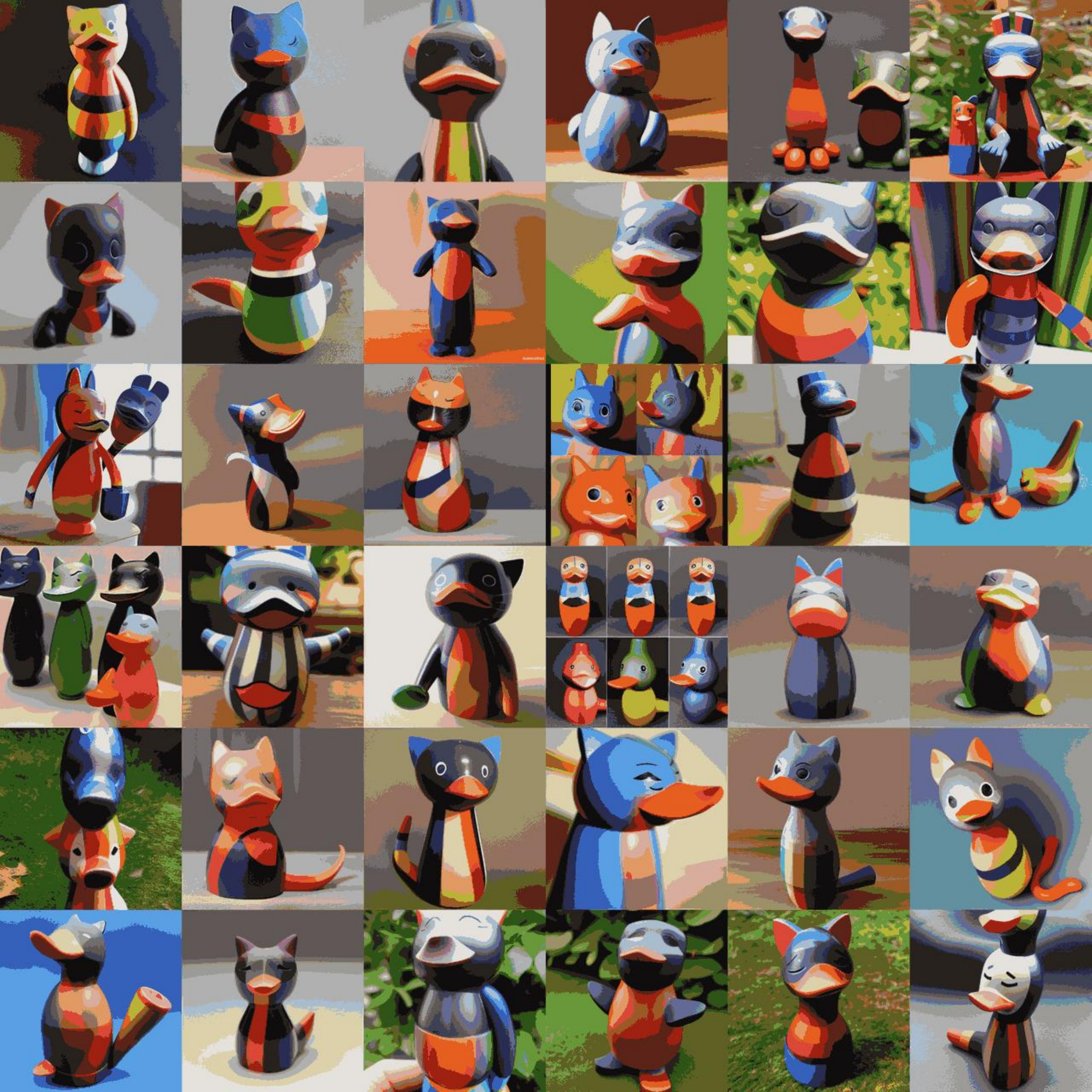}}\\
\vspace{-5pt}
\caption{Watermarked LoRA on stable diffusion model under the fine-tuning epoch of 100. The first row is in text-to-image task and the second row is in image-to-image task.}
\label{fig:retraindf}
\end{figure}

% \section*{Ethical Statement}

% There are no ethical issues.

\end{document}